\documentclass[preprint, sort&compress, 12pt]{elsarticle}
\usepackage[utf8]{inputenc}
\journal{Journal of Computational Science}
\usepackage{caption}
\usepackage{subcaption}
\usepackage{natbib}
\usepackage{graphicx}
\usepackage{amsmath}
\usepackage{booktabs}
\usepackage{bbold}
\usepackage{siunitx}
\DeclareSIUnit\RPM{rpm}
\DeclareSIUnit\BPM{bpm}
\usepackage{mathtools}
\usepackage{amsfonts}
\usepackage{latexsym}
\usepackage{isomath}
\usepackage{float}
\usepackage{bm}
\usepackage{xcolor}
\usepackage{nicefrac}
\usepackage[margin=1in]{geometry}
\usepackage{csquotes}
\usepackage[colorlinks = true,
            linkcolor = blue,
            urlcolor  = blue,
            citecolor = orange,
            anchorcolor = blue,
            pdfauthor=author]{hyperref}
\usepackage[nameinlink]{cleveref}
\usepackage{tabularx}   
\usepackage[linesnumbered,ruled]{algorithm2e}
\usepackage{multirow}
\usepackage{tikz}
\usetikzlibrary{spy}
\definecolor{inflow}{HTML}{3EA204}
\definecolor{outflow}{HTML}{FF9E01}
\usepackage[bottom, stable]{footmisc}

\newcommand{\tensor}[1]{\boldsymbol{#1}}             
\renewcommand{\vec}[1]{{\boldsymbol{#1}}}  

\newcommand{\dx}{\,\mathrm{d}\vec{x}}

\newcommand{\dsx}{\,\mathrm{d}s_{\vec{x}}}

\newcommand{\norm}[2]{{\left\lVert{#1}\right\rVert}_{#2}}

\newcommand{\figref}[1]{\figurename~\ref{#1}}
\newcommand{\tabref}[1]{\tablename~\ref{#1}}

\newcommand{\CASE}[1]{\STATE \textbf{case} #1\textbf{:} \begin{ALC@g}}
\newcommand{\ENDCASE}{\end{ALC@g}}

\newcommand{\DEFAULT}{\STATE \textbf{default:} \begin{ALC@g}}
\newcommand{\ENDDEFAULT}{\end{ALC@g}}
\newcommand{\DEFAULTLINE}[1]{\STATE \textbf{default:} }

\definecolor{darkgreen}{rgb}{0.13, 0.55, 0.13}
\newcommand{\reviewerOne}[1]{{\textcolor{red}{#1}}}
\newcommand{\reviewerTwo}[1]{{\textcolor{blue}{#1}}}
\newcommand{\reviewerThree}[1]{{\textcolor{darkgreen}{#1}}}
\newcommand{\generalChange}[1]{{\textcolor{orange}{#1}}}

\begin{document}

\begin{frontmatter}


\title{On the Incorporation of Obstacles in a Fluid Flow
 Problem Using a Navier-Stokes-Brinkman Penalization Approach}
\author[KFU]{Jana Fuchsberger\fnref{equal}}
\author[KFU]{Elias Karabelas\fnref{equal}\corref{mycorrespondingauthor}}
\cortext[mycorrespondingauthor]{Corresponding author}
\fntext[equal]{These authors contributed equally to this work.}
\ead{elias.karabelas@uni-graz.at}
\author[VIE,LBC]{Philipp Aigner}
\author[KCL]{Steven Niederer}
\author[MUG,BTMG]{Gernot Plank}
\author[VIE,LBC]{Heinrich Schima}
\author[KFU,BTMG]{Gundolf Haase}

\address[KFU]{Institute of Mathematics and
Scientific Computing, University of Graz, Heinrichstraße 36, A-8010 Graz, Austria}
\address[MUG]{Gottfried Schatz Research Center for Cell Signaling, Metabolism and Aging, Biophysics, Medical University of Graz, Neue Stiftingtalstraße 6/D04, A-8010 Graz, Austria}
\address[KCL]{Department of Biomedical Engineering, School of Biomedical Engineering \& Imaging Sciences, King’s College London, London, United Kingdom}
\address[VIE]{Center for Medical Physics and Biomedical Engineering, Medical University Vienna, Währinger Gürtel 18-20, A-1090 Vienna, Austria}
\address[LBC]{Ludwig Boltzmann Institute for Cardiovascular Research, Währinger Guertel 18-20, A-1090 Vienna, Austria}
\address[BTMG]{BioTechMed-Graz, Graz, Austria}
\begin{abstract}
Simulating the interaction of fluids with immersed moving solids
is playing an important role for gaining a better quantitative understanding 
of how fluid dynamics is altered by the presence of obstacles and,
vice versa, which forces are exerted on the solids by the moving fluid.
Such problems appear in various contexts, 
ranging from numerous technical applications such as e.g.\ turbines
to medical problems such as the regulation of cardiovascular hemodyamics by valves. 
Typically, the numerical treatment of such problems is posed 
within a fluid structure interaction (FSI) framework.
General FSI models are able to capture bidirectional interactions,
but are challenging to solve and computationally expensive.
Simplified methods offer a possible remedy by achieving better computational efficiency
to broaden the scope to demanding application problems 
with focus on understanding the effect of solids on altering fluid dynamics.  
In this study we report on the development of  a novel method for such applications. 
In our method rigid moving obstacles are incorporated in a fluid dynamics context 
using concepts from porous media theory.
Based on the Navier-Stokes-Brinkman equations 
which augments the Navier-Stokes equation with a Darcy drag term 
our method represents solid obstacles as time-varying regions 
containing a porous medium of vanishing permeability. 
Numerical stabilization and turbulence modeling is dealt with 
by using a residual based variational multiscale (RBVMS) formulation.
The additional Darcy drag term and its respective stabilization are easily accomodated 
in any existing finite-element based Navier-Stokes solver.
The key advantages of our approach 
-- computational efficiency and ease of implementation -- 
are demonstrated by solving a standard benchmark problem of a rotating blood pump
posed by the Food and Drug Administration Agency (FDA).
Validity is demonstrated by conducting a mesh convergence study 
and by comparison against the extensive set of experimental data provided for this benchmark.
\end{abstract}

\begin{keyword}
Computational fluid dynamics \sep variational multiscale methods \sep Hemodynamics \sep Penalization methods \sep large eddy simulation
\MSC[2010] 35Q30 \sep 74F10 \sep 76M10 \sep 76F65 \sep 76Z05
\end{keyword}

\end{frontmatter}
\section{Introduction}
\label{sec:introduction}
Simulating the interaction of fluids with immersed moving solids
is playing an important role for gaining a better quantitative understanding 
of how fluid dynamics is altered by the presence of obstacles and,
vice versa, which forces are exerted on the solids by the moving fluid.
Such problems appear in various contexts, 
ranging from numerous technical applications such as e.g.\ turbines
to medical problems such as the regulation of cardiovascular hemodyamics by valves. 
Typically, the numerical treatment of such problems is posed 
within a fluid structure interaction (FSI) framework.

General FSI models \reviewerTwo{\cite{2006}} are able to capture bidirectional interactions,
but are challenging to solve and computationally expensive.
However, there is a broad range of application scenarios 
that do not require to consider FSI.
For instance, if the relation of primal interest is the impact of a moving obstacle 
upon fluid flow a moving domain Navier-Stokes formulation may suffice \reviewerTwo{\cite{vanLoon2007}.}
This is the case in any situation where the motion of an obstacle can be considered to be imposed as the feedback of surrounding fluid flow upon the obstacle's motion is small.
The key mechanism at play in such a case is the solid acting as flow obstacle, 
that is, to impede any flow within the obstacle.  
For instance, the feedback of blood onto the motion of a rotating blood pump can be considered negligible.
The dynamics of the blood pump are know \emph{a priori} and can be considered as a source term to CFD.
Similarly, while a cardiac valve is moved by blood flow from a closed to an open configuration or vice versa,
this occurs within a very short transitional phase
apart from which the impact of a closed or open valvular configuration on hemodymanics 
will be of primal interest. 
In such scenarios 
where neglecting the feedback of fluid on the motion of an obstacle can be deemed
a sufficiently accurate approximation, 
simpler methods are applicable.
These may offer higher computational efficiency and, thus,
allow to broaden the scope to demanding application problems.


Capturing the behavior of obstacles and their impact on the computational model of a physical system overall 
can lead to challenging problems.
Resolving the motion of rotating objects like turbines or pumps induces topological changes in the computational domain, 
thus ruling out a number of numerical methods that rely on a fixed topology over time.  
Also, resolving the large displacements occurring during opening and closing of heart valves 
may require advanced non-trivial remeshing strategies further increasing complexity \cite{Behr2001, Chandran2010}.
This complexity and the incurring computational costs currently hinder
-- amongst other difficulties such as the patient-specific generation of valvular anatomical models --
the clinical translation of computational valve models based on fully coupled FSI formulations,
despite the significant recent advances achieved in this field, see \cite{Astorino2009,DinizdosSantos2008,Weinberg2007,vanLoon2004,McQueen2001,Maisano2005,Wenk2010,Terahara2020} \reviewerThree{as well as \cite{Antonietti2019,Alauzet2016,Zonca2018,Massing2015}} and references therein.
However, for a sufficiently accurate quantification of velocity and pressure fields 
a bidirectionally coupled FSI formulation may not be necessary.
In such cases the kinematics of an obstacle such as the rotor of a pump driven by an engine 
or the combined effect of wall motion and valvular anatomy can be imposed,
either based on geometric description of trajectories or, in the case of the heart,
by image-driven kinematic models which can be derived from tomographic imaging data \cite{Razeghi2020,Rueckert1999,Shi2013}.


%
For such applications immersed boundary methods (IBM) (and fictitious domain methods) have been proposed.
Based on the seminal work \cite{Peskin1972} the IBM has proven to be a viable approach 
that combines computational efficiency, ease of implementation and numerical stability \cite{Mittal2005}.
IB methods have also been applied in the context of heart valve modeling, see \cite{Astorino2012}, 
offering the advantage of reduced computational cost, increased robustness and stability 
with respect to classical FSI models \cite{Yao2012,Votta2013,Chnafa2014}.
However, if used in a unstructured finite element context specialized numerical routines 
for integrating over arbitrary polygonal surfaces in the computational domain are necessary.
This arises as a consequence of IBM methods that have to track surfaces of moving solids.


In this study we deal with IBM from a different perspective. 
\reviewerThree{In classical IBM, solids are embedded into the CFD grid and linked together by virtual body forces introduced in the discretized systems.}
Using the Navier-Stokes-Brinkman (NSB) equations for (moving) domains we arrive at a similar algorithm basing our approach on modeling obstacles as porous media with vanishing permeability. 
\reviewerThree{Here, the action of the solid is already included on the continuous level}.
Instead of tracking surfaces, we have to find methods for calculating suitable permeability distributions.
The domains covered by fluid and valve are blended together into a single domain
where the position of an obstacle is modeled by adapting an artificial permeability over the volume of the obstacle.
Hence, the problem of surface tracking is transformed into determining volumes of high and low permeability.
Avoiding the need for tracking actual surfaces of obstacles is advantageous 
as this facilitates the easier implementation within available CFD or FSI software.
This approach can \generalChange{be} seen as a different interpretation of feedback-forcing or direct-forcing methods, see \cite{Goldstein1993, Fadlun2000}.
Besides, rigorous convergence proofs for the NSB model exist, see for example \cite{arquis1984conditions,Khadra2000,carbou2003},
and the viability of using the NSB model in an HPC context has also been demonstrated previously 
in studies of insect flight, see \cite{Engels2016, Engels2016:2, Engels2018}.

In this study we present a modified stabilized finite element discretization of the NSB equations 
using the residual-based variational multiscale (RBVMS) formulation \cite{Bazilevs2007, Bazilevs2013}.
We introduce a suitable algorithm for determining permeability fields.
Validation results are given in the form of a mesh-convergence study 
conducted with a mock model of an arterial anastomosis where flow through branches is regulated by switching valves.
Further validation, highlighting the versatility of this approach to general moving objects, 
will be given by considering a standardized benchmark problem proposed by the FDA \cite{Malinauskas2017fda}
for which extensive experimental validation data are available.

\section{Methods}
\label{sec:methods}


\subsection{The Navier-Stokes-Brinkman Equations}
The Navier-Stokes-Brinkman (NSB) model, originating from porous media theory, can be employed with the purpose of simulating viscous flow including complex shaped solid obstacles in a fluid domain, see \cite{arquis1984conditions}, and \cite{Angot1999feb,carbou2003} for a in-depth mathematical analysis. 
\generalChange{The NSB model has a wide range of applications, i.e. insect flight \cite{Engels2016}, and geothermal engineering \cite{Blank2020}.}
In the present work, we use the NSB model including the adaptation for moving obstacles:

\begin{align}
    &\rho \left( \frac{\partial}{\partial t} \vec{u} +\vec{u} \cdot \nabla \vec{u} \right)- \nabla \cdot \tensor{\sigma}(\vec{u},p)+\frac{\mu}{K} (\vec{u}-\vec{u}_s) = 0  \quad  &\mathrm{in} \, \, \mathbb{R}^+ \times \Omega \label{eq:NSB}\\ 
    &\nabla \cdot \vec{u}=0  &\mathrm{in} \, \, \mathbb{R}^+ \times \Omega \label{eq:incomp}\\
    &\vec{u}=0 &\mathrm{on} \, \, \Gamma_{\mathrm{noslip}} \label{eq:noslip}\\
    &\tensor{\sigma}\vec{n}-\rho \beta (\vec{u}\cdot\vec{n})_- {\color{darkgreen}\vec u\color{black}} =\vec{h} &\mathrm{on} \, \, \Gamma_{\mathrm{outflow}}\label{eq:outflow}\\
    &\vec{u}=\vec{g} &\mathrm{on} \, \, \Gamma_{\mathrm{inflow}}\label{eq:inflow}\\
    &\left.\vec{u}\right|_{t=0}=\vec{u}_0
\end{align}

Here $p(\vec{x},t)$ and $\vec{u}(\vec{x},t)$ represent the fluid pressure and the flow velocity respectively, $\mu$ is the dynamic viscosity and $\rho$ the density. The volume penalization term $\frac{\mu}{K(t,\vec x)}\vec{u}(t,\vec x)$ is commonly known as \emph{Darcy drag} which is characterized by the permeability $K(t,\vec x)$.
\reviewerOne{The ability of the NSB model to reproduce Darcy's equation in low permeability regimes is investigated in \ref{sec:appendix:darcytest}.}
In \eqref{eq:NSB} the Darcy drag is modified to enforce correct no-slip conditions for obstacles moving with the obstacle velocity $\vec{u}_s(\vec{x},t)$. The fluid stress tensor $\tensor{\sigma}(\vec{u},p)$ and strain rate tensor $\tensor{\epsilon}(\vec{u},p)$ are defined as follows:
\begin{align}
    &\tensor{\sigma}(\vec{u},p)=-p \tensor{I}+ 2 \mu \, \tensor{\epsilon}(\vec{u},p),\\
    &\tensor{\epsilon}(\vec{u},p) =\frac{1}{2}\left(\nabla \vec u + \left(\nabla \vec u\right)^\top\right).
\end{align}
For $\vec h=\vec 0$, \eqref{eq:outflow} is known as a directional do-nothing boundary condition \cite{EsmailyMoghadam2011, Braack2014}, where $\vec{n}$ is the outward normal of the fluid domain, $\beta \leq \frac{1}{2}$ is a positive constant and \eqref{eq:backflowstab} is added for backflow stabilization with 
\begin{equation}
    (\vec{u}\cdot\vec{n})_-:=\frac{1}{2}(\vec{u}\cdot\vec{n}-|\vec{u}\cdot\vec{n}|). \label{eq:backflowstab} 
\end{equation}

The spatial domain $\Omega$ is split up into three \reviewerThree{time-dependent} sub-domains by means of the permeability $K(t,\vec x)$, namely the fluid sub-domain $\Omega_f(t)$, the porous sub-domain $\Omega_p(t)$ and the solid sub-domain $\Omega_s(t)$.
\begin{equation}
K(t,\vec x)= \label{eq:perm}
\begin{dcases}
K_f \rightarrow + \infty & \text{if } \vec x \in \Omega_f(t)  \\
K_p &\text{if } \vec x \in \Omega_p(t)  \\
K_s \rightarrow 0^+ &\text{if } \vec x \in \Omega_s(t)  
\end{dcases}
\end{equation}
\reviewerThree{Figure \ref{fig:volumefractions} shows an illustration of a possible distribution of the fractional volumes indicating the areas of low, high, and intermediate permeability.}
In $\Omega_f(t)$ the classical Navier–Stokes (NS) equations are recovered, while in $\Omega_p$ the full NSB equations describe fluid flowing trough a porous medium, $\vec{u}$ and $p$ are understood in an averaged sense in this context.
In $\Omega_s(t)$ the velocity $\vec{u}$ is approaching $\vec{u}_s$ and thus asymptotically satisfying the no-slip condition on the  $\Omega_f(t) / \Omega_s(t)$ interface.
Note that even in the case where $K \rightarrow 0^+$ the penalization term has a well defined limit, see \cite{Angot1999feb}.
\reviewerOne{Furthermore the convergence was shown to be of order $\mathcal{O}(K)$ for the Stokes-Brinkman equations \cite{Angot1999}.
More recently, in \cite{aguayo2020analysis} the same statement was proven for the NSB equations.
Additionally we refer to \cite{Ingram2011} for convergence results for the discretization of the NSB eqautions with finite elements.}

Looking at the NSB equations \eqref{eq:NSB} it is apparent, that they can be obtained by outright extending the NS equations with a linear penalization term. As a matter of fact, this facilitates the adaptation of existing NS solvers to feature time dependent obstacles in the fluid domain.

\subsection{Variational Formulation and Numerical Stabilization}
Following \cite{Bazilevs2007,Bazilevs2013} the discrete variational formulation of \eqref{eq:NSB} including the boundary conditions \eqref{eq:outflow}, \eqref{eq:inflow} and \eqref{eq:noslip} can be stated in the following abstract form:

Find $\vec{u}^h \in [\mathcal{S}^1_{h,\vec{g}}(\mathcal{T}_\mathrm{N})]^3, \, p^h \in \mathcal{S}^1_h (\mathcal{T}_\mathrm{N})$ such that, for all $\vec{w}^h \in [\mathcal{S}^1_{h,\vec{0}}(\mathcal{T}_\mathrm{N})]^3 $ and for all $q^h \in \mathcal{S}^1_h(\mathcal{T}_\mathrm{N})$
\begin{equation}
    A_{\mathrm{NS}}(\vec{w}^h,q^h;\vec{u}^h,p^h) + S_\mathrm{RBVMS}(\vec{w}^h,q^h;\vec{u}^h,p^h) = F_{\mathrm{NS}}(\vec w_h)
\end{equation}
with the bilinear form of the NSB equations
\begin{equation}
\begin{aligned}
    &A_{\mathrm{NS}}(\vec{w}^h,q^h;\vec{u}^h,p^h)=\\
    & \int\limits_{\Omega} \vec{w}^h \cdot \left[ \rho \left( \frac{\partial \vec{u}^h}{\partial t}+ \vec{u}^h \cdot \nabla \vec{u}^h \right)  +  \frac{\mu}{K} (\vec{u}^h -\vec{u}_s^h) \right]+ \tensor{\epsilon}(\vec{w}^h) : \tensor{\sigma}(\vec{u}^h,p^h) \, \dx \\ 
    & -\int\limits_{\Gamma_{\mathrm{outflow}}} \rho \beta (\vec{u}^h\cdot\vec{n})_- \vec{w}^h \cdot \vec{u}^h \, \dsx + \int_{\Omega} q^h \nabla \cdot \vec{u}^h \, \dx  \label{eq:varNSB},
\end{aligned}
\end{equation}
the bilinear form $S_\mathrm{RBVMS}$, which will be explained later in Equation~\eqref{eq:bilinearform:VMS}, and the right hand side contribution
\begin{align}
    F_\mathrm{NS} = \int\limits_{\Gamma_\mathrm{outflow}} \vec h \cdot \vec w_h\, \dsx.
\end{align}
We use standard notation to describe the finite element function space $\mathcal S^1_{h,*}(\mathcal T_N)$ as a conformal trial space of piece-wise linear, globally continuous basis functions $w_h$ over a decomposition $\mathcal T_N$ of $\Omega$ into $N$ finite elements constrained by $w_h=*$ on essential boundaries. 
The space $S^1_h(\mathcal T_N)$ denotes the same space without constraints.
For further details we refer to \cite{brenner2007mathematical, steinbach2007numerical}.
As previously described in \cite{Karabelas2018} we utilize the residual based variational multiscale (RBVMS) formulation as proposed in \citep{Bazilevs2007,Bazilevs2013}, providing turbulence modeling in addition to numerical stabilization.
In the following we give a short summary of the changes necessary to use RBVMS methods for the NSB equations.
Briefly, the RBVMS formulation is based on a decomposition of the solution and weighting function spaces into coarse and fine scale subspaces and the corresponding decomposition of the velocity and the pressure and their respective test functions. 
Henceforth the fine scale quantities and their respective test functions shall be denoted with the superscript $'$. 
We assume $\vec{u}_s =\vec{u}_s^h$ , quasi-static fine scales ($\frac{\partial \vec{u}'}{\partial t} = 0$), as well as $\frac{\partial \vec{w}^h}{\partial t} = 0$, $\vec{u}' = \vec 0$  on $\partial \Omega$ and incompressibility conditions for $\vec{u}^h$ and $\vec{u}'$.  
The fine scale pressure and velocity are approximated in an element-wise manner by means of the residuals $\vec{r}_M$ and $r_C$.
\begin{align}
\vec{u}' &= -\frac{\tau_{\mathrm{SUPS}}}{\rho} \; \vec{r}_M(\vec{u}^h,p^h) \label{app_u}\\
p' &= -\rho \;\nu_{\mathrm{LSIC}}\; r_C(\vec{u}^h)\label{app_p}
\end{align}

The residuals of the NSB equations and the incompressibility constraint are:
\begin{align}
\vec{r}_M(\vec{u}^h,p^h) &= \rho \partial_t \vec{u}^h + \rho \vec{u}^h \cdot \nabla \vec{u}^h - \nabla \cdot \tensor{\sigma}(\vec{u}^h,p^h)+ \frac{\mu}{K} (\vec{u}^h-\vec{u}^h_s) \label{eq:Mres}\\
r_C(\vec{u}^h) &= \nabla \cdot \vec{u}^h\label{eq:Cres}
\end{align}

Taking all assumptions into consideration and employing the scale decomposition followed by partial integration yields the bilinear form of the RBVMS formulation $S_\mathrm{RBVMS}(\vec{w}^h,q^h;\vec{u}^h,p^h)$,
\begin{align}
\label{eq:bilinearform:VMS}
S_\mathrm{RBVMS}(\vec{w}^h,q^h;\vec{u}^h,p^h)&:=\\
\label{eq:bilinarform:VMS:1}+ &\sum_{\Omega_e \in \mathcal T_N} \int_{\Omega_e} \tau_\mathrm{SUPS}\left( \vec{u}^h \cdot \nabla \vec{w}^h +  \frac{1}{\rho} \nabla q^h - \frac{\nu}{K} \vec{w}^h\right) \vec{r}_M(\vec{u}^h,p^h) \, \dx  \\
\label{eq:bilinarform:VMS:2}+ &\sum_{\Omega_e \in \mathcal T_N} \int_{\Omega_e} \rho \, \nu_\mathrm{LSIC}\, \nabla \cdot \vec{w}^h \, r_C(\vec{u}^h) \, \dx  \\
\label{eq:bilinarform:VMS:3}- &\sum_{\Omega_e  \in \mathcal T_N} \int_{\Omega_e} \tau_\mathrm{SUPS} \, \vec{w}^h \cdot \left(\vec{r}_M(\vec{u}^h,p^h) \cdot \nabla \vec{u}^h  \right) \dx  \\
\label{eq:bilinarform:VMS:4}- &\sum_{\Omega_e \in \mathcal T_N} \int_{\Omega_e}  \frac{\tau_{\mathrm{SUPS}}^2}{\rho} \nabla  \vec{w}^h : (\vec{r}_M(\vec{u}^h,p^h) \otimes \vec{r}_M(\vec{u}^h,p^h)) \dx.
\end{align}
The residuals \eqref{eq:Mres} and \eqref{eq:Cres} are evaluated for every element $\Omega_e \in \mathcal T_N$.
Following \citep{PhDPauli} the stabilization parameters $\tau_\mathrm{SUPS}$ and $\nu_\mathrm{LSIC}$ are defined as:

\begin{align}
\tau_{\mathrm{SUPS}} &:= \left( \frac{4}{\Delta t ^2} + \vec{u}^h \cdot \tensor{G} \vec{u}^h + \left(\frac{\nu}{K}\right)^2+C_I \nu^2 \tensor{G} : \tensor{G} \right)^{-\frac{1}{2}}\\
\nu_{\mathrm{LSIC}} &:= \frac{1}{\text{tr}(\tensor{G}) \, \tau_{\mathrm{SUPS}}}
\end{align}

Here $\tensor{G}$ is the three dimensional element metric tensor defined per finite element as
\begin{align*}
    \tensor G \lvert_{\tau_l} &:= \tensor J_l^{-1} \tensor J_l^{-\top},
\end{align*}
with $\tensor J_l$ being the Jacobian of the transformation of the reference element to the physical finite element $\tau_l \in \mathcal T_N$, $\Delta t$ denotes time step size and $C_I$ is a positive constant \generalChange{derived from an element-wise inverse estimate \cite{Bazilevs2013,Bazilevs2007}.} \reviewerThree{Specific values of $C_I$ depend on the finite element type, e.g.: tetrahedrons, prisms, etc., as well as polynomial degree see \cite{Harari1992}. For lowest order finite elements an arbitrary value can be chosen and different types of values have been reported in literature, see \cite{PhDPauli} for a more in-depth discussion. We chose a value of $C_I=\num{30}$ in agreement with the one reported in \cite{Forti2015}.}

\subsection{Obstacle Representation}
\label{sec:methods:obs_algo}
\begin{figure}[htp]
    \centering
    \begin{picture}(210,180)
    \put(0,0){ \includegraphics[width=9cm]{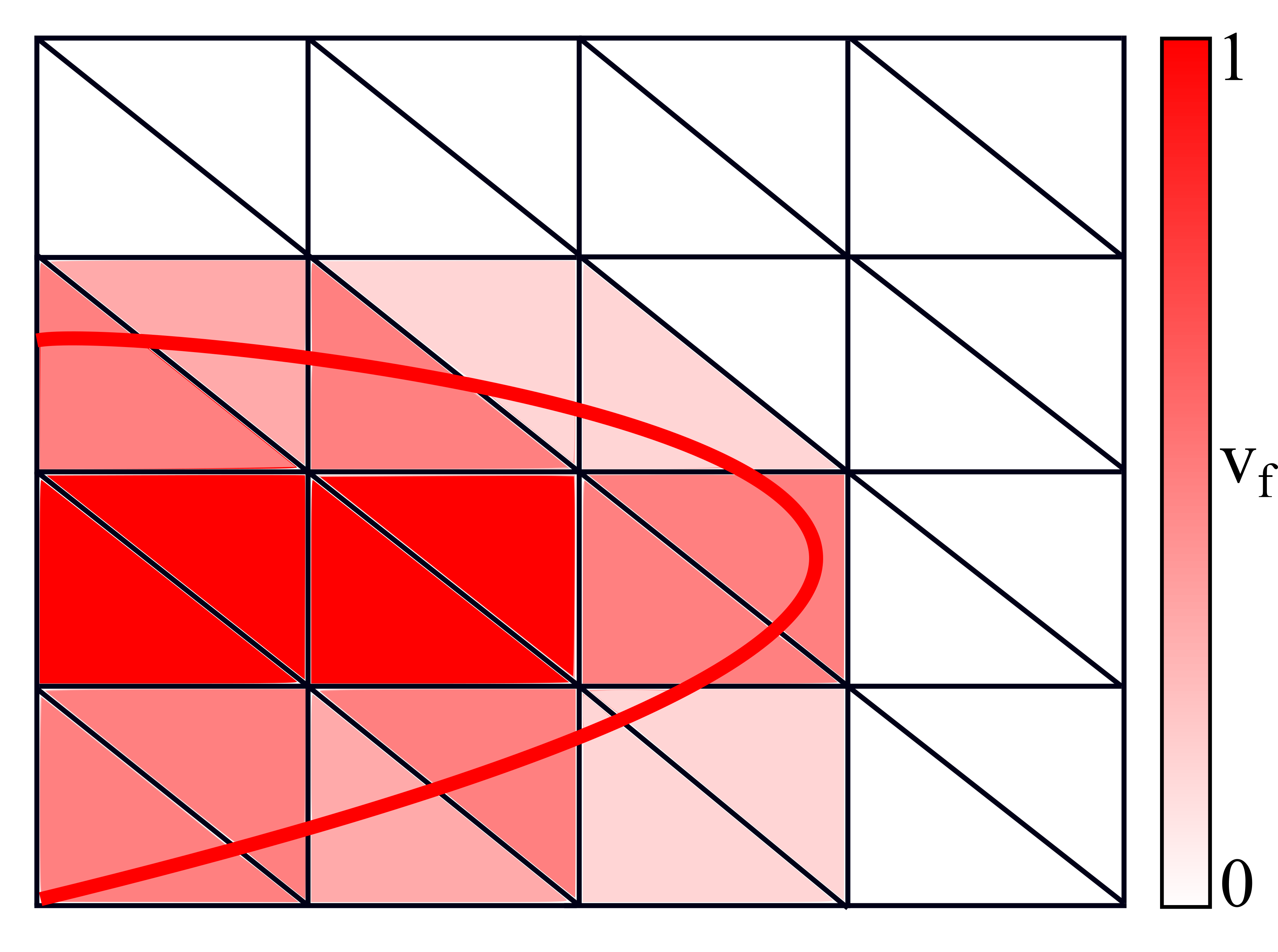}}
    \begin{Large}
    \put(70,60){$\Omega_s$}
    \put(180,150){$\Omega_f$}
    \put(20,107){$\Omega_p$}
    \end{Large}
    \end{picture}
    \caption{Schematic representation of the $v_f$ distribution associated to an obstacle,which is represented by the red line, at a fixed point time t. }
    \label{fig:volumefractions}
\end{figure}
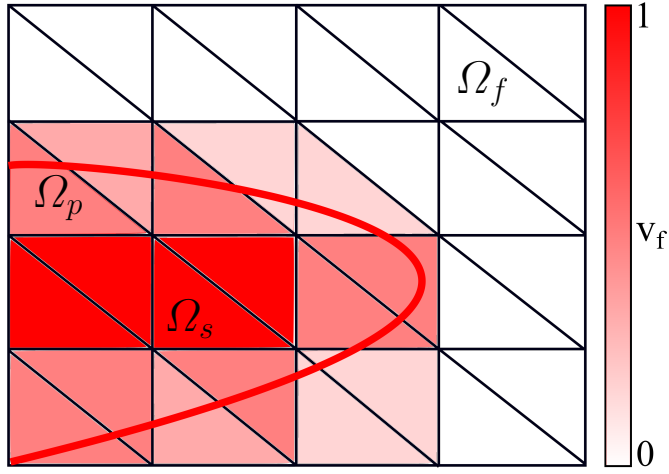

The remaining problem is to determine a permeability distribution $K$ 
that is suitable to model given flow obstacles. 
This task is solved by representing the obstacles using triangular surface meshes 
followed by element-wise calculation of the partial volume covered by the obstacle. 
In the first step, all nodes within the obstacle are identified using the ray casting algorithm \cite{Moeller1997, Haines1994}. 
Subsequently, all elements are split into three categories and receive a corresponding volume fraction value $v_f$, describing the partial volume covered by the obstacle:
\begin{itemize}
    \item Elements fully covered by the obstacle lie in $\Omega_s$, consequently $v_f=1$.
    \item Elements outside the obstacle lie in $\Omega_f$ and obtain $v_f=0$.
    \item Elements that are split by the element surface correspond to elements in $\Omega_p$, hence
    \begin{equation}\label{eq:frac_vol}
        v_f=\frac{V_{in}}{V_{tot}}
    \end{equation}
    where $V_{in}$ denotes the element volume covered by the obstacle and $V_{tot}$ is the total element volume.
\end{itemize} 
This procedure is carried out for every time step 
and yields a time-dependent, element-based volume fraction distribution $v_f(t,\tau)$, 
that serves as a basis to provide a suitable permeability distribution, see \figref{fig:volumefractions}.
In this work we define $\frac{1}{K(t,\tau)} :=\frac{v_f(t,\tau)}{\hat{K}}$ with $\hat{K}$ being a fixed penalization factor, e.g. $\hat{K} = 10^{-6}$.
All permeability distributions in this work have been generated 
using the open-source software \emph{Meshtool}\footnote{\url{https://bitbucket.org/aneic/meshtool/src/master/}}, see \cite{Neic2020}.
For details we refer to Algorithm\,\ref{algo_volfracs} in \ref{sec:appendix:pseudocode} summarizing the main workflow as implemented in \emph{Meshtool}.
For cyclic motions such as the movement of a rotary blood pump or a turbine the time-dependent permeability distribution can be determined in a preprocessing step.

\color{red}
\subsection{Evaluating Forces in the NSB equations}
\label{sec:fluid_forces}
The NSB equations offer a convenient way to evaluate surface forces.
Following \cite{Wan2004} we can use the volume fraction distribution $v_f$ \eqref{eq:frac_vol} to define an approximation to the surface normal vector as
\begin{equation}\label{eq:normal_obs}
    \widetilde{\vec n} := -\nabla v_f.
\end{equation}
This definition allows to replace surface integrals with volume integrals as $\widetilde{\vec n}$ is zero everywhere except close to the surface and an approximation to the normal close to the surface \cite{Brackbill1992}.
With that we can define approximations to standard forces:
\begin{itemize}
    \item Fluid force
    \begin{equation}\label{eq:drag}
        \vec F = \int_{\partial \Omega_s} \tensor{\sigma} \vec n \dsx \approx \int_{\Omega} \tensor{\sigma} \widetilde{\vec n} \dx
    \end{equation}
    \item Torque
    \begin{equation}\label{eq:torque}
        \vec T = \int_{\partial \Omega_s} \vec r(\vec x) \times \tensor{\sigma} \vec n \dsx \approx \int_{\Omega} \vec r(\vec x) \times \tensor{\sigma} \widetilde{\vec n} \dx
    \end{equation}
\end{itemize}
Accuracy of these approximations are dependent on $K$ as well as the mesh size $h$, see \cite{Angot1999, Angot1999feb, Khadra2000}.
\color{black}
\subsection{Numerical Solution Strategy}
\label{sec:num_ex}
CFD simulations often require highly resolved meshes and small time step sizes. 
Thus, efficient and massively parallel solution algorithms for the linearized system of equations become an important factor to deal with the resulting computational load. 
Spatio-temporal discretization of all PDEs and the solution of the arising systems of equations relied upon the Cardiac Arrhythmia Research Package (CARPentry), see \cite{Vigmond2003}.
For temporal discretization of the NSB equations we used the \reviewerOne{implicit} generalized-$\alpha$ method with a spectral radius $\rho_\infty \in [0,0.5]$.
\reviewerOne{It has been demonstrated in \cite{Jansen2000} that, for linear problems, second-order accuracy, unconditional stability, and optimal high frequency damping can be achieved.
More recently, it was shown in \cite{Liu2020} that the same holds for the Navier-Stokes equations.}
After discretization in space as described in Section \ref{sec:methods} and temporal discretization using the generalized-$\alpha$ integrator we obtain a nonlinear algebraic system to solve for advancing time from timestep $t_n$ to $t_{n+1}$.
A quasi inexact Newton-Raphson method is used to solve this system with linearization approach similar to \cite{Bazilevs2013} adapted to the NSB equations.
\reviewerThree{More specific, following \cite{Bazilevs2013} we dropped the terms \eqref{eq:bilinarform:VMS:3}--\eqref{eq:bilinarform:VMS:4} from Jacobian evaluation, approximated every occurrence of $\vec u_h^{k+1} \nabla \vec u_h^{k+1}$ or $\vec u_h^{k+1} \nabla \vec w_h$ as $\vec u_h^{k} \nabla \vec u_h^{k+1}$ and $\vec u_h^{k}\nabla \vec w_h$, and lagged values for $\tau_\mathrm{SUPS}$ and $\nu_\mathrm{LSIC}$ by one Newton iteration.}
\generalChange{Having that, at} each iteration a  block system of the form
\begin{align*}
    \begin{bmatrix}
     \tensor K_h & \tensor B_h \\
     \tensor C_h & \tensor D_h
    \end{bmatrix}
    \begin{bmatrix}
      \Delta \vec u \\
      \Delta \vec p
    \end{bmatrix}
    = -
    \begin{bmatrix}
     -\vec{R}_{\mathrm{upper}} \\
     -\vec{R}_{\mathrm{lower}}
    \end{bmatrix},
\end{align*}
\generalChange{must be} solved with $\tensor K_h$, $\tensor B_h$, $\tensor C_h$, and $\tensor D_h$ denoting the Jacobian matrices, $\Delta \vec u$, $\Delta \vec p$ representing the velocity and pressure updates and $\vec R_\mathrm{upper}$, $\vec R_\mathrm{lower}$ indicating the residual contributions.
In this regard we use the flexible generalized minimal residual method (fGMRES) and efficient preconditioning based on the \texttt{
PCFIELDSPLIT}\footnote{\url{https://www.mcs.anl.gov/petsc/petsc-current/docs/manualpages/PC/PCFIELDSPLIT.html}} package from the library \emph{PETSc} \cite{petsc-web-page, petsc-user-ref, petsc-efficient} and the incorporated suite \emph{HYPRE BoomerAMG} \cite{Henson2002}.
By extending our previous work \cite{Augustin2016, Karabelas2018, Karabelas2019} we implemented the methods in the finite element code \emph{Cardiac Arrhythmia Research Package} (CARPentry) \cite{Vigmond2003, Vigmond2008}.
\reviewerTwo{Parallel performance and scalability of CARPentry has been previously investigated in \cite{Augustin2016} and in \cite{Karabelas2018} for the CFD module. Additionally, in \ref{sec:mov_sphere} we included new scaling data.}
\section{Numerical Examples}

\subsection{\reviewerOne{Moving Sphere Benchmark}}
\label{sec:mov_sphere}
\color{red}
To assess the spatial accuracy of our method we performed a simple benchmark problem featured in \cite{Udaykumar2001,Seo2011}.
A sphere of diameter $D$ is submerged into a cube of dimensions $4D$ and oscillates along the $x$-axis according to \begin{equation}
    x_c(t) := x_c(0) + A_0 \left(1 - \cos\left(2\pi f_0 t\right)\right),
\end{equation}
with $x_c(t)$ being the $x$-coordinate of the sphere's center, $A_0$ being the amplitude, and $f_0$ frequency.
The velocity of the sphere is deduced as $v_c(t) := 2\pi f_0 A_0 \sin(2 \pi f_0 t)$.
Following \cite{Seo2011} we set $A_0=\num{0.125}$, $f_0=\num{0.1}$.
The simulation time was chosen as $[0,10]\si{\second}$, diameter of the sphere as $D=\SI{1}{\meter}$, density $\rho=\SI{1}{\kilogram\per\cubic\meter}$, and viscosity $\mu=\SI{1e-3}{\pascal\second}$.
This corresponds to a Reynolds number of $\num{78.54}$.
We generated successively refined hexahedral meshes with $N=6859$ (very coarse), $N=117649$ (coarse), $N=970299$ (medium), and $N=7880599$ (fine) elements.
As previously noted, convergence w.r.t. volume penalization is of order $\mathcal{O}(K)$.
Thus for assessing spatial accuracy we fix $K=\num{1e-10}$.
Similarly, for the second-order generalized-$\alpha$ time integrator we chose $\rho_\infty = \num{0.5}$ and a time step size of $\Delta t = \SI{1}{\milli\second}$.
Calculations were performed on the \emph{Vienna Scientific Cluster 4} (vsc4) as described in \ref{sec:num_ex} with a relative Newton-tolerance per time step of $\num{1e-6}$.
This resulted on average in fourNewton iterations per time step with an average of $\num{13.8}$ fGMRES iterations per Newton iteration for each discretization level.
From this simulations we calculated the fluid forces as described in \ref{sec:fluid_forces} and similar to \cite{Seo2011} evaluated the $2\delta$-discontinuity for the drag and lift force $C_{2\delta}^n := \lvert C^{n+1} - 2 C^n + C^{n-1} \rvert$ with $C$ being either the $x$-coordinate (drag) or $y$-coordinate of the force vector $\vec F$.
\figref{fig:mov_sphere}b) depicts a log-log plot of best-fit curves w.r.t. mesh size of the root mean square of $C_{2\delta}$.
Additionally we compared the $L_2$ norms of $\vec u$ and $p$ against the solutions on the finest grid and averaged the errors over all time steps to extract an average estimated order of convergence (\emph{eoc}).
\autoref{tab:movsphere_conv} and \figref{fig:mov_sphere}b) show the results obtained in this experiment.
Further, \figref{fig:mov_sphere}a) shows a strong scaling graph of the average time for performing one fGMRES iteration on the finest grid plotted against the number of MPI threads, and \figref{fig:mov_sphere}c) and d) show velocity streamlines colored with pressure at $t=\SI{2.5}{\second}$ amd $t=\SI{5}{\second}$.
{
\begin{table}[htp]
\color{red}
\caption{\textcolor{red}{Convergence for the moving sphere benchmark. Norms are taken as averages over all time steps.}}\label{tab:movsphere_conv}
\begin{center}
\begin{tabular}{l|cccc}
\toprule
 & $\norm{\vec u - \vec u_h}{2,\mathrm{avg}}$ & \emph{eoc} & $\norm{p-p_h}{2,\mathrm{avg}}$ & \emph{eoc}\\ 
\midrule
\textbf{very coarse} & $\num[scientific-notation=true]{0.0211}$ & & $\num[scientific-notation=true]{0.0000821}$ & \\ 
\textbf{coarse} & $\num[scientific-notation=true]{0.00654}$ & $\num{1.69}$ & $\num[scientific-notation=true]{0.00003297}$ & $\num{1.31}$\\ 
\textbf{medium} & $\num[scientific-notation=true]{0.00176}$ & $\num{1.89}$ & $\num[scientific-notation=true]{0.00001093}$ & $\num{1.59}$\\ 
\bottomrule
\end{tabular} 
\end{center}
\end{table}
}

{
\begin{figure}[htbp]
    \centering
    \includegraphics[height=5in, keepaspectratio]{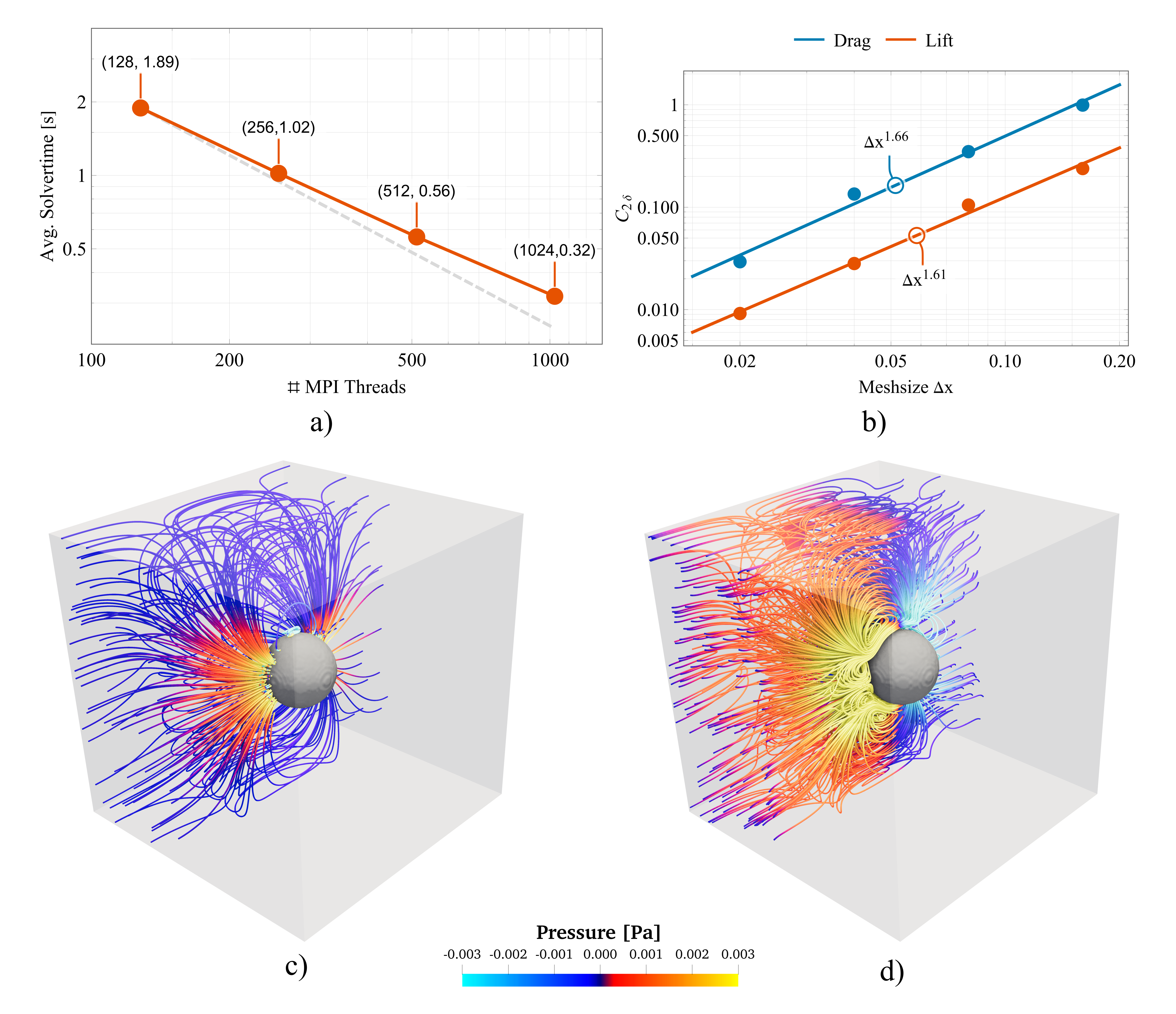}
    \caption{\textcolor{red}{Results for the moving sphere benchmark. Subfigure a) shows a strong scaling plot of the average time of performing one fGMRES iteration, subfigure b) shows the best-fit curves for the $2\delta$-discontinuity for drag and lift, subfigure c) shows the instantaneous velocity streamlines at $t=\SI{2.5}{\second}$, and subfigure d) shows the instantaneous velocity streamlines at $t=\SI{5}{\second}$ both colored with pressure.}}
    \label{fig:mov_sphere}
\end{figure}
}
\color{black}
\subsection{\generalChange{Torus Benchmark}}
\begin{figure}[h]
    \centering
    \includegraphics[width=\linewidth, keepaspectratio]{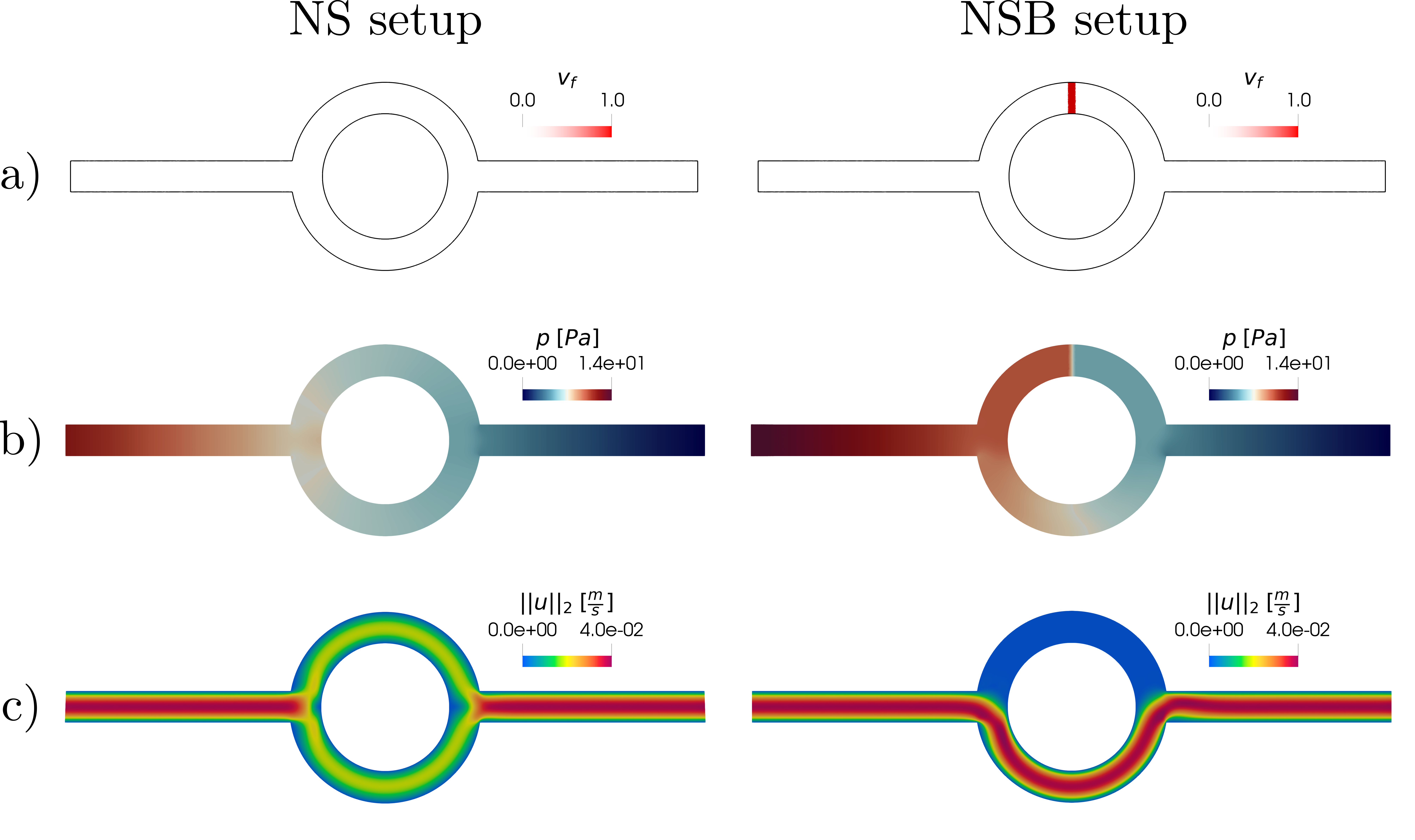}
    \caption{a) Three dimensional torus geometry with a total length of \SI{8}{\centi\metre} and a pipe diameter of \SI{0.4}{\centi\metre} featuring two different setups. NS setup: A pure Navier-Stokes simulation. NSB setup: A Navier-Stokes-Brinkman simulation including an obstacle with a thickness of \SI{0.1}{\metre} blocking the upper pipe of the torus geometry. b) Pressure field for NS and NSB setup. c) L2 norm of the velocity field for NS and NSB setup. }
    \label{fig:mcsetup}
\end{figure}

\generalChange{For emphasizing the differences between classic mesh convergence studies and LES type convergence studies, we conducted an experiment using a three dimensional torus geometry} representing a mock model of a human artery tract. 
Two different setups were used, see \figref{fig:mcsetup},
to compare mesh convergence properties of the NSB model to those of the pure NS model. 
For both setups three different finite element mesh resolutions, 
coarse (ref$_0$), medium (ref$_1$) and fine (ref$_2$) were used, see \tabref{tab:FEres}.
All simulations were carried out using a density of $\rho=\SI{1060}{\kilo\gram\per\metre\cubed}$ and a viscosity of $\mu=\SI{4.0e-3}{\pascal\second}$.
Two different cases were considered: A laminar case with a a constant inflow flow rate of $\SI{2.4e-7}{\metre\cubed\per\second}$, 
resulting in a Reynolds number of $\mathrm{Re} = 20$. A pulsatile case driven by a sinusoidal inflow with a peak flow rate of $\SI{4.7e-5}{\metre\cubed\per\second}$ and a period of $\SI{60}{\BPM}$ yielding a Reynolds number of $\mathrm{Re} = 4000$.
Both cases were simulated for a total duration of $\SI{4}{\second}$, which corresponds to 4 periods of the pulsatile inflow.
The time step size was $\SI{5e-4}{\second}$ in the laminar case and $\SI{1.25e-4}{\second}$ in the pulsatile case. 

To ascertain that numerical errors due to the iterative solution is negligible
relative to errors due to mesh discretization,
in every Newton-Raphson step iterations were carried out until full convergence was achieved.
As convergence criterion for Newton's method we used the relative residual of the nonlinear system.
Full convergence was declared when the relative residual was smaller than \SI{1.0e-5}{}.
As a first step, mesh convergence was assessed by calculating the ratio $R$ of solution changes between mesh refinement levels, as proposed in \citep{Stern2001}, see \eqref{eq:R}.
For a detailed discussion about the methods used to evaluate mesh convergence see \ref{sec:appendix:meshconv}.
Convergence behavior can be classified into three categories depending on the ratio $R$:
\begin{equation}
    \begin{array}{cl}\label{eq:Rinterpretation}
    R \in (0,1)  & \hspace{1cm}\text{monotonic convergence} \\
    R \in (-1,0) & \hspace{1cm}\text{oscillatory convergence}\\
    \text{else}  & \hspace{1cm}\text{divergence}.
    \end{array}
\end{equation}
Considering the total number of nodes from \tabref{tab:FEres} using \eqref{eq:reff}, or straightforward calculation from the average edge length yields an effective refinement ratio of $r_\mathrm{eff} \approx 2$.

\begin{table}[htp]
\caption{Different FE mesh resolutions used in mesh convergence study}\label{tab:FEres}
\begin{center}
\begin{tabular}{c|ccc}
\toprule
 & $\mathrm{ref}_0$ & $\mathrm{ref}_1$ & $\mathrm{ref}2$ \\ 
\midrule
\textbf{\# nodes} & 18844 & 126841 & 926394 \\ 
\textbf{\# elements} & 79025 & 632200 & 5057600 \\ 
\textbf{average edge length} [\si{\centi\metre}] & 0.054 & 0.028 & 0.014\\ 
\bottomrule
\end{tabular} 
\end{center}
\end{table}

To get a good overview, we chose to study convergence in terms of five different physical quantities:
\begin{itemize}
    \item The $L_2$ norm of the flow velocity, $\norm{\vec{u}(t,\vec{x})}{2}$.
    \item The pressure field, $p(t,\vec{x})$.
    \item The maximum pressure, $p_\mathrm{max}(t)=\norm{p(t)}{\infty}$.
    \item The pressure drop across the obstacle, $\Delta p(t)$.
    \item The flow \reviewerTwo{through} the obstacle, $q(t)=\int_A \vec{u}(t,\vec{x}) \cdot \vec{n}_A\,\dsx$ with $A$ being a surface\footnote{The surface used to calculate the flow \generalChange{through} the obstacle was placed closely behind the obstacle in flow direction. This is necessary, because the velocity inside the obstacle has to be interpreted as an averaged quantity (see \cref{sec:methods}) and a flow calculation in the solid or porous domain may not yield a flow in the traditional sense.} and $n_A$ its outer unit normal vector.
\end{itemize}

\begin{figure}[htp]
\begin{subfigure}{\linewidth}
\centering
\includegraphics[width=\linewidth,keepaspectratio,trim=0cm 0.25cm 0cm 0cm, clip]{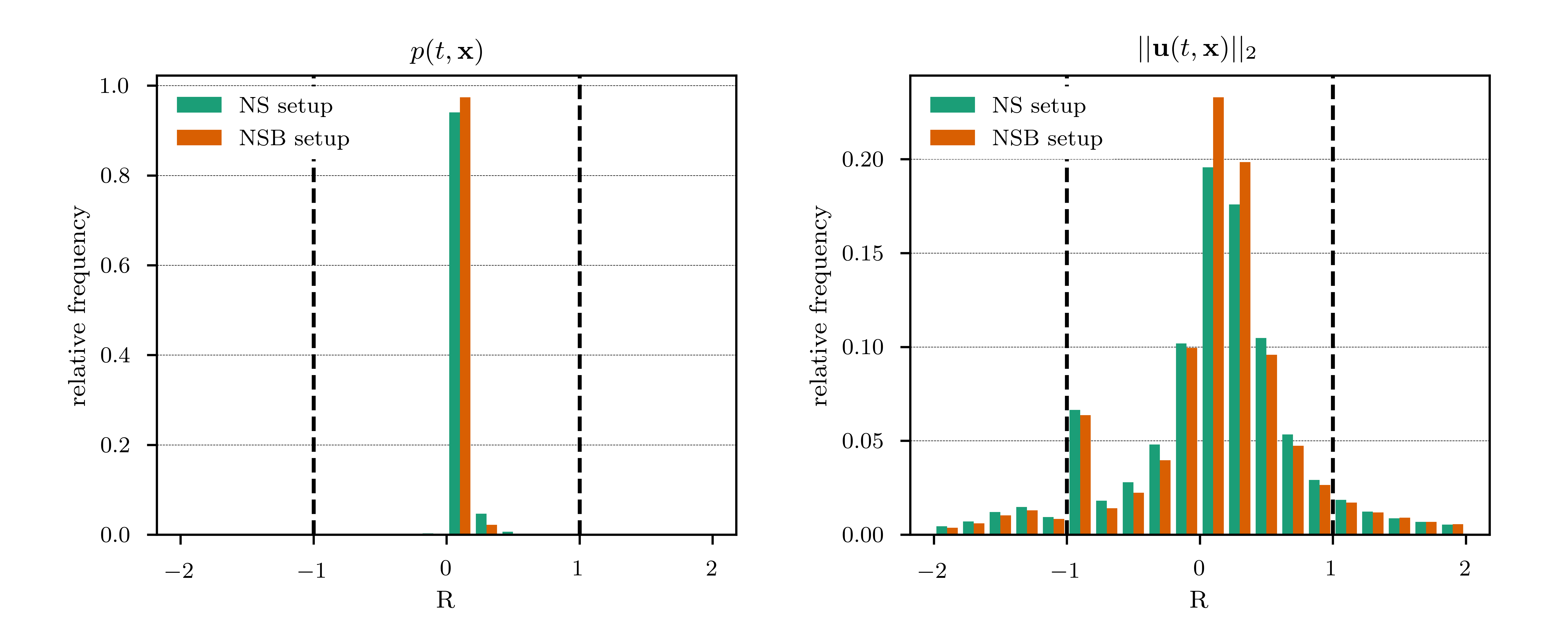} 
\caption{Laminar Case ($Re=20$)}
\end{subfigure}
\begin{subfigure}{\linewidth}
\centering
\includegraphics[width=\linewidth,keepaspectratio,trim=0cm 0.25cm 0cm 0cm, clip]{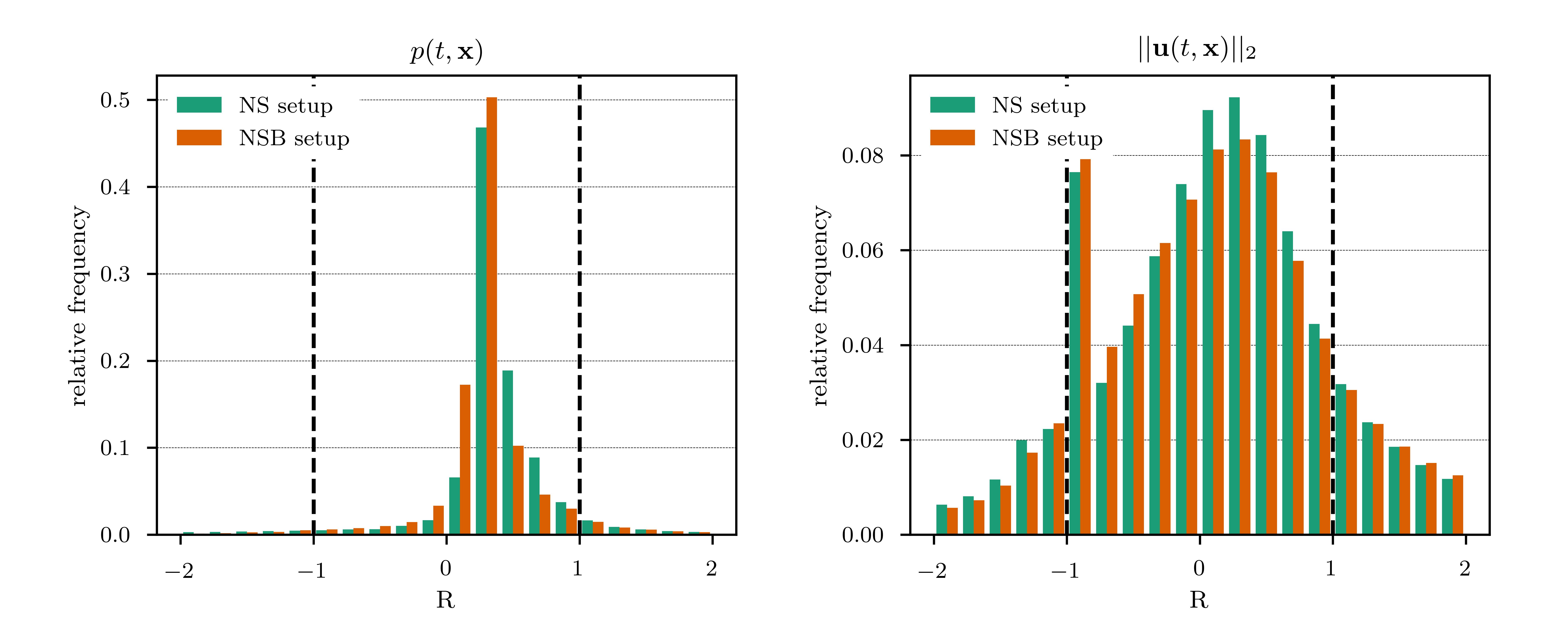} 
\caption{Pulsatile Case ($Re=4000$)}
\end{subfigure}
\caption{Histograms showing the distribution of point-wise $R$ values of all time steps, comparing the NS and NSB setup for $\norm{\vec{u}(t,\vec{x})}{2}$ and $p(t,\vec{x})$.}\label{fig:hist}
\end{figure}

Demonstrating point-wise convergence for every time step proved to be a challenging task, 
in particular with regard to fluid velocity. 
Nevertheless, histograms in \figref{fig:hist} and \tabref{tab:R} indicate 
that there are no relevant differences in mesh convergence between the NS and NSB setup. 

\begin{table}[htp]
\caption{Percentage of $R$ values from \cref{fig:hist} that fall within a $[-1,1]$ interval, rounded to whole numbers.}\label{tab:R}
\begin{center}
\begin{tabular}{c|cccc}
\toprule
&\multicolumn{2}{c}{laminar $Re=20$}&\multicolumn{2}{c}{pulsatile $Re=4000$}\\
&\textbf{pressure} & \textbf{velocity}&\textbf{pressure} & \textbf{velocity} \\ 
\midrule 
\textbf{NS setup} & 100\% & 82\%  & 92\% &  64\% \\ 
\textbf{NSB setup} & 100\% & 84\%  & 89\% & 66\% \\ 
\bottomrule 
\end{tabular} 
\end{center}
\end{table}

\begin{figure}[htp]
\begin{subfigure}{0.49\linewidth}
\centering
  \includegraphics[width=\linewidth,keepaspectratio]{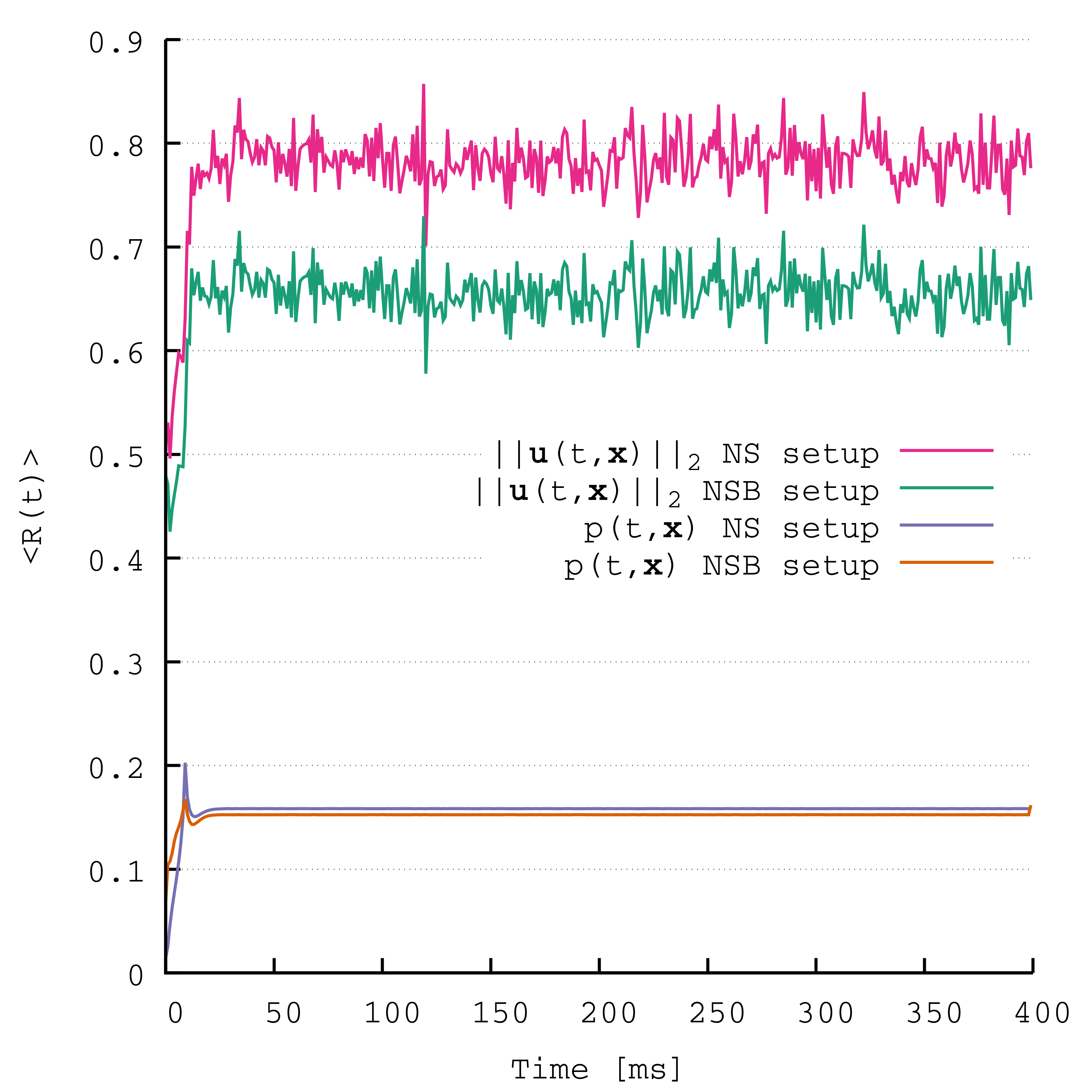}
  \caption{Laminar Case ($Re=20$)}
\end{subfigure}
\hfill
\begin{subfigure}{0.49\linewidth}
\centering
  \includegraphics[width=\linewidth,keepaspectratio]{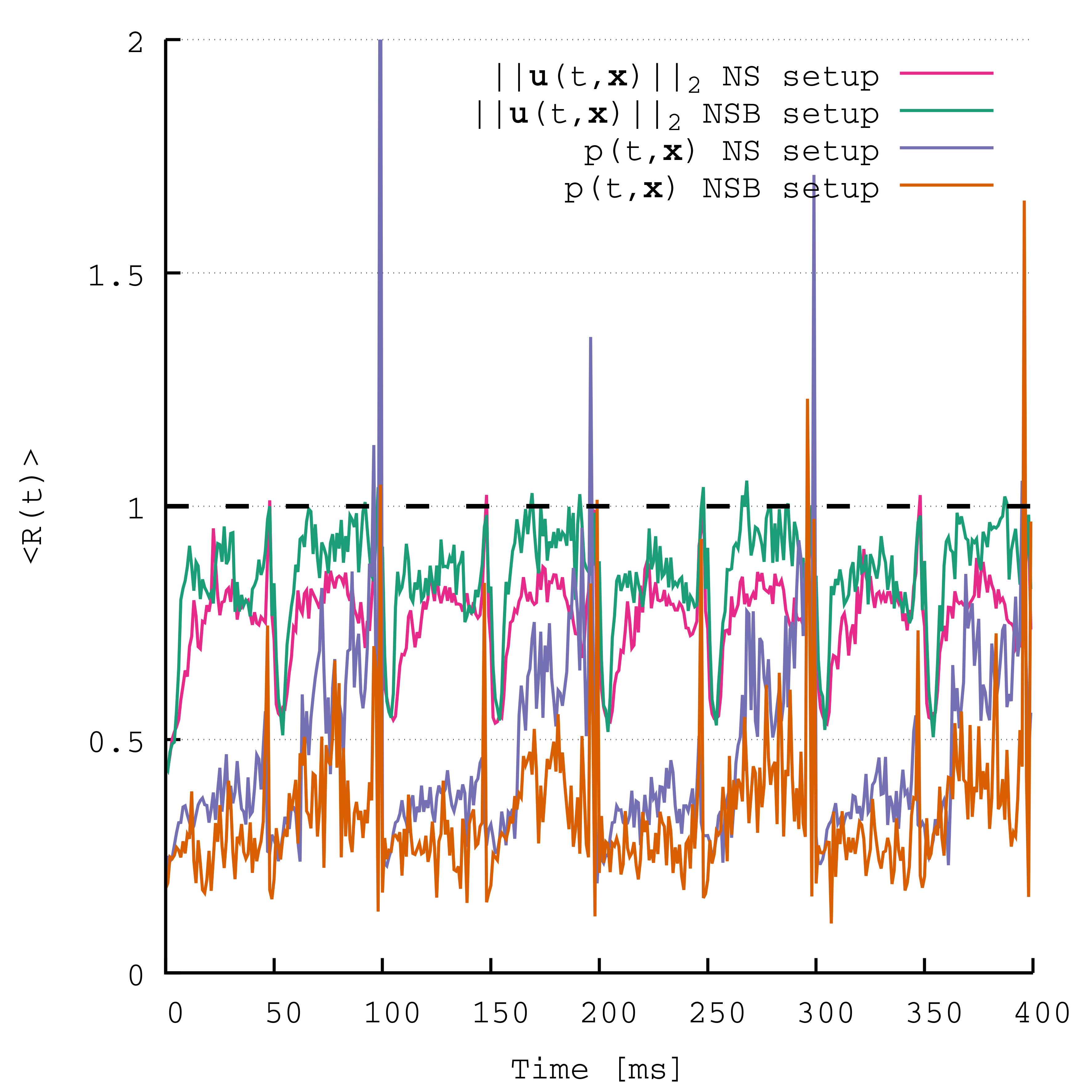} 
  \caption{Pulsatile Case ($Re=4000$)}
\end{subfigure}
\caption{The global convergence ratio $\langle R(t) \rangle$ plotted over time for $\norm{\vec{u}(t,\vec{x})}{2}$ and $p(t,\vec{x})$.}\label{fig:Rglob}
\end{figure}

Point-wise convergence is particularly challenging to obtain. 
Therefore we considered alternative metrics such as the assessment of convergence 
based on a global convergence ratio $\langle R(t) \rangle$ as proposed in \citep{Stern2001}, 
or the usage of \eqref{eq:R} based on a derived physical quantity.
The global convergence ratio regarding the pressure and velocity solutions, 
shown in \figref{fig:Rglob}, indicates convergence for every time step in the laminar case and for almost every time step in the pulsatile case.
Furthermore, NS and NSB setup perform equally well, 
which is consistent with the results shown in \figref{fig:hist} and \tabref{tab:R}.
However, from the definition of $\langle R(t) \rangle$ in \eqref{eq:Rglob} it is evident 
that $\langle R(t) \rangle \geq 0$.
Thus, using the global convergence ratio
monotonic convergence cannot be distinguished from oscillatory convergence.

Further analysis of accuracy dependence on mesh resolution was carried out 
by evaluating the grid convergence index (GCI), see \eqref{eq:GCI} \citep{Roache1994} 
and by determining the order of convergence following \citep{deVahlDavis1983}, see \eqref{eq:alpha}.
Results are presented in \tabref{tab:a_R_GCI} as mean values over time 
with the respective standard deviation indicating the total range of data.
According to these metrics for the laminar case the convergence rate was excellent, with $<2.7$ for all quantities and all time steps.
The convergence ratio indicated monotonic convergence for all cases 
and the GCI suggested very low uncertainty.
For the pulsatile case however, lower convergence rates with higher standard deviations were found.
The convergence \reviewerTwo{ratio} indicated monotone convergence for the flow rate and monotone/oscillatory convergence for the pressure drop.
The higher standard deviation of the convergence rate of the maximum pressure in both setups suggests not convergent behavior in some time steps.
The GCI suggests an increased uncertainty for all quantities.

\begin{table}[htp]
\caption{Mean values M and standard deviation SD of the order of convergence, $\alpha(t)$, the convergence ratio $R(t)$, and the grid convergence index ($\mathrm{GCI}(t)$)}\label{tab:a_R_GCI}
\begin{center}
\begin{tabular}{ccl|cccccc}
\toprule
&&&\multicolumn{2}{c}{$\alpha(t)$} & \multicolumn{2}{c}{$R(t)$} & \multicolumn{2}{c}{$\mathrm{GCI}(t)$}  \\ 
&&& M & SD & M & SD & M [\%] &SD [\%]   \\ 
\midrule
\multirow{4}{1.8cm}{laminar $\mathrm{Re}=20$}&$p_\mathrm{max}$ &NS s. & 2.8361 & 0.0001&  0.14003 & 0.00001& 0.1381 & 0.0001\\ 
&$p_\mathrm{max}$ &NSB s. & 2.8111 & 0.0001 & 0.14249 & 0.00001 & 0.2999 &0.0001\\ 
&$\Delta p$&NSB s.& 2.789 & 0.001 & 0.1447 & 0.0001& 0.316 & 0.001 \\ 
&$q$ &NSB s.& 3.18 & 0.01 & 0.111 & 0.001& 0.55 & 0.01 \\
\midrule
\multirow{4}{1.8cm}{pulsatile $\mathrm{Re}=4000$}&$p_\mathrm{max}$& NS s. & 1.2 & 1.5&  0.01 & 13.0& 389 & 4913 \\
&$p_\mathrm{max}$ &NSB s. & 2.0 & 1.9 & 0.3 & 0.8 & 0.1 & 101 \\ 
&$\Delta p$  &NSB s.& 2.5 & 1.2 & 0.2 & 0.3 & 9 & 38 \\ 
&$q$ &NSB s.& 2.3 & 1.2 & 0.2 & 0.2& 31 & 12 \\
\bottomrule
\end{tabular} 
\end{center}
\end{table}

\begin{figure}[htp]
\begin{subfigure}{0.49\linewidth}
\centering
  \includegraphics[width=\linewidth,keepaspectratio]{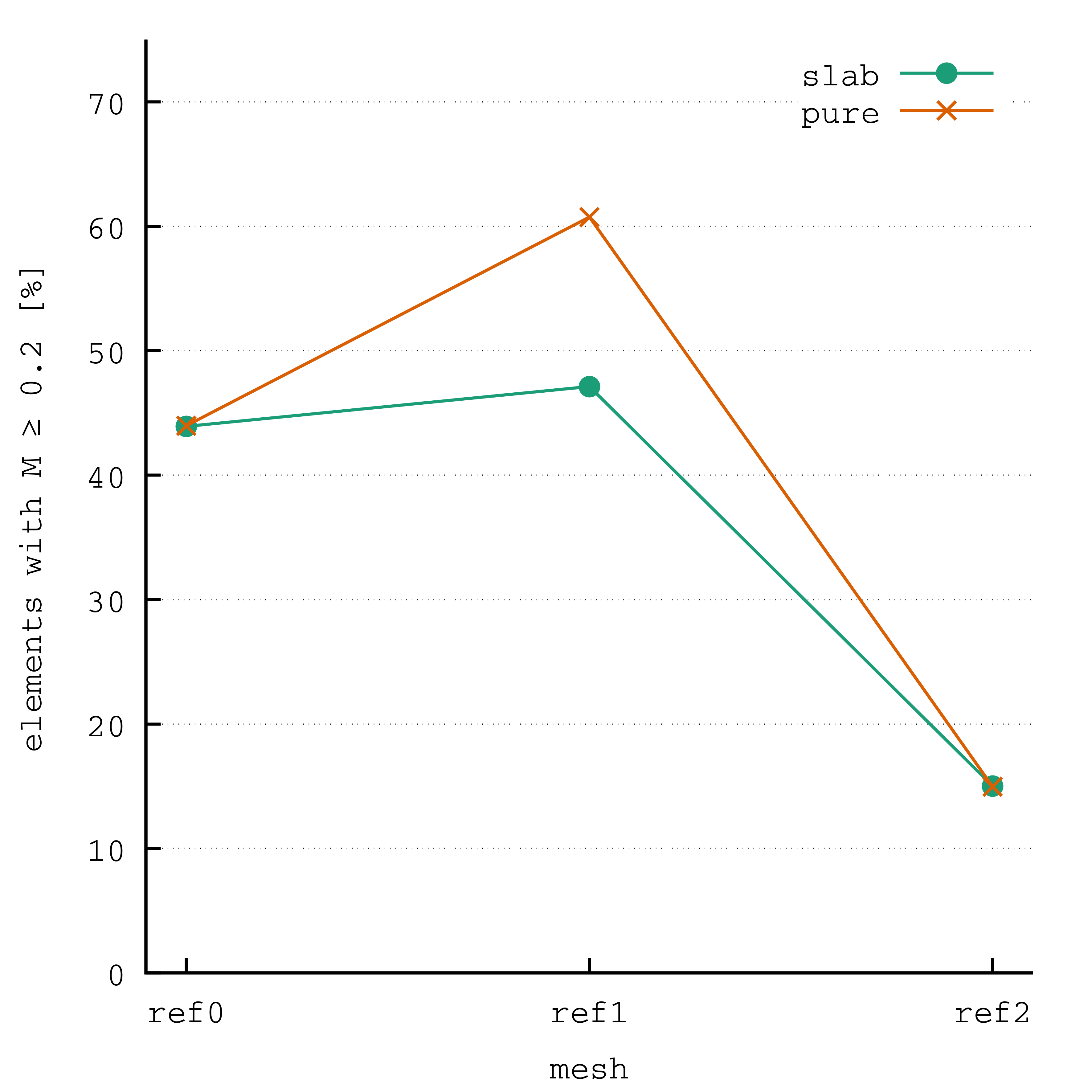}
  \caption{Elements with $M\geq0.2$  in percent.}
\end{subfigure}
\hfill
\begin{subfigure}{0.49\linewidth}
\centering
  \includegraphics[width=\linewidth,keepaspectratio]{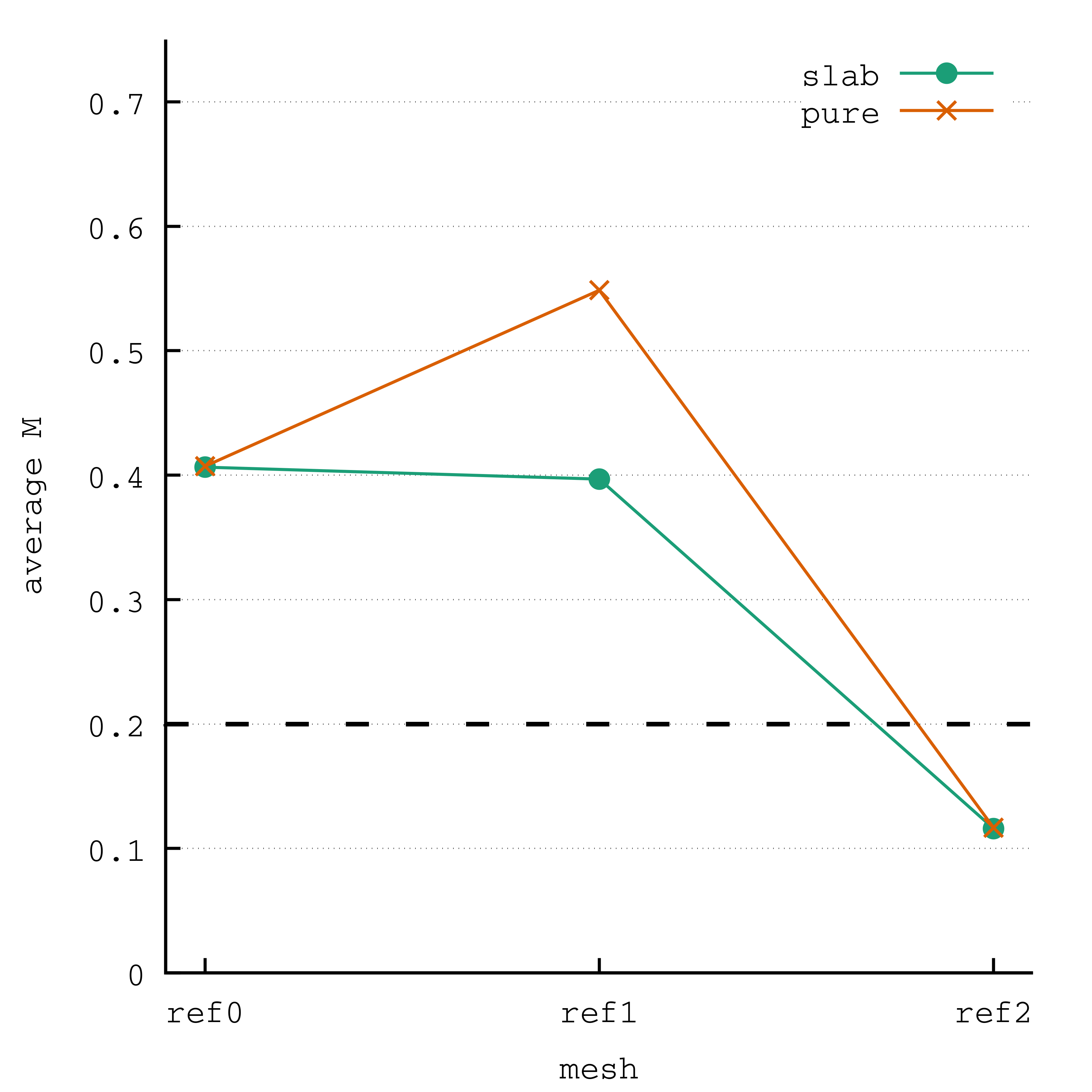} 
  \caption{$M$ averaged over all points in space and time.}
\end{subfigure}
\caption{Grid convergence analysis of the pulsatile case ($\mathrm{Re}=4000$).}\label{fig:pope}
\end{figure}

In LES type models, the resolution of solution quantities is fundamentally linked to the numerical method used, which can cause traditional mesh convergence criteria to fail as soon as turbulence occurs, see \ref{sec:appendix:meshconv}.
To remedy this problem \citep{Pope2004} proposes the use of a measure of turbulence resolution $M$, see \cref{eq:TKEfrac}, utilizing the fraction of turbulent kinetic energy resolved by the grid in question.
$M$ represents the amount of turbulent kinetic energy, that is not resolved by the computational grid.
In \citep{Pope2004} a threshold of $M < 0.2$ is used to classify a LES solution as well resolved.
\figref{fig:pope}a shows the percentage of elements that meet the criterion, and  \figref{fig:pope}b depicts the average value of $M$ for all three grid resolutions.
From that we deemed the solutions on the finest grid well resolved.

\subsection{FDA Round-Robin Benchmark}
\begin{figure}
    \centering
    \includegraphics[width=\textwidth,keepaspectratio]{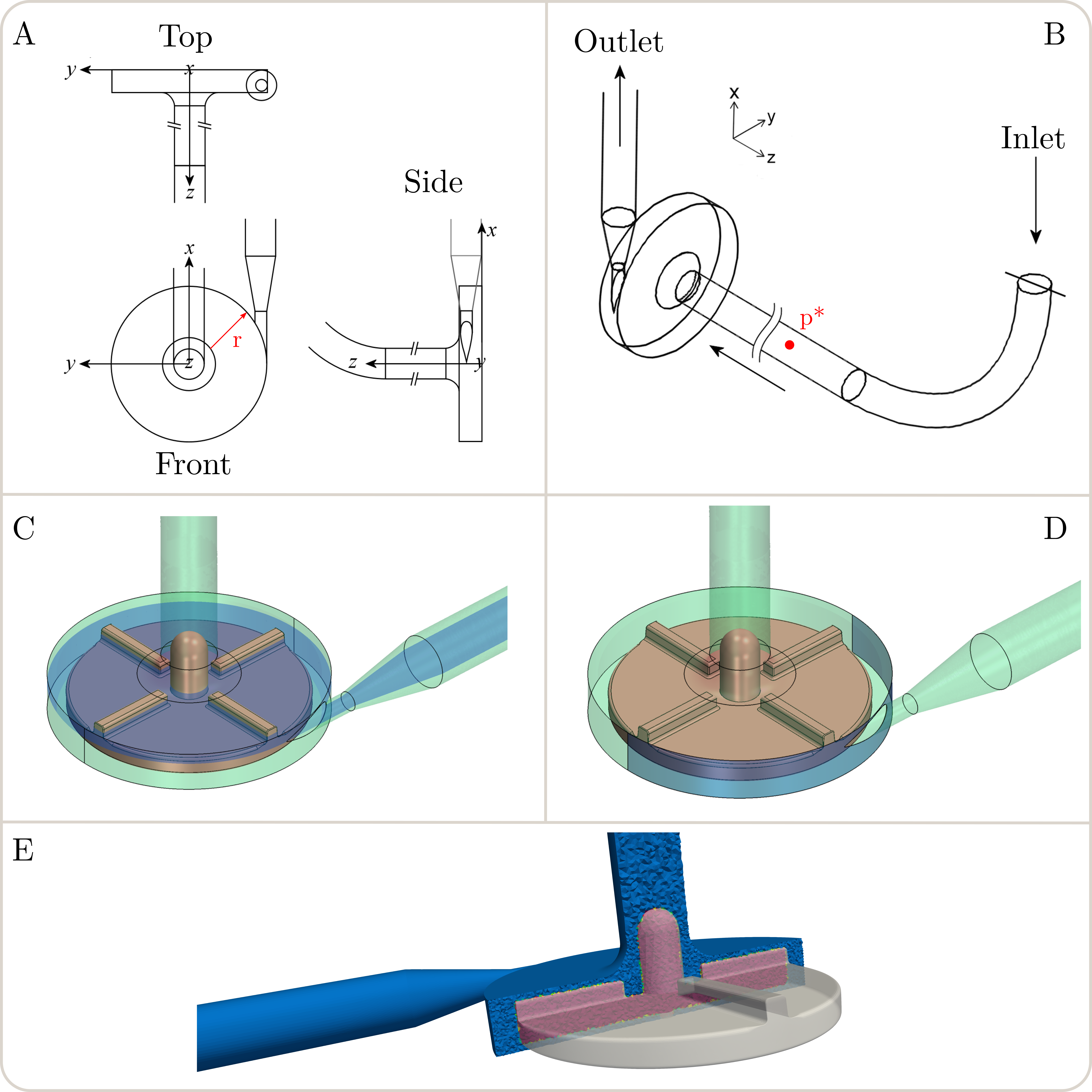}
    \caption{FDA round robin benchmark setup: Subfigure A depicts a Top/Front/Side view of the blood pump housing. Note that the geometry origin coincides with the rotor hub mounting point. Additionally, Subfigure A shows a radial line (in red) at angle $\phi=45^{\circ}$ with origin $(0,0,0.006562)$ and radius $r$ varying from \SIrange[range-units = single]{0.006}{0.03}{\metre}. This line is used for PIV data comparison. Subfigure B shows the overall shape of the blood pump housing marked velocity inlet and pressure outlet. Additionally Subfigure B indicates the point $p^*=(0,0,\SI{0.175}{\metre})$ used for calculating the pressure head $\Delta p$ between $p^*$ and the outlet. Subfigure C displays the blade passage slice defined as the plane $z=0.006562$\si{\metre}. Subfigure D shows the rim surface used to calculate the wall shear stress. Lastly, Subfigure E shows the immersed rotor (opaque gray) in the computaional mesh. The colors indicate the permeability distribution $K$. Subfigures A -- D have been taken and adapted from the \href{https://nciphub.org/collections/post/1246/download/Blood_Pump.zip}{data set} provided by the FDA \cite{FDABenchHP}.}
    \label{fig:fda_setup}
\end{figure}

In this section the applicability of our approach is demonstrated 
using a standard benchmark study initiated by the FDA \cite{Malinauskas2017fda} 
and also studied elsewhere \cite{Marinova2016,Good2020}
with the aim to evaluate the suitability of using CFD simulations 
in the regulatory safety evaluations.
We use the benchmark models of a typical centrifugal blood pump 
for which extensive experimental measurements on velocities and pressures
were provided to support CFD validation.
The centrifugal blood pump consists of the two main components housing and rotor. 
Blood enters the housing  through a curved inlet tube, 
where it meets a hub and rotor blades rotating the blood within the housing. 
Blood exits the pump through a diffuser and continues into the outlet.
A schematic is shown in Subfigure \ref{fig:fda_setup}A and Subfigure \ref{fig:fda_setup}B.
For a detailed description of the setup we refer to here \cite{FDABenchHP}.
Experimental data for six scenarios of differing rotor speed and inflow rates,
summarized in \tabref{tab:FDA:opc}, are available.
Using the provided CAD files of the setup a computational mesh of the housing 
along with a surface mesh of the rotor disk was created using \emph{Meshtool} \cite{Neic2020}.
Mesh convergence was investigated using Pope's criterion $M$ as described in \ref{sec:pope_crit}.
We performed simulation experiments on the most demanding case 6 on various refined meshes until $M < 0.2$.
Here, we calculated an space and time-averaged value of $M$ using the last 8 revolutions in time and the whole computational domain in space.
Values for $M$ on the \generalChange{finest} grid are depicted in \tabref{tab:FDA:PopeCriterion}.
The resulting mesh was then used for the entire benchmark.
The final computational mesh, consisting of tetrahedral and prismatic elements, 
comprised around six million finite elements and one million nodes.

\begin{figure}[hbtp]
\begin{center}
\includegraphics[width=\textwidth,keepaspectratio]{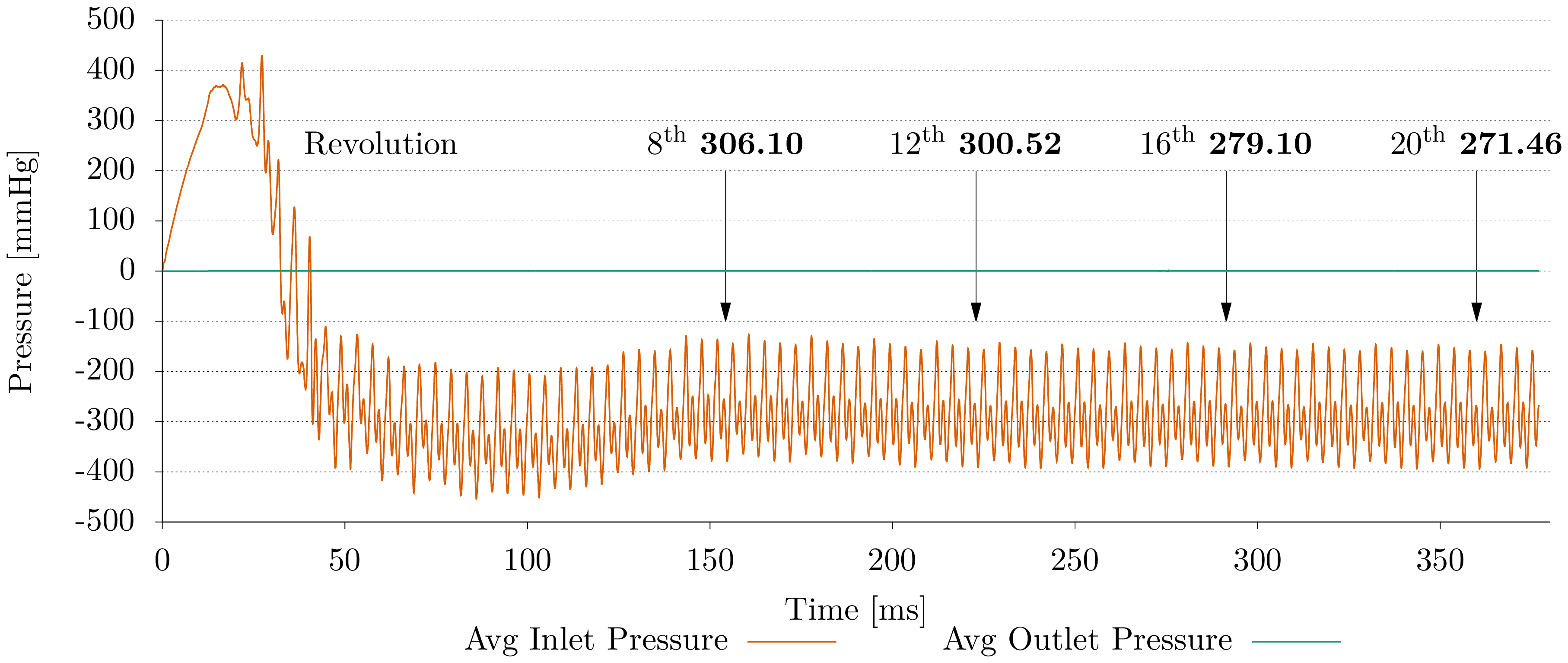}
\end{center}
\caption{Pressure evolution as a function of time for 22 revolutions (including 2 revolutions startup phase) for case 5. Four values for pressure heads at the beginning of a specific revolution are indicated with arrows. The green curve represents zero-pressure at the outlet 
and the orange curve shows the pressure at the inlet measurement probe located at $p^*$ as depicted in Subfigure \ref{fig:fda_setup}B}
\label{fig:fda3500rpm_pressure_trace}
\end{figure}

\begin{figure}[hbtp]
\begin{center}
\includegraphics[width=\textwidth,keepaspectratio]{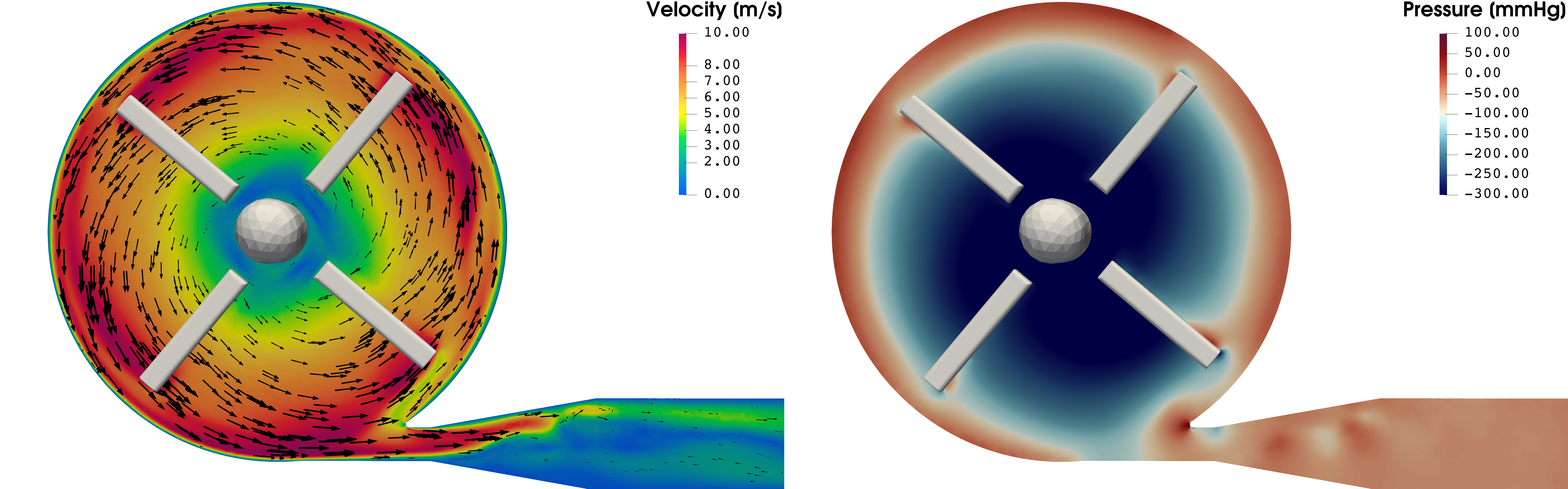}
\end{center}
\caption{Velocity and pressure fields for case 5 on the blade passage slice $z=\SI{6.562e-3}{\metre}$ at time $t=\SI{268.57}{\milli\second}$.}\label{fig:fda3500rpm_vel_press_comp}
\end{figure}

\begin{figure}[ht]
\begin{subfigure}{.5\textwidth}
  \centering
  \includegraphics[width=\linewidth]{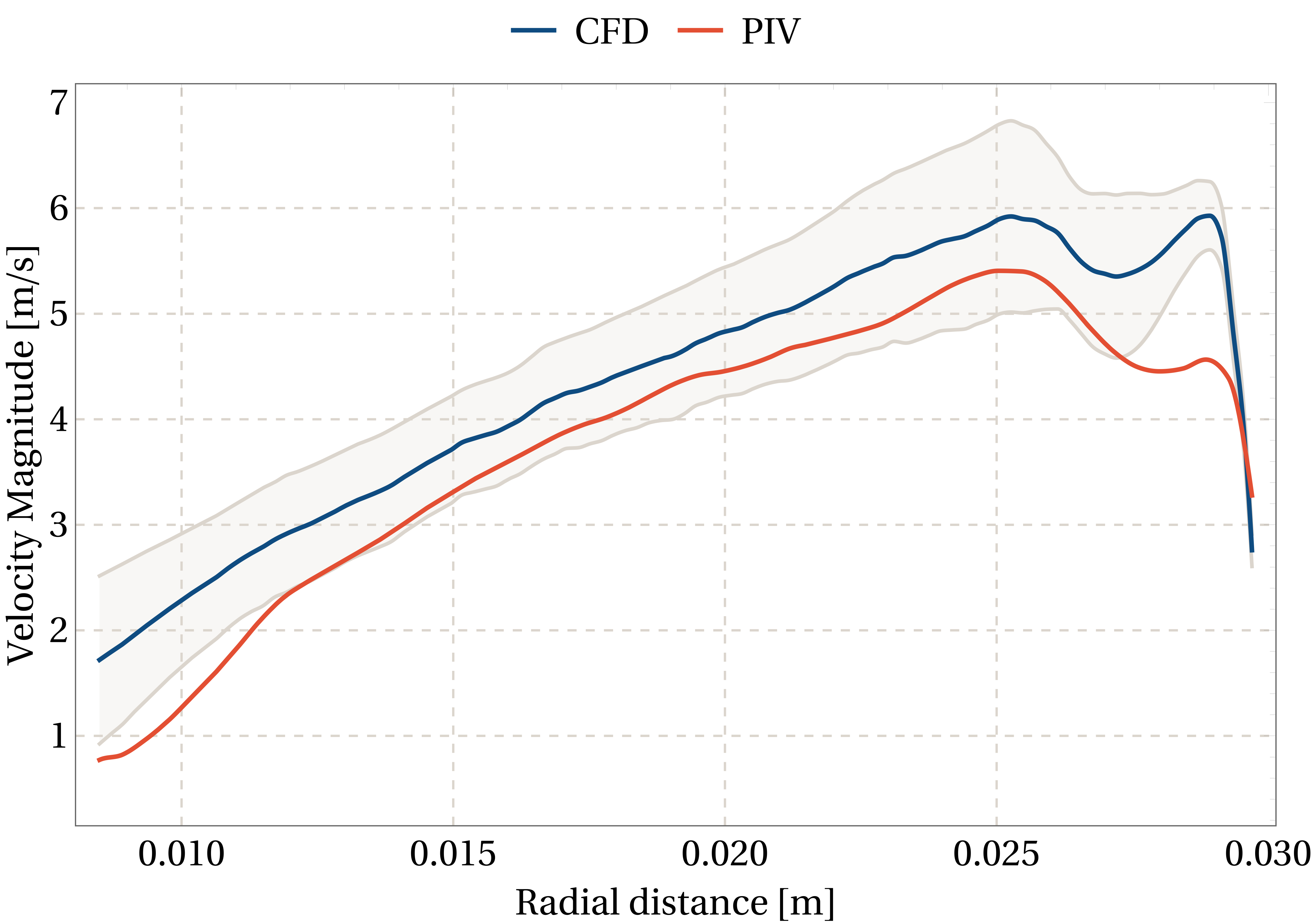}  
  \caption{Case 1}
  \label{fig:FDA:radial_plot:2500rpm:1}
\end{subfigure}
\begin{subfigure}{.5\textwidth}
  \centering
  \includegraphics[width=\linewidth]{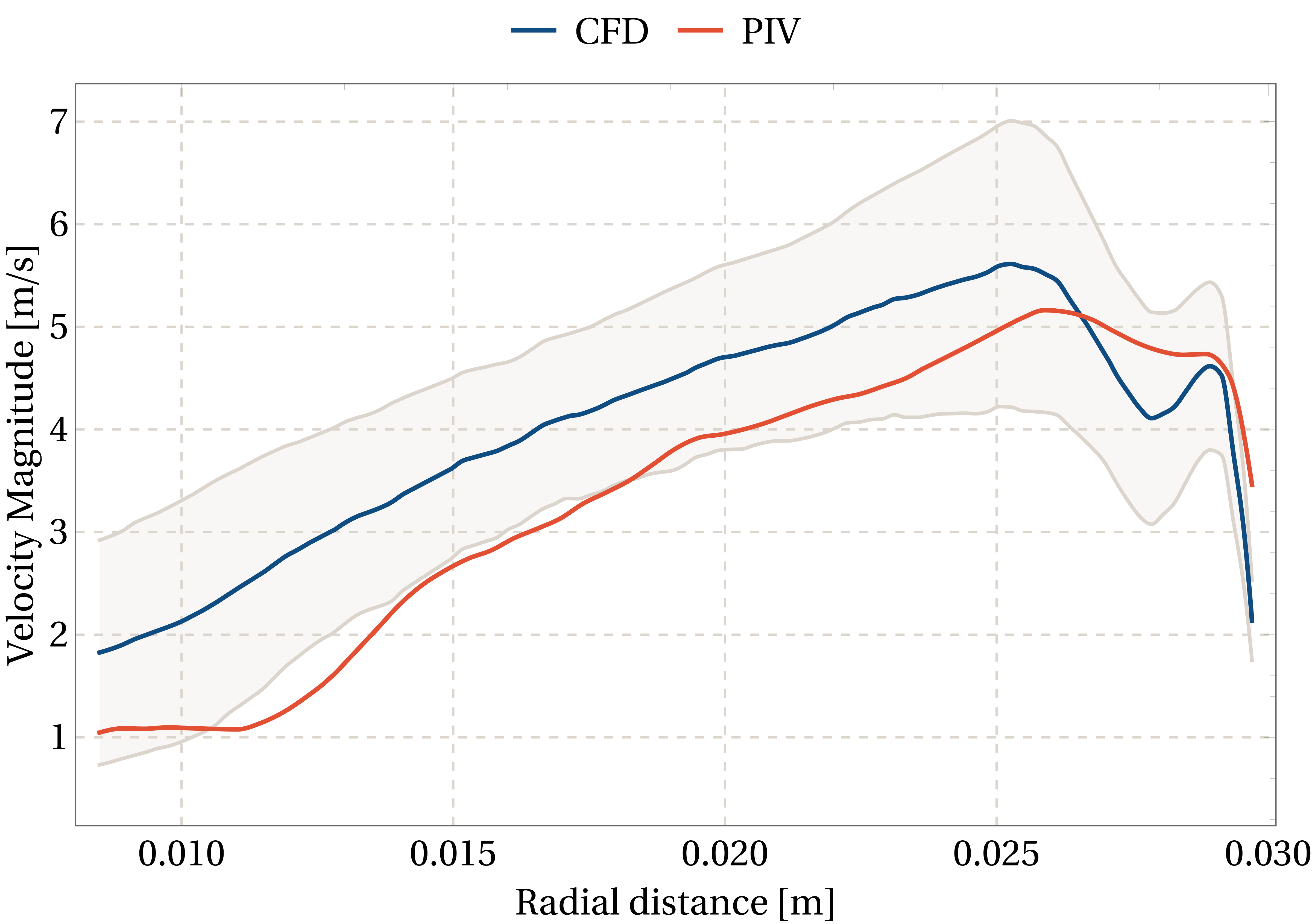}  
  \caption{Case 2}
  \label{fig:FDA:radial_plot:2500rpm:2}
\end{subfigure}
\caption{Comparison of PIV data with CFD data for \SI{2500}{\RPM}. Blue line indicates $\overline{u}_\mathrm{abs}$, orange line indicates the PIV data, and the shaded gray area corresponds to $\overline{u}_\mathrm{abs}\pm \sigma$ defined in \eqref{eq:mean} and \eqref{eq:stddev}.}
\label{fig:FDA:radial_plot:2500rpm}
\end{figure}

\begin{figure}
\begin{subfigure}{.5\textwidth}
  \centering
  \includegraphics[width=\linewidth]{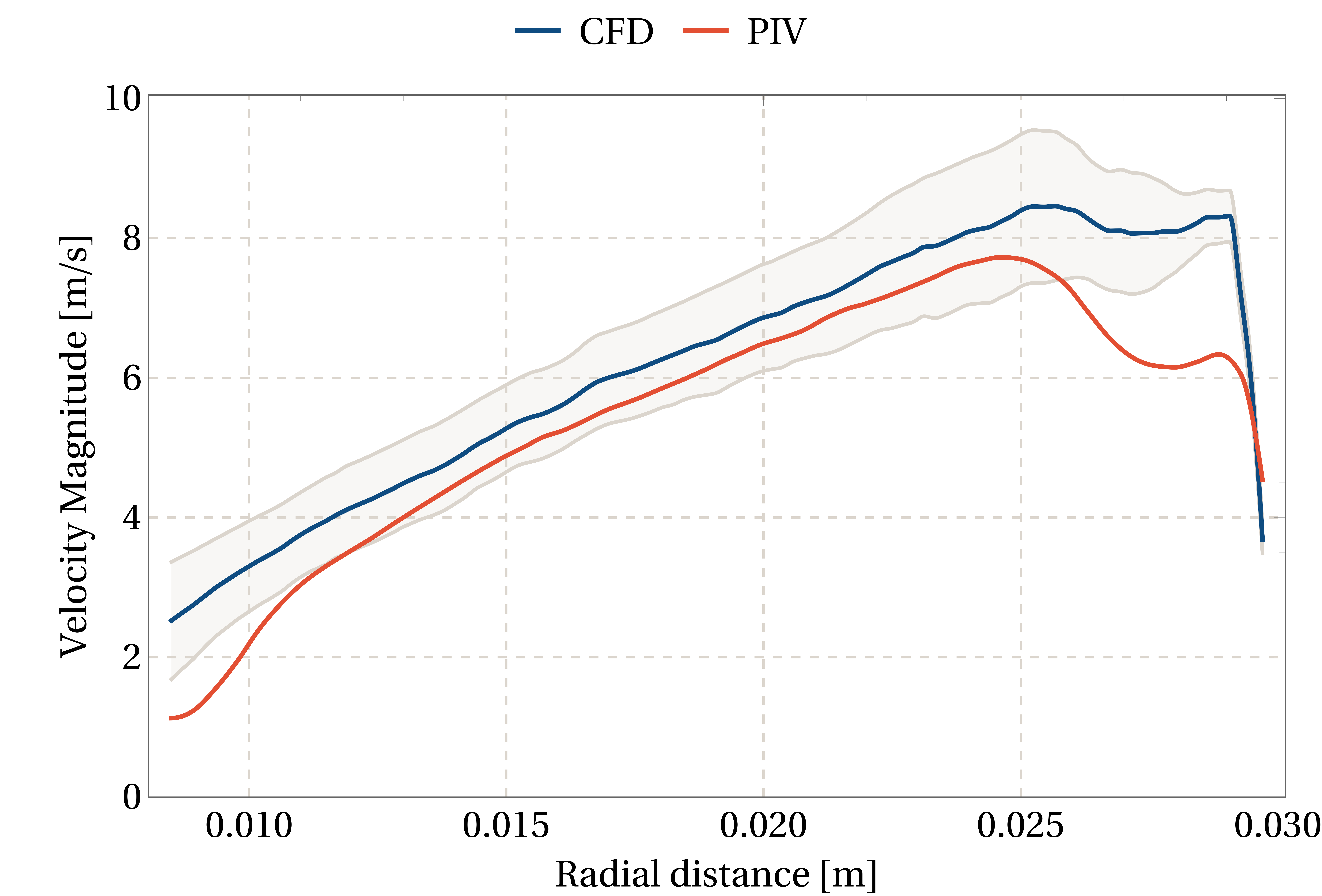}  
  \caption{Case 3}
  \label{fig:FDA:radial_plot:3500rpm:1}
\end{subfigure}
\begin{subfigure}{.5\textwidth}
  \centering
  \includegraphics[width=\linewidth]{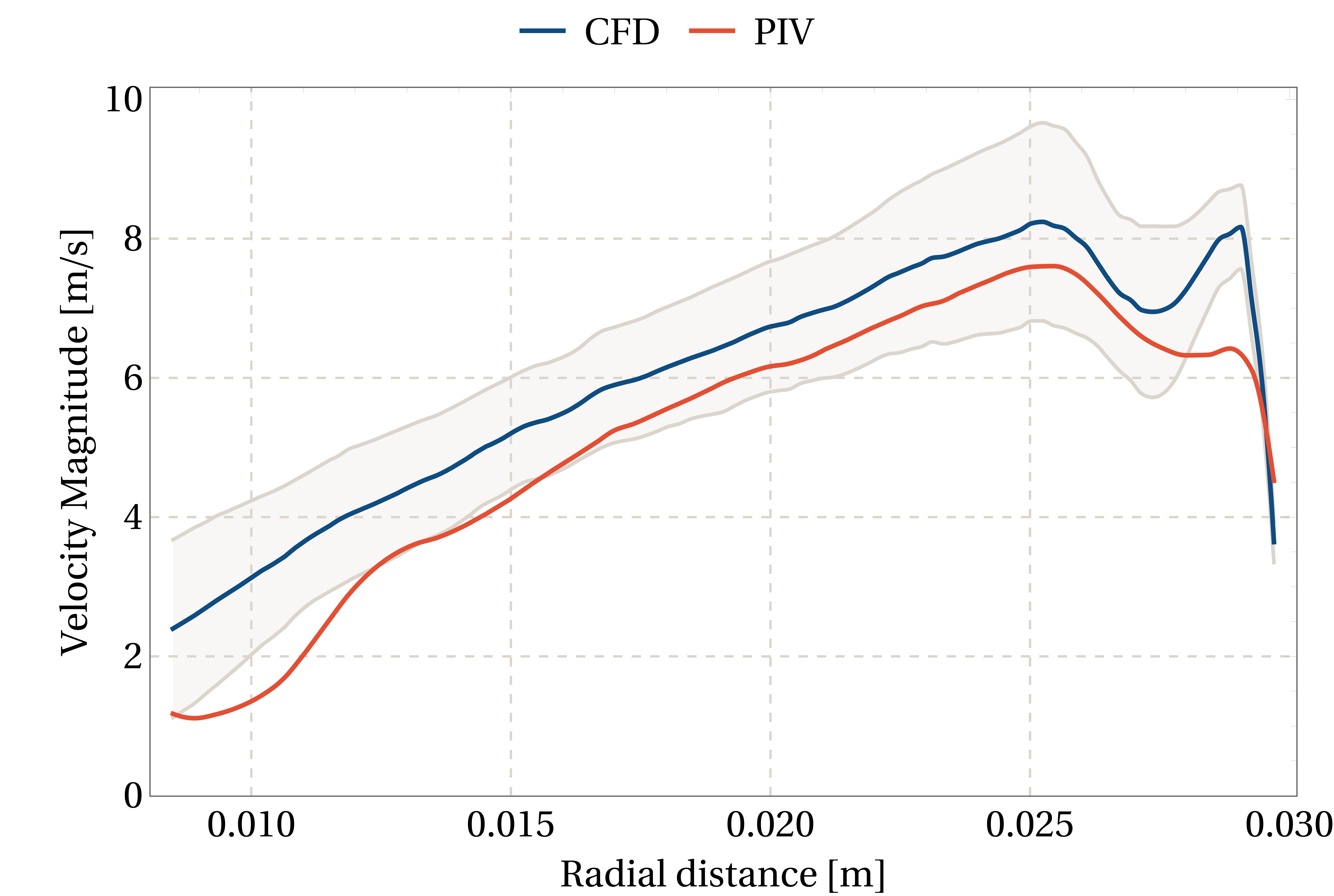}  
  \caption{Case 4}
  \label{fig:FDA:radial_plot:3500rpm:2}
\end{subfigure}
\begin{subfigure}{.5\textwidth}
  \centering
  \includegraphics[width=\linewidth]{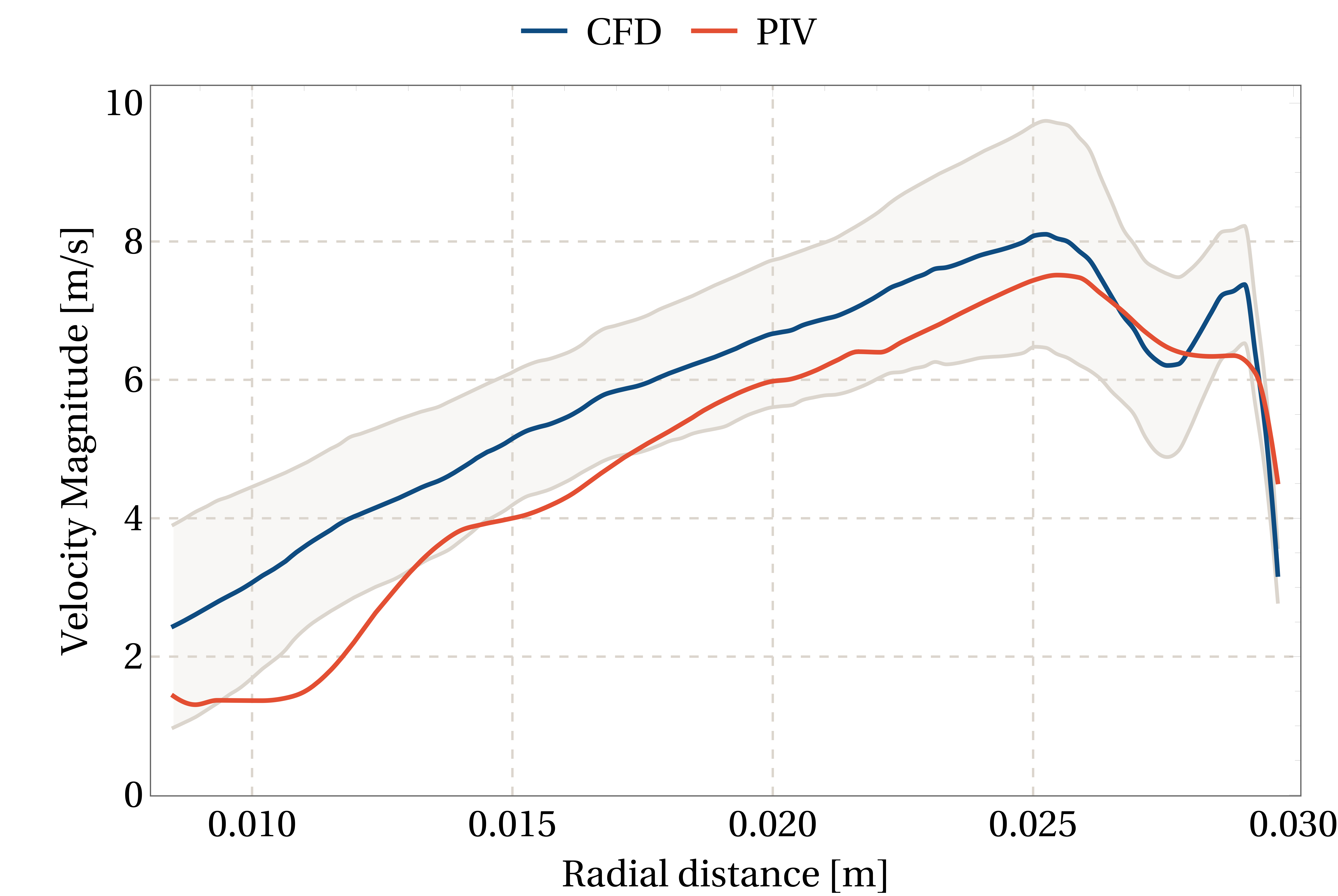}  
  \caption{Case 5}
  \label{fig:FDA:radial_plot:3500rpm:3}
\end{subfigure}
\begin{subfigure}{.5\textwidth}
  \centering
  \includegraphics[width=\linewidth]{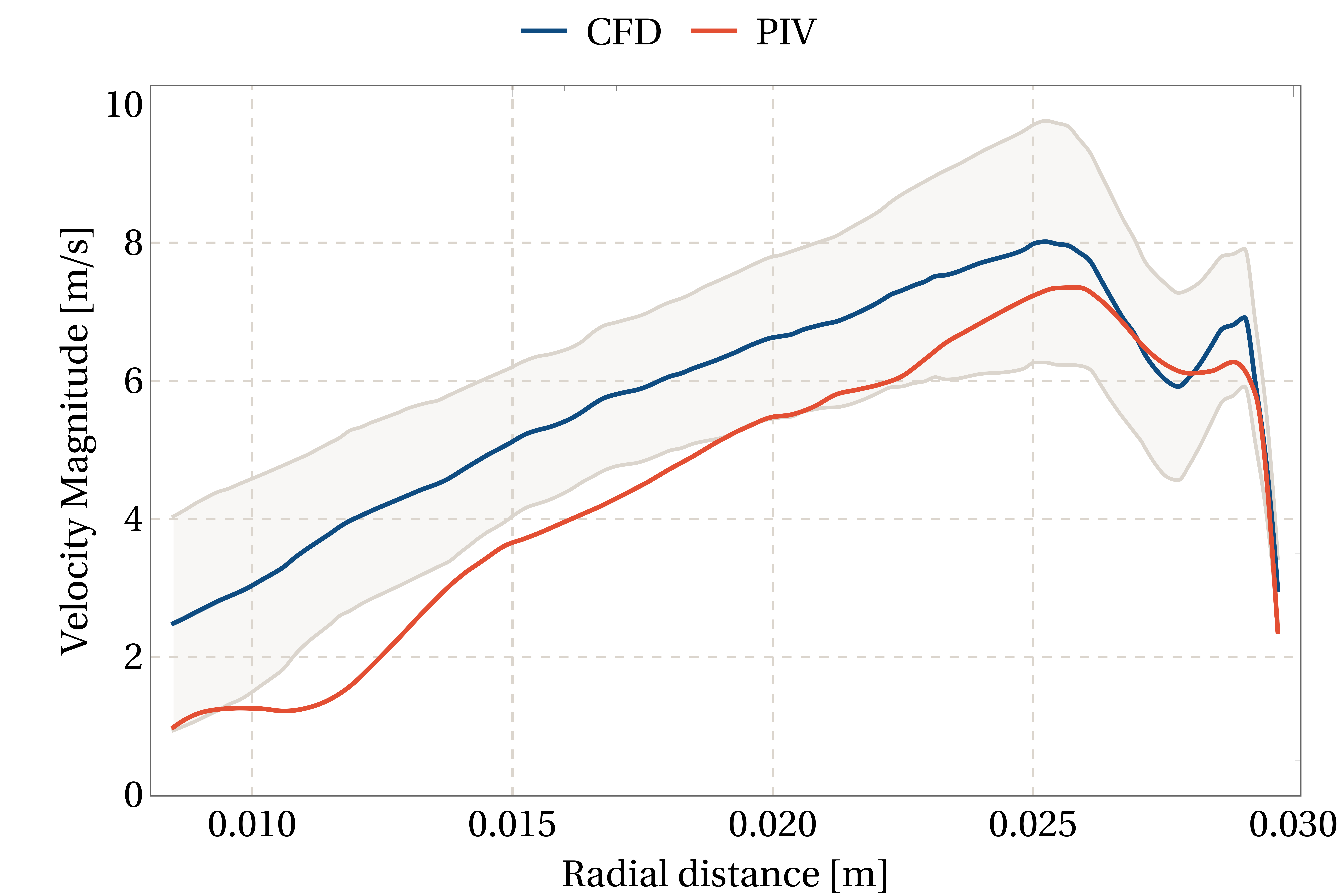}  
  \caption{Case 6}
  \label{fig:FDA:radial_plot:3500rpm:4}
\end{subfigure}
\caption{Comparison of PIV data with CFD data for \SI{3500}{\RPM}. Depicted are the same graphs as in Figure \ref{fig:FDA:radial_plot:2500rpm}.}
\label{fig:FDA:radial_plot:3500rpm}
\end{figure}

The rigid body movement of the rotor disk was pre-calculated. 
For every point $\vec x$ on the rotor surface a rotation around the $z$-axis is performed with the angle $\theta(t)$ defined as
\begin{align*}
 \theta(t) := \begin{cases}
 2 \pi \frac{t^3\left(6t^2-3t\left(5+ft)T_a+\left(10+7ft\right)T_a^2-4fT_a^3\right)\right)}{T_a^5} & \text{for } t < T_a, \\
 2\left(\pi +f \pi (t-T_a)\right) & \text{else}
 \end{cases},
\end{align*}
where $T_a$ denotes the end time of ramp-up phase and $f$ denotes the frequency.
The velocity at a given radius was calculated then as $\vec v := \boldsymbol{\omega} \times \vec r$ 
with $\vec \omega = \frac{\mathrm{d}\theta}{\mathrm{d}t} \vec e_z$ and $\vec r$ 
denoting the distance from the point $\vec x$ on the rotor surface to the barycenter of the disk.
For calculating the permeability areas we adapted the procedure outlined in Section~\ref{sec:methods:obs_algo} as follows:
\begin{itemize}
    \item At every time instant we update the nodes of the surface mesh with the precalculated new positions,
    \item The obstacle velocity $\vec u_s$ is calculated on the fly 
    as projection of the velocity $\vec v$ onto the computational mesh 
    using a radial basis function projector\,\cite{Lazzaro2002}.
\end{itemize}
Due to the periodicity of the movement we only calculated the permeability distribution for the ramp-up phase plus one \generalChange{revolution} and then extended the permeability distribution periodically.
The obtained permeability distribution is illustrated in Subfigure \ref{fig:fda_setup}E.
While blood is known to display non-Newtonian behavior \cite{Easthope80},
experimental studies \cite{Chien1970} showed 
that at high shear rates, higher than $\SI{100}{\per\second}$ as in this benchmark, the viscosity of human blood 
with physiological hematocrit reaches a constant value.
Thus, the choice of a Newtonian model for the benchmark is well justified.
In our benchmark simulations values of $\rho = \SI{1035}{\kilo\gram\per\cubic\metre}$ 
and $\mu=\SI{3.5e-3}{\pascal\second}$ were chosen 
for fluid density and dynamic viscosity, respectively in accordance with the simulation parameters given out by the FDA.
No-slip boundary conditions were applied on the housing wall. 
Across the cross section of the inlet a parabola-type inflow condition scaled to the inflow rate in Table\,\ref{tab:FDA:opc}
was defined and smoothly increased to its nominal value at $t=T_a$.
Shapes of the inflow parabolas were inferred from data sets published in \cite{Hariharan2018}.
At the outlet of the pump housing a directional-do-nothing boundary condition \cite{Braack2014} was imposed.
For cases 1 and 2 a fixed time step size of $\Delta t = \SI{0.048}{\milli\second}$ was chosen 
while in all other cases a value of $\Delta t = \SI{0.0286}{\milli\second}$ was used.
This temporal discretization led to 500 time steps per revolution in cases 1 and 2, 
and to 600 time steps per revolution otherwise.
The penalization parameter $\hat K$ was chosen as $10^{-9}$ in all cases 
and a spectral radius of $\rho_{\infty} = 0.2$ was chosen 
for the generalized-$\alpha$ integrator leading to a stable performance of the simulator over all simulation setups, see also \cite{Jansen2000}.
Overall, a total of  11000 and 13200  time steps were computed for cases 1 and 2 
and all other cases, respectively.
This corresponded to 2 full revolutions of the ramp-up phase 
and 20 revolutions with a constant angular velocity in all cases.
Computations were carried out on the \emph{Vienna Scientific Cluster 4} (VSC4) 
using 1200 MPI processes.
On average, 5 Newton-Raphson iterations per time-step were needed 
to obtain a relative residual of $<10^{-5}$ 
and \SI{20}{\second} to complete one time step.
The total compute times for the cases ranged between \SIrange{60}{70}{\hour} 
for the different cases.
In a post-processing step the following derived quantities were calculated:
\begin{enumerate}
    \item The pressure head $\Delta p$ between the outflow and the point $p^*$ depicted in Subfigure \ref{fig:fda_setup}B based on the time averaged pressure over the last two revolutions.
    \item The shaft torque defined as
    \color{red}
    \begin{align*}
        T := \norm{\int\limits_{\Omega} \vec r(\vec x) \times \tensor \sigma(\vec u, p)\widetilde{\vec n} ~\dx}{}
    \end{align*}
    \color{black}
    \generalChange{with $\vec r$ denoting the distance of a point $\vec x$ on the rotor surface 
    and the shaft mount, coinciding with the origin,
    and $\widetilde{\vec n}$ denoting the approximation to the outer unit normal of the rotor surface as described in \ref{sec:fluid_forces}.} Again, velocity and pressure were take as time-averaged over the last two revolutions.
    \item The wall shear stress over the pump housing rim, see Subfigure \ref{fig:fda_setup}D, based on the time-averaged velocity over the last two revolutions.
\end{enumerate}
\figref{fig:fda3500rpm_pressure_trace} shows the time evolution of the pressure at point $p^*$ and the outlet for case 5.
It can be seen that a quasi steady state was reached after the $8^{\mathrm{th}}$ revolution.
Velocity and pressure fields for case 5 over a cross-section 
are shown in \figref{fig:fda3500rpm_vel_press_comp}.
From the representation by means of velocity vectors it can be observed that once the fluid has left the low-velocity inflow region, fluid particles are accelerated by the rotor blades and start to follow a circumferential path. 
The rotor blades create regions of high pressure in front of them, 
especially at their tips where the velocity of the blade is the highest, 
leaving regions of lower pressure behind them. 
The velocity field behaves accordingly. 
Higher velocity values are observed in the wake of the blades.
The velocity vectors at the outflow are well aligned with the outflow direction,
thus reducing the amount of turbulence in this critical area.
In the extended outflow region the tube radius increases, 
the flow velocity drops and emerging turbulence rises, as expected.
Post-processing results are shown in \tabref{tab:FDA:results}.
Comparing with existing results, see \cite{Malinauskas2017fda}, 
we conclude that the computed values for the pressure head 
are in agreement with measured values for cases 1 -- 5.
The values for case 6 are not in agreement with measured values, possibly due to insufficient mesh resolution.
Additionally, we performed a quantitative comparison between a particle image velocimetry (PIV) data set, released as part of the benchmark by the FDA, and our simulations.
\reviewerOne{For PIV comparisons we relied on data published as \texttt{.xslx} files on the FDA website \url{https://ncihub.org/wiki/FDA_CFD/ComputationalRoundRobin2Pump/PumpData}.
We compared against PIV data for the first quadrant of the blade passage plane by extracting the velocity components along the red line depicted in \figref{fig:fda_setup}A.
}
\figref{fig:FDA:radial_plot:2500rpm} and \figref{fig:FDA:radial_plot:3500rpm} show a comparison of the velocity magnitudes along the radial line in the pump housing as depicted in Subfigure \ref{fig:fda_setup}A.
We used the following definitions for mean velocity magnitude and standard deviation:
\begin{align}
    \label{eq:mean}\overline{u}_\mathrm{abs}(\vec x) &:= \frac{1}{t_\mathrm{end}-t_\mathrm{start}} \int\limits_{t_\mathrm{start}}^{t_\mathrm{end}} \norm{\vec u(t,\vec x)}{}\mathrm{d}t,\\
    \label{eq:stddev}\sigma(\vec x) &:= \sqrt{\frac{1}{t_\mathrm{end}-t_\mathrm{start}} \int\limits_{t_\mathrm{start}}^{t_\mathrm{end}} (\norm{\vec u(t,\vec x)}{} - \overline{u}_\mathrm{abs}(\vec x))^2\mathrm{d}t},
\end{align}
where $t_\mathrm{end}$ was chose as end point of the last revolution and $t_\mathrm{start}$ was chosen as starting point of the 18th revolution.
From the comparison we can conclude, that the CFD simulations and the PIV measurements agree reasonably well within the error bound of the CFD simulation for cases 1 -- 5, while case 6 shows less agreement.
This may again be attributed to insufficient mesh resolution.
Values for torque or wall shear stresses have not been published yet 
and as such could not be used for validation.
Videos showing the flow field evolution have been generated for all six cases 
and are provided in the supplementary material.

\begin{table}[h]
\caption{Operating Conditions for the FDA benchmark.}
\label{tab:FDA:opc}
\centering
\begin{tabular}{c|SS}
\toprule
\textbf{Case} & \textbf{Inflow} [\si{\litre\per\minute}] &
\textbf{Rotational Speed} [\si{\RPM}] \\
\midrule
1 & 2.5 &  2500  \\
2 & 6.0 &  2500  \\
3 & 2.5 &  3500  \\
4 & 4.5 &  3500  \\
5 & 6.0 &  3500  \\
6 & 7.0 &  3500  \\
\bottomrule
\end{tabular}
\end{table}

\begin{table}[h]
\caption{Time-averaged pressure head over last two revolutions, time-averaged wall shear stress magnitude over the housing rim and shaft torque for all 6 FDA simulation cases.}
\label{tab:FDA:results}
\centering
\begin{tabular}{c|SSS}
\toprule
\textbf{Case} & \textbf{Pressure Head} [\si{\mmHg}] & \textbf{Torque} [\si{\newton\metre}] & \textbf{Wall Shear Stress} [\si{\pascal}] \\
\midrule
$1$ & 179.68 & 5.1e-3 & 20.26\\
$2$ & 75.10 & 1.09e-2 & 28.24\\
$3$ & 424.41 & 4.00e-3 & 23.96\\
$4$ & 353.49 & 1.20e-2 & 29.04\\
$5$ & 297.52 & 1.49e-2 & 32.51\\
$6$ & 280.72 & 1.65e-2 & 34.95\\
\bottomrule
\end{tabular}
\end{table}

\begin{table}[h]
\caption{Space and time-averaged values of Pope's criterion $M$, averaging over the whole domain spatially and the last 8 revolutions in time.}
\label{tab:FDA:PopeCriterion}
\centering
\begin{tabular}{c|c}
\toprule
\textbf{Case} & $M$ \\
\midrule
1 & 0.124  \\
2 & 0.137  \\
3 & 0.151  \\
4 & 0.162  \\
5 & 0.192  \\
6 & 0.201  \\
\bottomrule
\end{tabular}
\end{table}


%

\section{Discussion}

In this study we report on the development of a novel method  
for flow obstacles in context of CFD simulations based on a stabilized finite element discretization of the NSB equations.
We evaluated numerical efficiency and accuracy of the method 
by carrying out simulations of \generalChange{one standard benchmark and two relevant application scenarios.}


We introduced suitable modifications of the RBVMS formulation 
and implemented a procedure for immersing (moving) rigid objects 
into an existing Eulerian computational mesh.
\reviewerOne{To the best of the authors knowledge, there has not been an attempt in existing literature to solve the NSB equations with an RBVMS formulation.}

\reviewerOne{The NSB equations are strongly linked to immersed boundary methods.
However, for NSB equations the penalization of obstacles is introduced on the continuous level using discontinuous masking functions. 
Furthermore, as opposed to IBM they can be derived from first principles in physics, obeying Darcy's law inside of the obstacle and the NS equations outside, and rigorous mathematical proofs exists showing existence and behavior of solutions in the case of $\lim_{K\to 0}$ with a model error independent of discretization \cite{Angot1999, Angot1999feb, aguayo2020analysis}.
Additionally, immersed boundary methods introduce penalization on the discrete level by generating suitable body forces mimicking the action of the obstacle.
They are tailored to Cartesian grids, stemming from their use with finite volume methods. 
Care must be taken in choosing the right interpolation algorithm \cite{Piquet2016, Specklin2018, Seo2011}.
The translation to unstructured grids while straightforward with the NSB equations is still under development for IBM methods \cite{Abgrall2014, Ouro2015}}.
\reviewerThree{
On the other hand, the NSB equations bear similarity between the resistive immersed surface method (RIS) \cite{Astorino2012} or resisitive immersed implicit surface method (RIIS) \cite{Fedele2017}.
The important difference lies in penalizing on the continuous level through the introduction of a Dirac-Delta function acting on the \emph{surface} alone.
and can be seen as a limiting case of the NSB equations, see \cite{Fernndez2008}.
While being a very competitive choice for thin structures like heart valves, these methods can not be used for modeling thicker objects like rotary blood pumps.
Furthermore, the introduction of Dirac-Delta functions, being not a function but a distribution, limits the regularity of solutions, and appropriate approximations of the Dirac functions have to be implemented see \cite{Engquist2005, Fedele2017}.}

As the NSB equations are obtained by adding a Darcy drag term 
to the classical Navier Stokes equations 
the approach lends itself easily to extend existing finite element based CFD solver.
All required steps are implemented purely on the element level 
and as such are well suited for single-core as well as highly parallel HPC simulations.
The presented approach offers benefits in \reviewerTwo{senarii} where the kinematics of the moving object is cyclic, 
such as the repetitive closing of heart valves or the rotation of a blood pump 
at regular angular velocity. 
In such scenarios Arbitrary Eulerian Lagrangian methods would fail 
-- without any additional potentially expensive remeshing --
due to a change in topology of the domain over time.
The proposed RBVMS formulation offers stabilization and turbulence modeling 
without introducing any additional equations 
and without the need for tuning parameters.
For generating the time varying permeability fields 
we implemented an algorithm based on ray tracing.
This can be done either on-the-fly or in a pre-processing step.
When dealing with cyclic motions of an obstacle 
or when using image-driven kinematics models of cyclic motions 
such as a heart beat,
it is computationally advantageous to compute permeability fields for one cycle
in a pre-processing step and reuse the pre-computed fields in subsequent cycles. 

\reviewerOne{Our first benchmark shows, that almost optimal convergence orders, $h^2$ for velocity and $h^\frac{3}{2}$ for pressure in the $L_2$-norm, can be achieved, see \cite{Masud2009}.
For further improving convergence rate it is easy to incorporate a smoothed penalization field as has been recently shown in \cite{Hester2021}.}
Our numerical experiments and validation studies, like the round-robin FDA benchmark, indicate 
that our method is robust and sufficiently accurate.
Our method performed particularly well in the medium-range Reynolds number regime, 
but accuracy degraded when applied in a higher Reynolds number regime,
as indicated in \ref{sec:akh:numeric}.
In our view this does not indicate a fundamental limitation of the method \emph{per se}.
Rather, we believe that degradation in accuracy is attributable to the mesh resolution used
which might be too coarse for a higher Reynolds number regime. 
However, owing to the cost of repeating simulations at even higher spatial resolution
this has not been thoroughly investigated and will be addressed in further works.

Another major limitation relates to the underlying assumption
of negligible coupling between immersed obstacle and fluid.
In our formulation the kinematics of obstacles over time is assumed to be given \emph{a priori}
and is not influenced in any way by the motion of the surrounding fluid.
Thus the motion of obstacles can be prescribed.
Such a non-FSI approach having a given prescribed motion acting on the surrounding fluid
is suitable mostly when considering stiff objects like 
wind turbines, rotary blood pumps, or stiff prosthetic heart valves.
When soft biological tissues are considered such as heart valves 
the method is limited, but may still be applicable depending 
on the specifics of the physical effects under investigation.
For instance, if the influence of a heart valve in a given configuration,
i.e.\ in a closed or open state,
upon hemodynamics is under investigation, our method is perfectly suitable.
In pathological cases valves can be approximate quite well as stiff objects 
as they tend to be very stiff and do not undergo any large deformations.
On the other hand, the behavior of healthy valves is close to a perfect diode,
that is, in the open state the valve does not impose any obstacle to flow
and in the close state any flow is impeded.
If the transient flow-driven motion of the valve 
when switching from one configuration, the surface traction on the leaflets of the valves due to the flow or the high frequency wave-like motion of the valves within an outflow jet 
are of interest, our method in its current implementation is not suitable. 
If one is therefore interested in the effect of a fluid on an obstacle, 
that is deformed or co-transported by the fluid, 
and in resolving detailed multi-physics mechanisms around the fluid-obstacle interface, 
other fully coupled standard FSI approaches are the way to go in our opinion.

In subsequent works we plan to extend our current methodology to moving fluid domains 
in the context of image-driven kinematic models of four chamber heart simulations.
A further focus will be on investigating methods towards achieving a bidirectional coupling
where kinematics of an obstacle can be governed by pressure gradients,
for instance, where motion of a heart valve is governed by the pressure difference
between cardiac chamber and outflow tract.
%
\section*{Acknowledgment}
This research was supported by the Grants F3210-N18 and I2760-B30 from the Austrian Science Fund
(FWF) awarded to GP and by BioTechMed-Graz (Grant No. Flagship Project: ILearnHeart) 
awarded to GH and GP. 
We further acknowledge support by NAWI Graz
and by the PRACE project \enquote{71138: Image-based Learning in Predictive Personalized Models of Total Heart Function} for awarding us access to the Austrian HPC resources VSC3 and VSC4.

\section*{CRediT Author Statement}
\textbf{Jana Fuchsberger:} Conceptualization, Methodology, Software, Data Curation, Validation, Formal analysis, Writing - Original Draft, Writing - Review \& Editing, Visualization \textbf{Elias Karabelas:} Conceptualization, Methodology, Software, Data Curation, Validation, Supervision, Writing - Original Draft, Writing - Review \& Editing, Visualization; \textbf{Gundolf Haase:} Conceptualization, Resources, Writing - Review \& Editing, Supervision, Project administration, Funding acquisition; \textbf{Gernot Plank:} Conceptualization, Resources, Writing - Review \& Editing, Supervision, Project administration, Funding acquisition; \textbf{Steve Niederer:} Resources, Writing - Review \& Editing, Supervision, Funding acquisition; \textbf{Philipp Aigner:} Validation, Data Curation, Investigation, Visualization, Writing - Review \& Editing; \textbf{Heinrich Schima:} Supervision, Resources, Investigation, Writing - Review \& Editing
\section*{Declaration of Interests}
The authors declare that they have no known competing financial interests or personal relationships that could have appeared to influence the work reported in this paper.
\bibliographystyle{elsarticle-num}
\bibliography{references}

\begin{thebibliography}{10}
\expandafter\ifx\csname url\endcsname\relax
  \def\url#1{\texttt{#1}}\fi
\expandafter\ifx\csname urlprefix\endcsname\relax\def\urlprefix{URL }\fi
\expandafter\ifx\csname href\endcsname\relax
  \def\href#1#2{#2} \def\path#1{#1}\fi

\bibitem{2006}
H.-J. Bungartz, M.~Sch\"{a}fer (Eds.), Fluid-Structure Interaction, Springer
  Berlin Heidelberg, 2006.
\newblock \href {http://dx.doi.org/10.1007/3-540-34596-5}
  {\path{doi:10.1007/3-540-34596-5}}.

\bibitem{vanLoon2007}
R.~van Loon, P.~Anderson, F.~van~de Vosse, S.~Sherwin, Comparison of various
  fluid{\textendash}structure interaction methods for deformable bodies,
  Computers {\&} Structures 85~(11-14) (2007) 833--843.
\newblock \href {http://dx.doi.org/10.1016/j.compstruc.2007.01.010}
  {\path{doi:10.1016/j.compstruc.2007.01.010}}.

\bibitem{Behr2001}
M.~Behr, T.~Tezduyar, Shear-slip mesh update in 3d computation of complex flow
  problems with rotating mechanical components, Computer Methods in Applied
  Mechanics and Engineering 190~(24-25) (2001) 3189--3200.
\newblock \href {http://dx.doi.org/10.1016/s0045-7825(00)00388-1}
  {\path{doi:10.1016/s0045-7825(00)00388-1}}.

\bibitem{Chandran2010}
K.~B. Chandran, Role of computational simulations in heart valve dynamics and
  design of valvular prostheses, Cardiovascular Engineering and Technology
  1~(1) (2010) 18--38.
\newblock \href {http://dx.doi.org/10.1007/s13239-010-0002-x}
  {\path{doi:10.1007/s13239-010-0002-x}}.

\bibitem{Astorino2009}
M.~Astorino, J.-F. Gerbeau, O.~Pantz, K.-F. Traor{\'{e}},
  Fluid{\textendash}structure interaction and multi-body contact: Application
  to aortic valves, Computer Methods in Applied Mechanics and Engineering
  198~(45-46) (2009) 3603--3612.
\newblock \href {http://dx.doi.org/10.1016/j.cma.2008.09.012}
  {\path{doi:10.1016/j.cma.2008.09.012}}.

\bibitem{DinizdosSantos2008}
N.~D. dos Santos, J.-F. Gerbeau, J.-F. Bourgat, A partitioned
  fluid{\textendash}structure algorithm for elastic thin valves with contact,
  Computer Methods in Applied Mechanics and Engineering 197~(19-20) (2008)
  1750--1761.
\newblock \href {http://dx.doi.org/10.1016/j.cma.2007.03.019}
  {\path{doi:10.1016/j.cma.2007.03.019}}.

\bibitem{Weinberg2007}
E.~J. Weinberg, M.~R.~K. Mofrad, A finite shell element for heart mitral valve
  leaflet mechanics, with large deformations and 3d constitutive material
  model, Journal of Biomechanics 40~(3) (2007) 705--711.
\newblock \href {http://dx.doi.org/10.1016/j.jbiomech.2006.01.003}
  {\path{doi:10.1016/j.jbiomech.2006.01.003}}.

\bibitem{vanLoon2004}
R.~van Loon, P.~D. Anderson, J.~de~Hart, F.~P.~T. Baaijens, A combined
  fictitious domain/adaptive meshing method for fluid{\textendash}structure
  interaction in heart valves, International Journal for Numerical Methods in
  Fluids 46~(5) (2004) 533--544.
\newblock \href {http://dx.doi.org/10.1002/fld.775}
  {\path{doi:10.1002/fld.775}}.

\bibitem{McQueen2001}
D.~M. McQueen, C.~S. Peskin, Heart simulation by an immersed boundary method
  with formal second-order accuracy and reduced numerical viscosity, in:
  Mechanics for a New Mellennium, Kluwer Academic Publishers, 2001, pp.
  429--444.
\newblock \href {http://dx.doi.org/10.1007/0-306-46956-1_27}
  {\path{doi:10.1007/0-306-46956-1_27}}.

\bibitem{Maisano2005}
F.~Maisano, A.~Redaelli, M.~Soncini, E.~Votta, L.~Arcobasso, O.~Alfieri, An
  annular prosthesis for the treatment of functional mitral regurgitation:
  Finite element model analysis of a dog bone{\textendash}shaped ring
  prosthesis, The Annals of Thoracic Surgery 79~(4) (2005) 1268--1275.
\newblock \href {http://dx.doi.org/10.1016/j.athoracsur.2004.04.014}
  {\path{doi:10.1016/j.athoracsur.2004.04.014}}.

\bibitem{Wenk2010}
J.~F. Wenk, Z.~Zhang, G.~Cheng, D.~Malhotra, G.~Acevedo-Bolton, M.~Burger,
  T.~Suzuki, D.~A. Saloner, A.~W. Wallace, J.~M. Guccione, M.~B. Ratcliffe,
  First finite element model of the left ventricle with mitral valve: Insights
  into ischemic mitral regurgitation, The Annals of Thoracic Surgery 89~(5)
  (2010) 1546--1553.
\newblock \href {http://dx.doi.org/10.1016/j.athoracsur.2010.02.036}
  {\path{doi:10.1016/j.athoracsur.2010.02.036}}.

\bibitem{Terahara2020}
T.~Terahara, K.~Takizawa, T.~E. Tezduyar, Y.~Bazilevs, M.-C. Hsu, Heart valve
  isogeometric sequentially-coupled {FSI} analysis with the
  space{\textendash}time topology change method, Computational Mechanics 65~(4)
  (2020) 1167--1187.
\newblock \href {http://dx.doi.org/10.1007/s00466-019-01813-0}
  {\path{doi:10.1007/s00466-019-01813-0}}.

\bibitem{Antonietti2019}
P.~Antonietti, M.~Verani, C.~Vergara, S.~Zonca, Numerical solution of
  fluid-structure interaction problems by means of a high order discontinuous
  galerkin method on polygonal grids, Finite Elements in Analysis and Design
  159 (2019) 1--14.
\newblock \href {http://dx.doi.org/10.1016/j.finel.2019.02.002}
  {\path{doi:10.1016/j.finel.2019.02.002}}.

\bibitem{Alauzet2016}
F.~Alauzet, B.~Fabr{\`{e}}ges, M.~A. Fern{\'{a}}ndez, M.~Landajuela,
  Nitsche-{XFEM} for the coupling of an incompressible fluid with immersed
  thin-walled structures, Computer Methods in Applied Mechanics and Engineering
  301 (2016) 300--335.
\newblock \href {http://dx.doi.org/10.1016/j.cma.2015.12.015}
  {\path{doi:10.1016/j.cma.2015.12.015}}.

\bibitem{Zonca2018}
S.~Zonca, C.~Vergara, L.~Formaggia, An unfitted formulation for the interaction
  of an incompressible fluid with a thick structure via an {XFEM}/{DG}
  approach, {SIAM} Journal on Scientific Computing 40~(1) (2018) B59--B84.
\newblock \href {http://dx.doi.org/10.1137/16m1097602}
  {\path{doi:10.1137/16m1097602}}.

\bibitem{Massing2015}
A.~Massing, M.~Larson, A.~Logg, M.~Rognes, A nitsche-based cut finite element
  method for a fluid-structure interaction problem, Communications in Applied
  Mathematics and Computational Science 10~(2) (2015) 97--120.
\newblock \href {http://dx.doi.org/10.2140/camcos.2015.10.97}
  {\path{doi:10.2140/camcos.2015.10.97}}.

\bibitem{Razeghi2020}
O.~Razeghi, J.~A. Sol{\'{\i}}s-Lemus, A.~W. Lee, R.~Karim, C.~Corrado, C.~H.
  Roney, A.~de~Vecchi, S.~A. Niederer, {CemrgApp}: An interactive medical
  imaging application with image processing, computer vision, and machine
  learning toolkits for cardiovascular research, {SoftwareX} 12 (2020) 100570.
\newblock \href {http://dx.doi.org/10.1016/j.softx.2020.100570}
  {\path{doi:10.1016/j.softx.2020.100570}}.

\bibitem{Rueckert1999}
D.~Rueckert, L.~Sonoda, C.~Hayes, D.~Hill, M.~Leach, D.~Hawkes, Nonrigid
  registration using free-form deformations: application to breast {MR} images,
  {IEEE} Transactions on Medical Imaging 18~(8) (1999) 712--721.
\newblock \href {http://dx.doi.org/10.1109/42.796284}
  {\path{doi:10.1109/42.796284}}.

\bibitem{Shi2013}
W.~Shi, M.~Jantsch, P.~Aljabar, L.~Pizarro, W.~Bai, H.~Wang, D.~O'Regan,
  X.~Zhuang, D.~Rueckert, Temporal sparse free-form deformations, Medical Image
  Analysis 17~(7) (2013) 779--789.
\newblock \href {http://dx.doi.org/10.1016/j.media.2013.04.010}
  {\path{doi:10.1016/j.media.2013.04.010}}.

\bibitem{Peskin1972}
C.~S. Peskin, Flow patterns around heart valves: A numerical method, Journal of
  Computational Physics 10~(2) (1972) 252--271.
\newblock \href {http://dx.doi.org/10.1016/0021-9991(72)90065-4}
  {\path{doi:10.1016/0021-9991(72)90065-4}}.

\bibitem{Mittal2005}
R.~Mittal, G.~Iaccarino, {IMMERSED} {BOUNDARY} {METHODS}, Annual Review of
  Fluid Mechanics 37~(1) (2005) 239--261.
\newblock \href {http://dx.doi.org/10.1146/annurev.fluid.37.061903.175743}
  {\path{doi:10.1146/annurev.fluid.37.061903.175743}}.

\bibitem{Astorino2012}
M.~Astorino, J.~Hamers, S.~C. Shadden, J.-F. Gerbeau, A robust and efficient
  valve model based on resistive immersed surfaces, International Journal for
  Numerical Methods in Biomedical Engineering 28~(9) (2012) 937--959.
\newblock \href {http://dx.doi.org/10.1002/cnm.2474}
  {\path{doi:10.1002/cnm.2474}}.

\bibitem{Yao2012}
J.~Yao, G.~R. Liu, D.~A. Narmoneva, R.~B. Hinton, Z.-Q. Zhang, Immersed
  smoothed finite element method for fluid{\textendash}structure interaction
  simulation of aortic valves, Computational Mechanics 50~(6) (2012) 789--804.
\newblock \href {http://dx.doi.org/10.1007/s00466-012-0781-z}
  {\path{doi:10.1007/s00466-012-0781-z}}.

\bibitem{Votta2013}
E.~Votta, T.~B. Le, M.~Stevanella, L.~Fusini, E.~G. Caiani, A.~Redaelli,
  F.~Sotiropoulos, Toward patient-specific simulations of cardiac valves:
  State-of-the-art and future directions, Journal of Biomechanics 46~(2) (2013)
  217--228.
\newblock \href {http://dx.doi.org/10.1016/j.jbiomech.2012.10.026}
  {\path{doi:10.1016/j.jbiomech.2012.10.026}}.

\bibitem{Chnafa2014}
C.~Chnafa, S.~Mendez, F.~Nicoud, Image-based large-eddy simulation in a
  realistic left heart, Computers {\&} Fluids 94 (2014) 173--187.
\newblock \href {http://dx.doi.org/10.1016/j.compfluid.2014.01.030}
  {\path{doi:10.1016/j.compfluid.2014.01.030}}.

\bibitem{Goldstein1993}
D.~Goldstein, R.~Handler, L.~Sirovich, Modeling a no-slip flow boundary with an
  external force field, Journal of Computational Physics 105~(2) (1993)
  354--366.
\newblock \href {http://dx.doi.org/10.1006/jcph.1993.1081}
  {\path{doi:10.1006/jcph.1993.1081}}.

\bibitem{Fadlun2000}
E.~Fadlun, R.~Verzicco, P.~Orlandi, J.~Mohd-Yusof, Combined immersed-boundary
  finite-difference methods for three-dimensional complex flow simulations,
  Journal of Computational Physics 161~(1) (2000) 35--60.
\newblock \href {http://dx.doi.org/10.1006/jcph.2000.6484}
  {\path{doi:10.1006/jcph.2000.6484}}.

\bibitem{arquis1984conditions}
E.~Arquis, J.~Caltagirone, Sur les conditions hydrodynamiques au voisinage
  d’une interface milieu fluide-milieu poreux: application a la convection
  naturelle., CR Acad. Sci. Paris II 299 (1984) 1--4.

\bibitem{Khadra2000}
K.~Khadra, P.~Angot, S.~Parneix, J.-P. Caltagirone, Fictitious domain approach
  for numerical modelling of navier–stokes equations, International Journal
  for Numerical Methods in Fluids 34~(8) (2000) 651--684.
\newblock \href
  {http://dx.doi.org/10.1002/1097-0363(20001230)34:8<651::AID-FLD61>3.0.CO;2-D}
  {\path{doi:10.1002/1097-0363(20001230)34:8<651::AID-FLD61>3.0.CO;2-D}}.

\bibitem{carbou2003}
G.~Carbou, P.~Fabrie,
  \href{https://projecteuclid.org:443/euclid.ade/1355867981}{Boundary layer for
  a penalization method for viscous incompressible flow}, Adv. Differential
  Equations 8~(12) (2003) 1453--1480.
\newline\urlprefix\url{https://projecteuclid.org:443/euclid.ade/1355867981}

\bibitem{Engels2016}
T.~Engels, D.~Kolomenskiy, K.~Schneider, J.~Sesterhenn, {FluSI}: A novel
  parallel simulation tool for flapping insect flight using a fourier method
  with volume penalization, {SIAM} Journal on Scientific Computing 38~(5)
  (2016) S3--S24.
\newblock \href {http://dx.doi.org/10.1137/15m1026006}
  {\path{doi:10.1137/15m1026006}}.

\bibitem{Engels2016:2}
T.~Engels, D.~Kolomenskiy, K.~Schneider, F.-O. Lehmann, J.~Sesterhenn,
  Bumblebee flight in heavy turbulence, Physical Review Letters 116~(2).
\newblock \href {http://dx.doi.org/10.1103/physrevlett.116.028103}
  {\path{doi:10.1103/physrevlett.116.028103}}.

\bibitem{Engels2018}
T.~Engels, D.~Kolomenskiy, K.~Schneider, M.~Farge, F.-O. Lehmann,
  J.~Sesterhenn, \href{https://doi.org/10.1088/1873-7005/aa908f}{Helical
  vortices generated by flapping wings of bumblebees}, Fluid Dynamics Research
  50~(1) (2018) 011419.
\newblock \href {http://dx.doi.org/10.1088/1873-7005/aa908f}
  {\path{doi:10.1088/1873-7005/aa908f}}.
\newline\urlprefix\url{https://doi.org/10.1088/1873-7005/aa908f}

\bibitem{Bazilevs2007}
Y.~Bazilevs, V.~Calo, J.~Cottrell, T.~Hughes, A.~Reali, G.~Scovazzi,
  Variational multiscale residual-based turbulence modeling for large eddy
  simulation of incompressible flows, Computer Methods in Applied Mechanics and
  Engineering 197~(1-4) (2007) 173--201.
\newblock \href {http://dx.doi.org/10.1016/j.cma.2007.07.016}
  {\path{doi:10.1016/j.cma.2007.07.016}}.

\bibitem{Bazilevs2013}
Y.~Bazilevs, K.~Takizawa, T.~Tezduyar, Computational Fluid-Structure
  Interaction: Methods and Applications, John Wiley and Sons, 2013.
\newblock \href {http://dx.doi.org/10.1002/9781118483565}
  {\path{doi:10.1002/9781118483565}}.

\bibitem{Malinauskas2017fda}
R.~A. Malinauskas, P.~Hariharan, S.~W. Day, L.~H. Herbertson, M.~Buesen,
  U.~Steinseifer, K.~I. Aycock, B.~C. Good, S.~Deutsch, K.~B. Manning, B.~A.
  Craven, {FDA} benchmark medical device flow models for {CFD} validation,
  {ASAIO} Journal 63~(2) (2017) 150--160.
\newblock \href {http://dx.doi.org/10.1097/mat.0000000000000499}
  {\path{doi:10.1097/mat.0000000000000499}}.

\bibitem{Angot1999feb}
P.~Angot, C.-H. Bruneau, P.~Fabrie, A penalization method to take into account
  obstacles in incompressible viscous flows, Numerische Mathematik 81~(4)
  (1999) 497--520.
\newblock \href {http://dx.doi.org/10.1007/s002110050401}
  {\path{doi:10.1007/s002110050401}}.

\bibitem{Blank2020}
L.~Blank, E.~M. Rioseco, A.~Caiazzo, U.~Wilbrandt, Modeling, simulation, and
  optimization of geothermal energy production from hot sedimentary aquifers,
  Computational Geosciences 25~(1) (2020) 67--104.
\newblock \href {http://dx.doi.org/10.1007/s10596-020-09989-8}
  {\path{doi:10.1007/s10596-020-09989-8}}.

\bibitem{EsmailyMoghadam2011}
M.~E. Moghadam, , Y.~Bazilevs, T.-Y. Hsia, I.~E. Vignon-Clementel, A.~L.
  Marsden, A comparison of outlet boundary treatments for prevention of
  backflow divergence with relevance to blood flow simulations, Computational
  Mechanics 48~(3) (2011) 277--291.
\newblock \href {http://dx.doi.org/10.1007/s00466-011-0599-0}
  {\path{doi:10.1007/s00466-011-0599-0}}.

\bibitem{Braack2014}
M.~Braack, P.~B. Mucha, Directional do-nothing condition for the navier-stokes
  equations, Journal of Computational Mathematics 32~(5) (2014) 507--521.
\newblock \href {http://dx.doi.org/10.4208/jcm.1405-m4347}
  {\path{doi:10.4208/jcm.1405-m4347}}.

\bibitem{Angot1999}
P.~Angot, Analysis of singular perturbations on the brinkman problem for
  fictitious domain models of viscous flows, Mathematical Methods in the
  Applied Sciences 22~(16) (1999) 1395--1412.
\newblock \href
  {http://dx.doi.org/10.1002/(sici)1099-1476(19991110)22:16<1395::aid-mma84>3.0.co;2-3}
  {\path{doi:10.1002/(sici)1099-1476(19991110)22:16<1395::aid-mma84>3.0.co;2-3}}.

\bibitem{aguayo2020analysis}
J.~Aguayo, H.~Carrillo, Analysis of obstacles immersed in viscous fluids using
  brinkman's law for steady stokes and navier-stokes equations (2020).
\newblock \href {http://arxiv.org/abs/2012.08635} {\path{arXiv:2012.08635}}.

\bibitem{Ingram2011}
R.~Ingram, Finite element approximation of nonsolenoidal, viscous flows around
  porous and solid obstacles, {SIAM} Journal on Numerical Analysis 49~(2)
  (2011) 491--520.
\newblock \href {http://dx.doi.org/10.1137/090765341}
  {\path{doi:10.1137/090765341}}.

\bibitem{brenner2007mathematical}
S.~Brenner, R.~Scott, The mathematical theory of finite element methods,
  Vol.~15, Springer Science \& Business Media, 2007.

\bibitem{steinbach2007numerical}
O.~Steinbach, Numerical approximation methods for elliptic boundary value
  problems: finite and boundary elements, Springer Science \& Business Media,
  2007.

\bibitem{Karabelas2018}
E.~Karabelas, M.~A.~F. Gsell, C.~M. Augustin, L.~Marx, A.~Neic, A.~J. Prassl,
  L.~Goubergrits, T.~Kuehne, G.~Plank, Towards a computational framework for
  modeling the impact of aortic coarctations upon left ventricular load,
  Frontiers in Physiology 9.
\newblock \href {http://dx.doi.org/10.3389/fphys.2018.00538}
  {\path{doi:10.3389/fphys.2018.00538}}.

\bibitem{PhDPauli}
L.~H. Pauli, Stabilized finite element methods for computational design of
  blood-handling devices, Ph.D. thesis, RWTH Aachen University (6 2016).

\bibitem{Harari1992}
I.~Harari, T.~J. Hughes, What are c and h?: Inequalities for the analysis and
  design of finite element methods, Computer Methods in Applied Mechanics and
  Engineering 97~(2) (1992) 157--192.
\newblock \href {http://dx.doi.org/10.1016/0045-7825(92)90162-d}
  {\path{doi:10.1016/0045-7825(92)90162-d}}.

\bibitem{Forti2015}
D.~Forti, L.~Ded{\`{e}}, Semi-implicit {BDF} time discretization of the
  navier{\textendash}stokes equations with {VMS}-{LES} modeling in a high
  performance computing framework, Computers {\&} Fluids 117 (2015) 168--182.
\newblock \href {http://dx.doi.org/10.1016/j.compfluid.2015.05.011}
  {\path{doi:10.1016/j.compfluid.2015.05.011}}.

\bibitem{Moeller1997}
T.~M\"{o}ller, B.~Trumbore, Fast, minimum storage ray-triangle intersection,
  Journal of Graphics Tools 2~(1) (1997) 21--28.
\newblock \href {http://dx.doi.org/10.1080/10867651.1997.10487468}
  {\path{doi:10.1080/10867651.1997.10487468}}.

\bibitem{Haines1994}
E.~Haines, Point in polygon strategies, in: Graphics Gems, Elsevier, 1994, pp.
  24--46.
\newblock \href {http://dx.doi.org/10.1016/b978-0-12-336156-1.50013-6}
  {\path{doi:10.1016/b978-0-12-336156-1.50013-6}}.

\bibitem{Neic2020}
A.~Neic, M.~A. Gsell, E.~Karabelas, A.~J. Prassl, G.~Plank, Automating
  image-based mesh generation and manipulation tasks in cardiac modeling
  workflows using meshtool, {SoftwareX} 11 (2020) 100454.
\newblock \href {http://dx.doi.org/10.1016/j.softx.2020.100454}
  {\path{doi:10.1016/j.softx.2020.100454}}.

\bibitem{Wan2004}
D.~Wan, S.~Turek, L.~S. Rivkind, An efficient multigrid {FEM} solution
  technique for incompressible flow with moving rigid bodies, in: Numerical
  Mathematics and Advanced Applications, Springer Berlin Heidelberg, 2004, pp.
  844--853.
\newblock \href {http://dx.doi.org/10.1007/978-3-642-18775-9_83}
  {\path{doi:10.1007/978-3-642-18775-9_83}}.

\bibitem{Brackbill1992}
J.~Brackbill, D.~Kothe, C.~Zemach, A continuum method for modeling surface
  tension, Journal of Computational Physics 100~(2) (1992) 335--354.
\newblock \href {http://dx.doi.org/10.1016/0021-9991(92)90240-y}
  {\path{doi:10.1016/0021-9991(92)90240-y}}.

\bibitem{Vigmond2003}
E.~J. Vigmond, M.~Hughes, G.~Plank, L.~Leon, Computational tools for modeling
  electrical activity in cardiac tissue, Journal of Electrocardiology 36 (2003)
  69--74.
\newblock \href {http://dx.doi.org/10.1016/j.jelectrocard.2003.09.017}
  {\path{doi:10.1016/j.jelectrocard.2003.09.017}}.

\bibitem{Jansen2000}
K.~E. Jansen, C.~H. Whiting, G.~M. Hulbert, A generalized-$\alpha$ method for
  integrating the filtered navier{\textendash}stokes equations with a
  stabilized finite element method, Computer Methods in Applied Mechanics and
  Engineering 190~(3-4) (2000) 305--319.
\newblock \href {http://dx.doi.org/10.1016/s0045-7825(00)00203-6}
  {\path{doi:10.1016/s0045-7825(00)00203-6}}.

\bibitem{Liu2020}
J.~Liu, I.~S. Lan, O.~Z. Tikenogullari, A.~L. Marsden, A note on the accuracy
  of the generalized-$\alpha$ scheme for the incompressible {N}avier-{S}tokes
  equations, International Journal for Numerical Methods in Engineering 122~(2)
  (2020) 638--651.
\newblock \href {http://dx.doi.org/10.1002/nme.6550}
  {\path{doi:10.1002/nme.6550}}.

\bibitem{petsc-web-page}
S.~Balay, S.~Abhyankar, M.~F. Adams, J.~Brown, P.~Brune, K.~Buschelman,
  L.~Dalcin, A.~Dener, V.~Eijkhout, W.~D. Gropp, D.~Karpeyev, D.~Kaushik, M.~G.
  Knepley, D.~A. May, L.~C. McInnes, R.~T. Mills, T.~Munson, K.~Rupp, P.~Sanan,
  B.~F. Smith, S.~Zampini, H.~Zhang, H.~Zhang,
  \href{https://www.mcs.anl.gov/petsc}{{PETS}c {W}eb page},
  \url{https://www.mcs.anl.gov/petsc} (2019).
\newline\urlprefix\url{https://www.mcs.anl.gov/petsc}

\bibitem{petsc-user-ref}
S.~Balay, S.~Abhyankar, M.~F. Adams, J.~Brown, P.~Brune, K.~Buschelman,
  L.~Dalcin, A.~Dener, V.~Eijkhout, W.~D. Gropp, D.~Karpeyev, D.~Kaushik, M.~G.
  Knepley, D.~A. May, L.~C. McInnes, R.~T. Mills, T.~Munson, K.~Rupp, P.~Sanan,
  B.~F. Smith, S.~Zampini, H.~Zhang, H.~Zhang,
  \href{https://www.mcs.anl.gov/petsc}{{PETS}c users manual}, Tech. Rep.
  ANL-95/11 - Revision 3.13, Argonne National Laboratory (2020).
\newline\urlprefix\url{https://www.mcs.anl.gov/petsc}

\bibitem{petsc-efficient}
S.~Balay, W.~D. Gropp, L.~C. McInnes, B.~F. Smith, Efficient management of
  parallelism in object oriented numerical software libraries, in: E.~Arge,
  A.~M. Bruaset, H.~P. Langtangen (Eds.), Modern Software Tools in Scientific
  Computing, Birkh{\"{a}}user Press, 1997, pp. 163--202.

\bibitem{Henson2002}
V.~E. Henson, U.~M. Yang, {BoomerAMG}: A parallel algebraic multigrid solver
  and preconditioner, Applied Numerical Mathematics 41~(1) (2002) 155--177.
\newblock \href {http://dx.doi.org/10.1016/s0168-9274(01)00115-5}
  {\path{doi:10.1016/s0168-9274(01)00115-5}}.

\bibitem{Augustin2016}
C.~M. Augustin, A.~Neic, M.~Liebmann, A.~J. Prassl, S.~A. Niederer, G.~Haase,
  G.~Plank, Anatomically accurate high resolution modeling of human whole heart
  electromechanics: A strongly scalable algebraic multigrid solver method for
  nonlinear deformation, Journal of Computational Physics 305 (2016) 622--646.
\newblock \href {http://dx.doi.org/10.1016/j.jcp.2015.10.045}
  {\path{doi:10.1016/j.jcp.2015.10.045}}.

\bibitem{Karabelas2019}
E.~Karabelas, G.~Haase, G.~Plank, C.~M. Augustin, Versatile stabilized finite
  element formulations for nearly and fully incompressible solid mechanics,
  Computational Mechanics 65~(1) (2019) 193--215.
\newblock \href {http://dx.doi.org/10.1007/s00466-019-01760-w}
  {\path{doi:10.1007/s00466-019-01760-w}}.

\bibitem{Vigmond2008}
E.~Vigmond, R.~W. dos Santos, A.~Prassl, M.~Deo, G.~Plank, Solvers for the
  cardiac bidomain equations, Progress in Biophysics and Molecular Biology
  96~(1-3) (2008) 3--18.
\newblock \href {http://dx.doi.org/10.1016/j.pbiomolbio.2007.07.012}
  {\path{doi:10.1016/j.pbiomolbio.2007.07.012}}.

\bibitem{Udaykumar2001}
H.~Udaykumar, R.~Mittal, P.~Rampunggoon, A.~Khanna, A sharp interface cartesian
  grid method for simulating flows with complex moving boundaries, Journal of
  Computational Physics 174~(1) (2001) 345--380.
\newblock \href {http://dx.doi.org/10.1006/jcph.2001.6916}
  {\path{doi:10.1006/jcph.2001.6916}}.

\bibitem{Seo2011}
J.~H. Seo, R.~Mittal, A sharp-interface immersed boundary method with improved
  mass conservation and reduced spurious pressure oscillations, Journal of
  Computational Physics 230~(19) (2011) 7347--7363.
\newblock \href {http://dx.doi.org/10.1016/j.jcp.2011.06.003}
  {\path{doi:10.1016/j.jcp.2011.06.003}}.

\bibitem{Stern2001}
F.~Stern, R.~Wilson, H.~Coleman, E.~Paterson, Comprehensive approach to
  verification and validation of cfd simulations—part 1: Methodology and
  procedures, Journal of Fluids Engineering 123 (2001) 792.
\newblock \href {http://dx.doi.org/10.1115/1.1412235}
  {\path{doi:10.1115/1.1412235}}.

\bibitem{Roache1994}
P.~J. Roache, {Perspective: A Method for Uniform Reporting of Grid Refinement
  Studies}, Journal of Fluids Engineering 116~(3) (1994) 405--413.
\newblock \href {http://dx.doi.org/10.1115/1.2910291}
  {\path{doi:10.1115/1.2910291}}.

\bibitem{deVahlDavis1983}
G.~de~Vahl~Davis, {Natural convection of air in a square cavity: a bench mark
  numerical solution.}, International Journal for Numerical Methods in Fluids
  3~(3) (1983) 249--264.

\bibitem{Pope2004}
S.~B. Pope, \href{https://doi.org/10.1088/1367-2630/6/1/035}{Ten questions
  concerning the large-eddy simulation of turbulent flows}, New Journal of
  Physics 6 (2004) 35--35.
\newblock \href {http://dx.doi.org/10.1088/1367-2630/6/1/035}
  {\path{doi:10.1088/1367-2630/6/1/035}}.
\newline\urlprefix\url{https://doi.org/10.1088/1367-2630/6/1/035}

\bibitem{Marinova2016}
V.~Marinova, I.~Kerroumi, A.~Lintermann, J.~H. Göbbert, C.~Moulinec, S.~Rible,
  Y.~Fournier, M.~Behbahani,
  \href{http://hdl.handle.net/2128/10345}{{N}umerical {A}nalysis of the {FDA}
  {C}entrifugal {B}lood {P}ump}, in: NIC Symposium 2016, Vol.~48 of NIC Series,
  NIC Symposium 2016, Jülich (Germany), 11 Feb 2016 - 12 Feb 2016,
  Forschungszentrum Jülich GmbH, Zentralbibliothek, Jülich, 2016, pp.
  355--364.
\newline\urlprefix\url{http://hdl.handle.net/2128/10345}

\bibitem{Good2020}
B.~C. Good, K.~B. Manning, Computational modeling of the food and drug
  administration's benchmark centrifugal blood pump, Artificial Organs 44~(7).
\newblock \href {http://dx.doi.org/10.1111/aor.13643}
  {\path{doi:10.1111/aor.13643}}.

\bibitem{FDABenchHP}
Computational fluid dynamics round robin study,
  \url{https://ncihub.org/wiki/FDA_CFD/ComputationalRoundRobin2Pump}, accessed:
  2020-09-07.

\bibitem{Lazzaro2002}
D.~Lazzaro, L.~B. Montefusco, Radial basis functions for the multivariate
  interpolation of large scattered data sets, Journal of Computational and
  Applied Mathematics 140~(1-2) (2002) 521--536.
\newblock \href {http://dx.doi.org/10.1016/s0377-0427(01)00485-x}
  {\path{doi:10.1016/s0377-0427(01)00485-x}}.

\bibitem{Easthope80}
P.~Easthope, D.~Brooks, \href{http://europepmc.org/abstract/MED/7213990}{A
  comparison of rheological constitutive functions for whole human blood},
  Biorheology 17~(3) (1980) 235—247.
\newline\urlprefix\url{http://europepmc.org/abstract/MED/7213990}

\bibitem{Chien1970}
S.~Chien, Shear dependence of effective cell volume as a determinant of blood
  viscosity, Science 168~(3934) (1970) 977--979.
\newblock \href {http://dx.doi.org/10.1126/science.168.3934.977}
  {\path{doi:10.1126/science.168.3934.977}}.

\bibitem{Hariharan2018}
P.~Hariharan, K.~I. Aycock, M.~Buesen, S.~W. Day, B.~C. Good, L.~H. Herbertson,
  U.~Steinseifer, K.~B. Manning, B.~A. Craven, R.~A. Malinauskas,
  Inter-laboratory characterization of the velocity field in the {FDA} blood
  pump model using particle image velocimetry ({PIV}), Cardiovascular
  Engineering and Technology 9~(4) (2018) 623--640.
\newblock \href {http://dx.doi.org/10.1007/s13239-018-00378-y}
  {\path{doi:10.1007/s13239-018-00378-y}}.

\bibitem{Piquet2016}
A.~Piquet, O.~Roussel, A.~Hadjadj, A comparative study of brinkman penalization
  and direct-forcing immersed boundary methods for compressible viscous flows,
  Computers {\&} Fluids 136 (2016) 272--284.
\newblock \href {http://dx.doi.org/10.1016/j.compfluid.2016.06.001}
  {\path{doi:10.1016/j.compfluid.2016.06.001}}.

\bibitem{Specklin2018}
M.~Specklin, Y.~Delaur{\'{e}}, A sharp immersed boundary method based on
  penalization and its application to moving boundaries and turbulent rotating
  flows, European Journal of Mechanics - B/Fluids 70 (2018) 130--147.
\newblock \href {http://dx.doi.org/10.1016/j.euromechflu.2018.03.003}
  {\path{doi:10.1016/j.euromechflu.2018.03.003}}.

\bibitem{Abgrall2014}
R.~Abgrall, H.~Beaugendre, C.~Dobrzynski, An immersed boundary method using
  unstructured anisotropic mesh adaptation combined with level-sets and
  penalization techniques, Journal of Computational Physics 257 (2014) 83--101.
\newblock \href {http://dx.doi.org/10.1016/j.jcp.2013.08.052}
  {\path{doi:10.1016/j.jcp.2013.08.052}}.

\bibitem{Ouro2015}
P.~Ouro, L.~Cea, L.~Ram{\'{\i}}rez, X.~Nogueira, An immersed boundary method
  for unstructured meshes in depth averaged shallow water models, International
  Journal for Numerical Methods in Fluids 81~(11) (2015) 672--688.
\newblock \href {http://dx.doi.org/10.1002/fld.4201}
  {\path{doi:10.1002/fld.4201}}.

\bibitem{Fedele2017}
M.~Fedele, E.~Faggiano, L.~Ded{\`{e}}, A.~Quarteroni, A patient-specific aortic
  valve model based on moving resistive immersed implicit surfaces,
  Biomechanics and Modeling in Mechanobiology 16~(5) (2017) 1779--1803.
\newblock \href {http://dx.doi.org/10.1007/s10237-017-0919-1}
  {\path{doi:10.1007/s10237-017-0919-1}}.

\bibitem{Fernndez2008}
M.~A. Fern{\'{a}}ndez, J.-F. Gerbeau, V.~Martin, Numerical simulation of blood
  flows through a porous interface, {ESAIM}: Mathematical Modelling and
  Numerical Analysis 42~(6) (2008) 961--990.
\newblock \href {http://dx.doi.org/10.1051/m2an:2008031}
  {\path{doi:10.1051/m2an:2008031}}.

\bibitem{Engquist2005}
B.~Engquist, A.-K. Tornberg, R.~Tsai, Discretization of dirac delta functions
  in level set methods, Journal of Computational Physics 207~(1) (2005) 28--51.
\newblock \href {http://dx.doi.org/10.1016/j.jcp.2004.09.018}
  {\path{doi:10.1016/j.jcp.2004.09.018}}.

\bibitem{Masud2009}
A.~Masud, R.~Calderer, A variational multiscale stabilized formulation for the
  incompressible navier{\textendash}stokes equations, Computational Mechanics
  44~(2) (2009) 145--160.
\newblock \href {http://dx.doi.org/10.1007/s00466-008-0362-3}
  {\path{doi:10.1007/s00466-008-0362-3}}.

\bibitem{Hester2021}
E.~W. Hester, G.~M. Vasil, K.~J. Burns, Improving accuracy of volume penalised
  fluid-solid interactions, Journal of Computational Physics 430 (2021) 110043.
\newblock \href {http://dx.doi.org/10.1016/j.jcp.2020.110043}
  {\path{doi:10.1016/j.jcp.2020.110043}}.

\bibitem{Stoiber2013}
M.~Stoiber, T.~Schl\"{o}glhofer, P.~Aigner, C.~Grasl, H.~Schima, An alternative
  method to create highly transparent hollow models for flow visualization, The
  International Journal of Artificial Organs 36~(2) (2013) 131--134.
\newblock \href {http://dx.doi.org/10.5301/ijao.5000171}
  {\path{doi:10.5301/ijao.5000171}}.

\bibitem{raffel2018particle}
M.~Raffel, C.~E. Willert, F.~Scarano, C.~J. K{\"a}hler, S.~T. Wereley,
  J.~Kompenhans, Particle image velocimetry: a practical guide, Springer, 2018.

\bibitem{Fuchsberger2019}
J.~Fuchsberger, E.~Karabelas, P.~Aigner, H.~Schima, G.~Haase, G.~Plank,
  Validation study of computational fluid dynamics models of hemodynamics in
  the human aorta, {PAMM} 19~(1).
\newblock \href {http://dx.doi.org/10.1002/pamm.201900472}
  {\path{doi:10.1002/pamm.201900472}}.

\bibitem{Kao2015}
D.~L. Kao, J.~U. Ahmad, T.~Holst, Visualization and Quantification of Rotor Tip
  Vortices in Helicopter Flows, American Institute of Aeronautics and
  Astronautics, 2015.
\newblock \href {http://dx.doi.org/10.2514/6.2015-1369}
  {\path{doi:10.2514/6.2015-1369}}.

\bibitem{Longest2007}
P.~W. Longest, S.~Vinchurkar,
  \href{http://www.sciencedirect.com/science/article/pii/S135045330600107X}{Effects
  of mesh style and grid convergence on particle deposition in bifurcating
  airway models with comparisons to experimental data}, Medical Engineering \&
  Physics 29~(3) (2007) 350 -- 366.
\newblock \href
  {http://dx.doi.org/https://doi.org/10.1016/j.medengphy.2006.05.012}
  {\path{doi:https://doi.org/10.1016/j.medengphy.2006.05.012}}.
\newline\urlprefix\url{http://www.sciencedirect.com/science/article/pii/S135045330600107X}

\bibitem{Scuro2018}
N.~Scuro, E.~Angelo, G.~Angelo, D.~Andrade,
  \href{http://www.sciencedirect.com/science/article/pii/S0029549318300244}{A
  cfd analysis of the flow dynamics of a directly-operated safety relief
  valve}, Nuclear Engineering and Design 328 (2018) 321 -- 332.
\newblock \href
  {http://dx.doi.org/https://doi.org/10.1016/j.nucengdes.2018.01.024}
  {\path{doi:https://doi.org/10.1016/j.nucengdes.2018.01.024}}.
\newline\urlprefix\url{http://www.sciencedirect.com/science/article/pii/S0029549318300244}

\bibitem{Jin2017}
Y.~Jin, S.~Chai, J.~Duffy, C.~Chin, N.~Bose,
  \href{http://www.sciencedirect.com/science/article/pii/S0889974616307605}{Urans
  predictions of wave induced loads and motions on ships in regular head and
  oblique waves at zero forward speed}, Journal of Fluids and Structures 74
  (2017) 178 -- 204.
\newblock \href
  {http://dx.doi.org/https://doi.org/10.1016/j.jfluidstructs.2017.07.009}
  {\path{doi:https://doi.org/10.1016/j.jfluidstructs.2017.07.009}}.
\newline\urlprefix\url{http://www.sciencedirect.com/science/article/pii/S0889974616307605}

\bibitem{Hodis2012}
S.~Hodis, S.~Uthamaraj, A.~L. Smith, K.~D. Dennis, D.~F. Kallmes,
  D.~Dragomir-Daescu,
  \href{http://www.sciencedirect.com/science/article/pii/S0021929012004459}{Grid
  convergence errors in hemodynamic solution of patient-specific cerebral
  aneurysms}, Journal of Biomechanics 45~(16) (2012) 2907 -- 2913.
\newblock \href
  {http://dx.doi.org/https://doi.org/10.1016/j.jbiomech.2012.07.030}
  {\path{doi:https://doi.org/10.1016/j.jbiomech.2012.07.030}}.
\newline\urlprefix\url{http://www.sciencedirect.com/science/article/pii/S0021929012004459}

\bibitem{Dede2019}
L.~Ded{\`{e}}, F.~Menghini, A.~Quarteroni, Computational fluid dynamics of
  blood flow in an idealized left human heart, International Journal for
  Numerical Methods in Biomedical Engineering\href
  {http://dx.doi.org/10.1002/cnm.3287} {\path{doi:10.1002/cnm.3287}}.

\bibitem{Roache1997}
P.~J. Roache,
  \href{https://doi.org/10.1146/annurev.fluid.29.1.123}{Quantification of
  uncertainty in computational fluid dynamics}, Annual Review of Fluid
  Mechanics 29~(1) (1997) 123--160.
\newblock \href
  {http://arxiv.org/abs/https://doi.org/10.1146/annurev.fluid.29.1.123}
  {\path{arXiv:https://doi.org/10.1146/annurev.fluid.29.1.123}}, \href
  {http://dx.doi.org/10.1146/annurev.fluid.29.1.123}
  {\path{doi:10.1146/annurev.fluid.29.1.123}}.
\newline\urlprefix\url{https://doi.org/10.1146/annurev.fluid.29.1.123}

\bibitem{Scheidegger1974}
A.~Scheidegger, The Physics of Flow Through Porous Media, University of Toronto
  Press, 1974.

\bibitem{Ochoa2009}
I.~Ochoa, J.~A. Sanz-Herrera, J.~M. Garcia-Aznar, M.~Doblare, D.~M. Yunos,
  A.~R. Boccaccini, Permeability evaluation of 45s5 bioglass-based scaffolds
  for bone tissue engineering, Journal of Biomechanics.

\end{thebibliography}
\appendix
\section{Validation Study CFD}
In collaboration with the Medical University of Vienna, Austria we conducted an experimental validation study to show the correctness of our in-house CFD solver.
\subsection{Experimental PIV Setup}
\label{appendix:piv:exp}
A planar Particle Image Velocimetry (PIV) system (Dantec Dynamics, Skovlunde, Denmark) was used to capture flow patterns within a transparent human aortic block model (root diameter $\SI{19.5}{\milli\metre}$) created using a lost core technique \cite{Stoiber2013}. 
Three fields of view were used: A global view of the aortic arch and two vertical close-up views of the aortic root. 
For controlled inflow conditions into the model a straight inflow cross section (diameter $d=\SI{30}{\milli\metre}$, length $l=\SI{670}{\milli\metre}\approx 22d$) was applied. 
A continuous (Medtronic HVAD, Medtronic Inc., Dublin) and a piston pump (SuperPump, ViVitro Labs Inc., Victoria, BC, Canada) were used to generate stationary and pulsating inflows.
A 40\% glycerol-water mixture (density $\rho = \SI{1060}{\kilo\gram\per\cubic\metre}$ , dynamic viscosity $\mu = \SI{4.0e-3}{\pascal\second}$) seeded with PIV particles (PSP-20, medium diameter $\SI{20}{\micro\metre}$, polyamide 12, Dantec Dynamics A/S, Skovlunde, Denmark; seeding density 10-25 particles/interrogation area) was used as blood mimicking fluid. 
The particles inside the article were illuminated by a pulsed laser (NANO L 20-100 PIV Nd:YAG double oscillator laser system, Litron Lasers, Rugby, UK), and the image data was recorded with a high speed camera (SpeedSense9020 ,Vision Research, NJ, USA) with a resolution of 1152 x 896 pixels at a rate of $\SI{50}{\Hz}$. 
\begin{figure}[hbtp]
\begin{center}
\includegraphics[width=\textwidth,keepaspectratio]{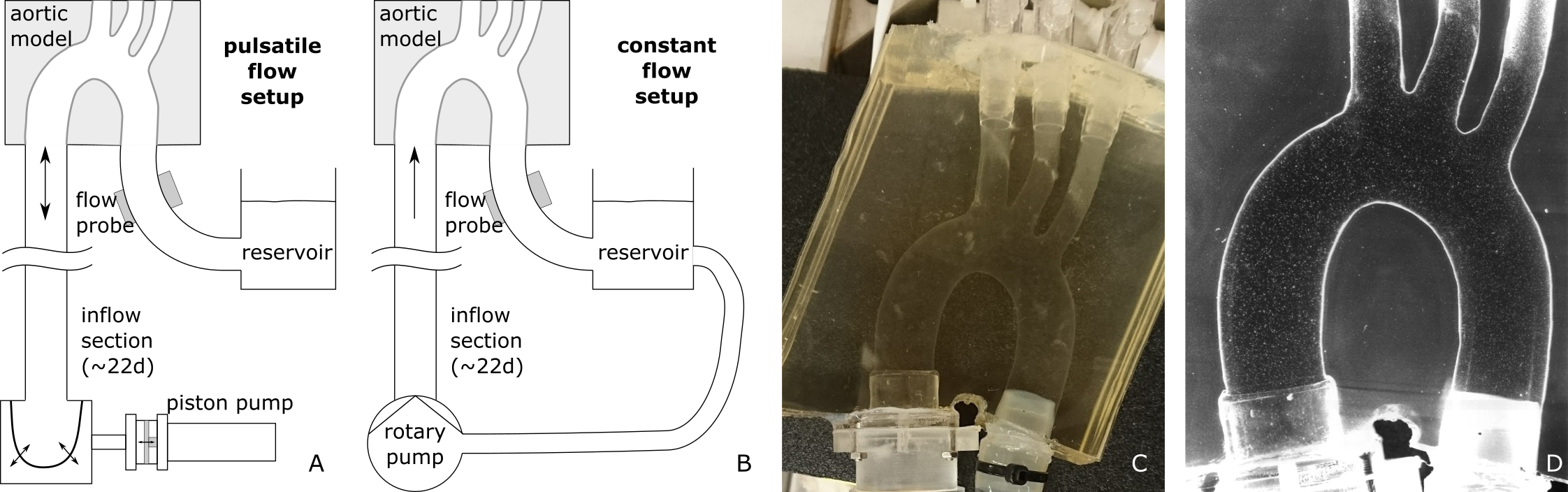}
\end{center}
\caption{Sketch of the pulsatile (A), continuous (B) flow setup used for PIV measurements and images of the actual model (C) and during PIV measurement (D)}\label{fig:akh:piv_setup}
\end{figure}

The steady state flow was averaged over $\SI{4}{\second}$ and for pulsating flows a phased averaging over 12 beats (100 images per beat) was applied. The timing of the laser pulses was individually selected based on the one-quarter displacement rule \cite{raffel2018particle} with values in the range of \SIrange{200}{1400}{\micro\second}. 
Flow velocity calculations were performed in DynamicStudio (v3.41, Dantec Dynamics, Skovlunde, Denmark) using adaptive correlation algorithms with a 32x32 pixel interrogation area and 50\% overlap. The vector maps were exported to Matlab (The MathWorks, Inc. Natick, MS, USA) for further analysis.
A controller board (DS1103 PPC Controller Board, dSPACE GmbH, Paderborn, Germany) was used to record hemodynamic parameters at $\SI{100}{\Hz}$. 
Disposable pressure transducers (TruWave, Edwards Lifesciences LLC, Irvine, CA, USA) and transducer amplifiers (TAM-A, Hugo Sachs Elektronik - Harvard Apparatus GmbH, March-Hugstetten, Germany) were used to record pressures; ultrasonic transit time clamp on flowmeters (Transonic H16XL flow probes, HT110R flowmeters, Transonic Systems Inc., Ithaca, NY, USA) were used to measure flows.

\begin{table}[h]
\caption{Reynolds numbers and flow rates of the applied conditions. Last two cases were only available under pulsatile conditions.}
\label{tab:AKH:cond}
\centering
\begin{tabular}{c|SS}
\toprule
\textbf{Reynolds Number} & \textbf{Peak flow} [\si{\litre\per\minute}] & \textbf{Piston Pump Stroke Volume} [\si{\milli\litre}] \\
\midrule
$750$ & 1.8 &  4.0  \\
$1500$ & 3.5 & 8.7  \\
$2000$ & 4.8 & 11.7 \\
$2300$ & 5.4 & 13.9 \\
$3000$ & 7.3 & 18.4 \\
$4000$ & 9.5 & 24.9 \\
$6000$* & 14.2 & 37.1 \\
$8000$* & 19.1 & 50.4 \\
\bottomrule
\end{tabular}
\end{table}

\subsection{Numerical Validation}
\label{sec:akh:numeric}
\begin{figure}[ht]
    \centering
    \includegraphics[width=\textwidth, keepaspectratio]{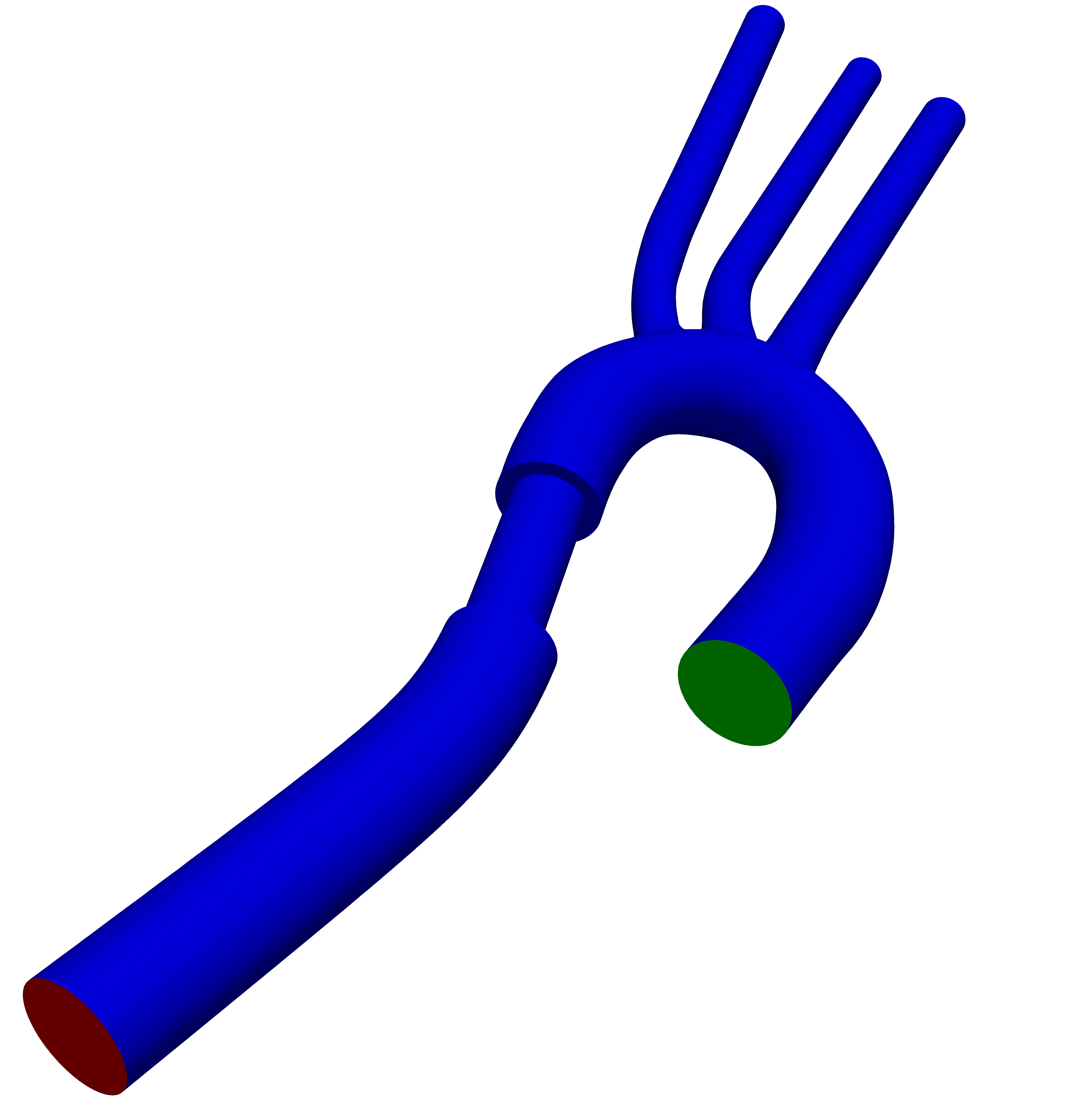}
    \caption{Geometry of the PIV benchmark study. No-slip boundaries are colored blue, inhomogenous Dirichlet surface is colored red, and directional do-nothing boundary is colored green.}
    \label{fig:AKH:geometry}
\end{figure}

\begin{figure}[ht]
\begin{subfigure}{0.5\textwidth}
  \centering
  \includegraphics[width=\linewidth]{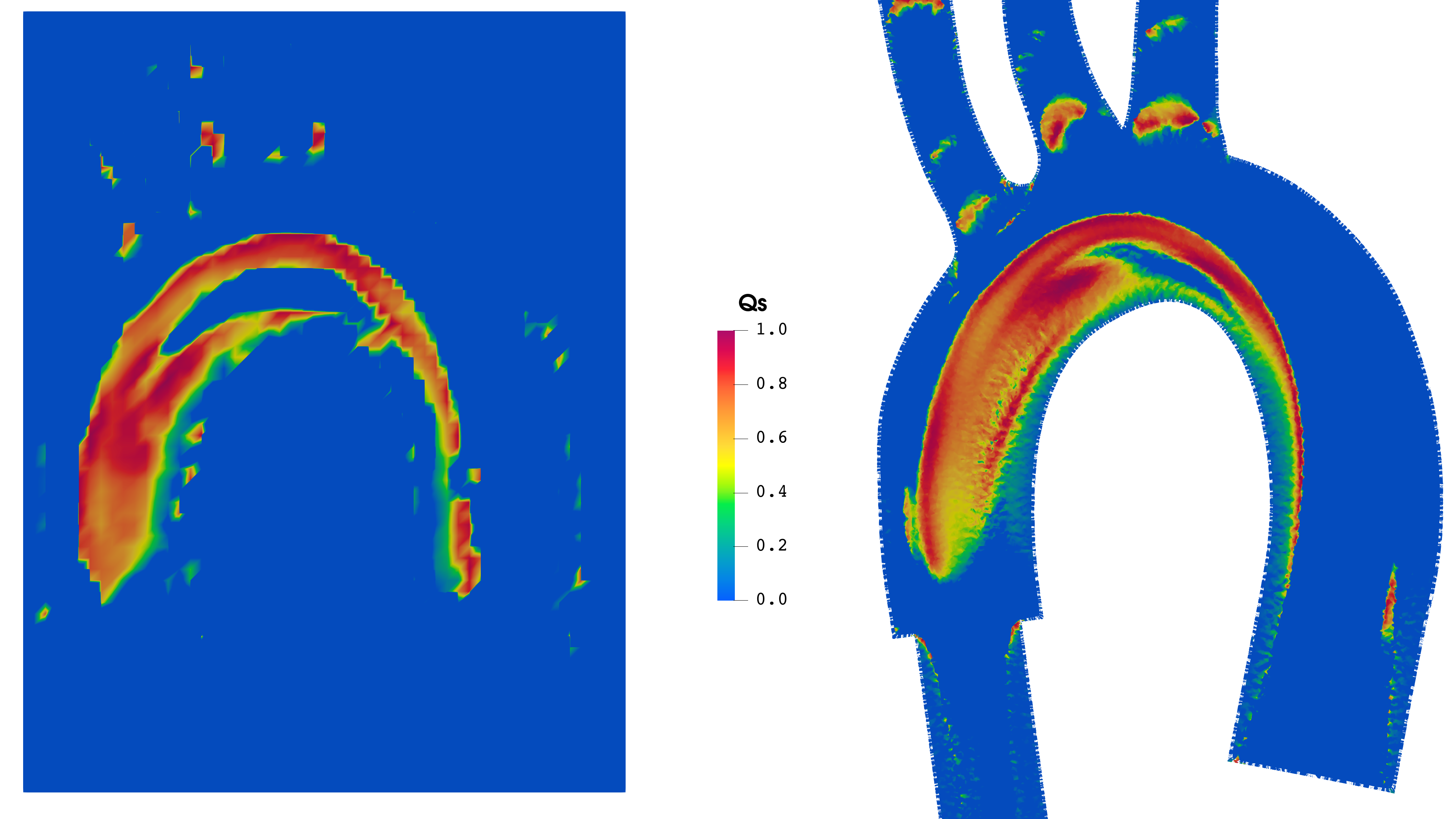}  
  \caption{$\mathrm{Re} = 750$}
  \label{fig:AKH:0750}
\end{subfigure}
\begin{subfigure}{0.5\textwidth}
  \centering
  \includegraphics[width=\linewidth]{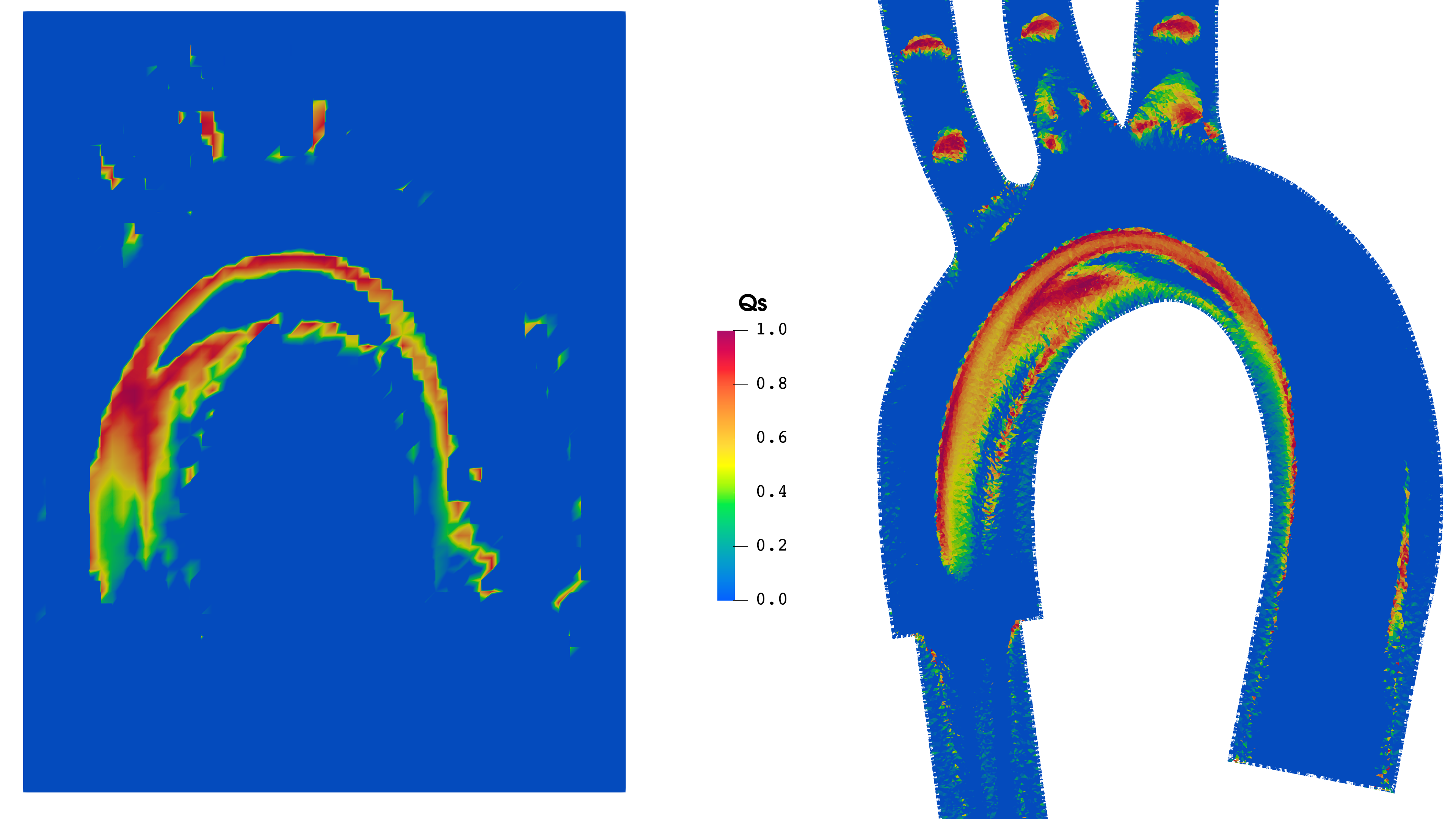}  
  \caption{$\mathrm{Re} = 1500$}
  \label{fig:AKH:1500}
\end{subfigure}
\begin{subfigure}{0.5\textwidth}
  \centering
  \includegraphics[width=\linewidth]{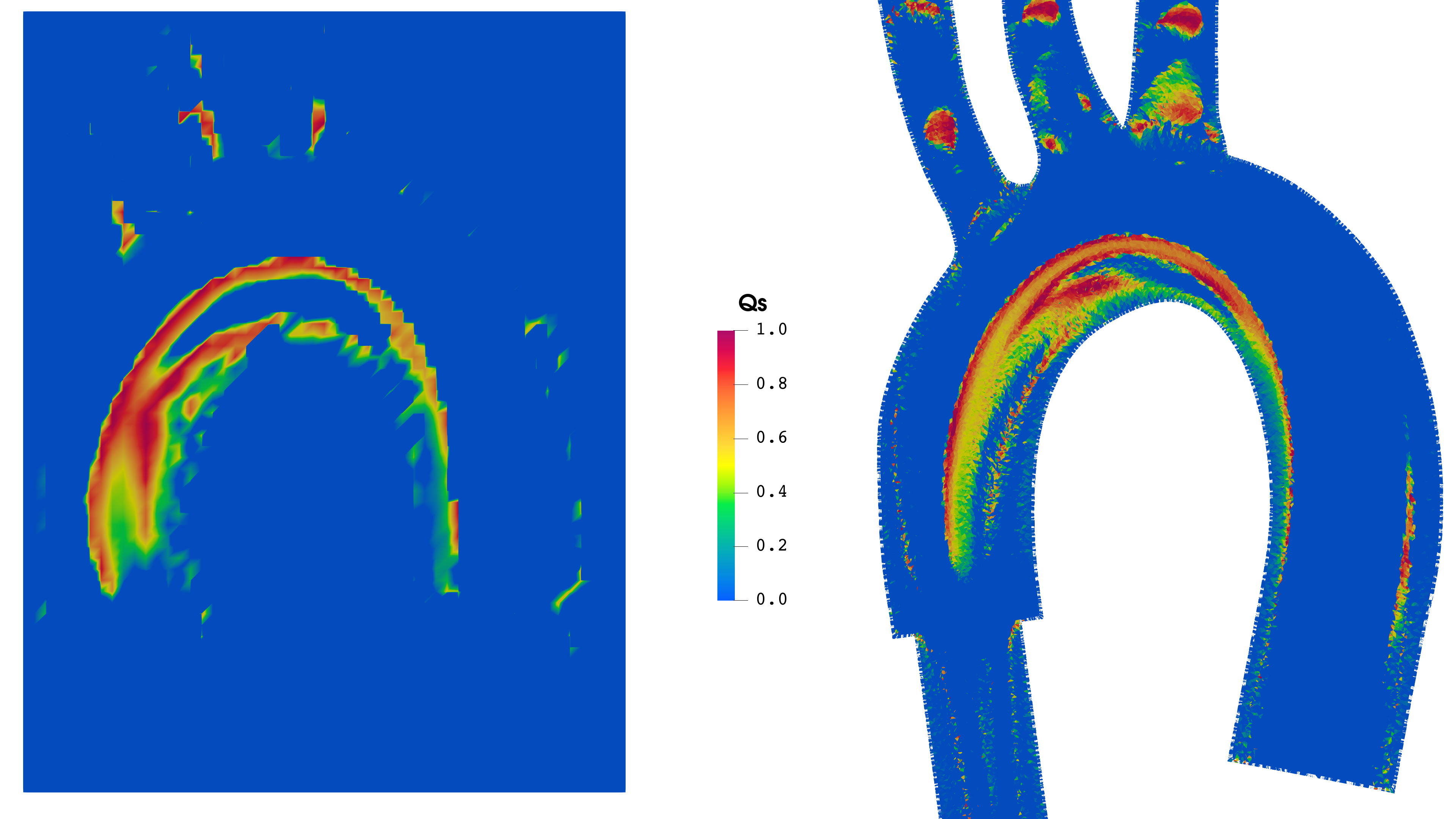}  
  \caption{$\mathrm{Re} = 2000$}
  \label{fig:AKH:2000}
\end{subfigure}
\begin{subfigure}{0.5\textwidth}
  \centering
  \includegraphics[width=\linewidth]{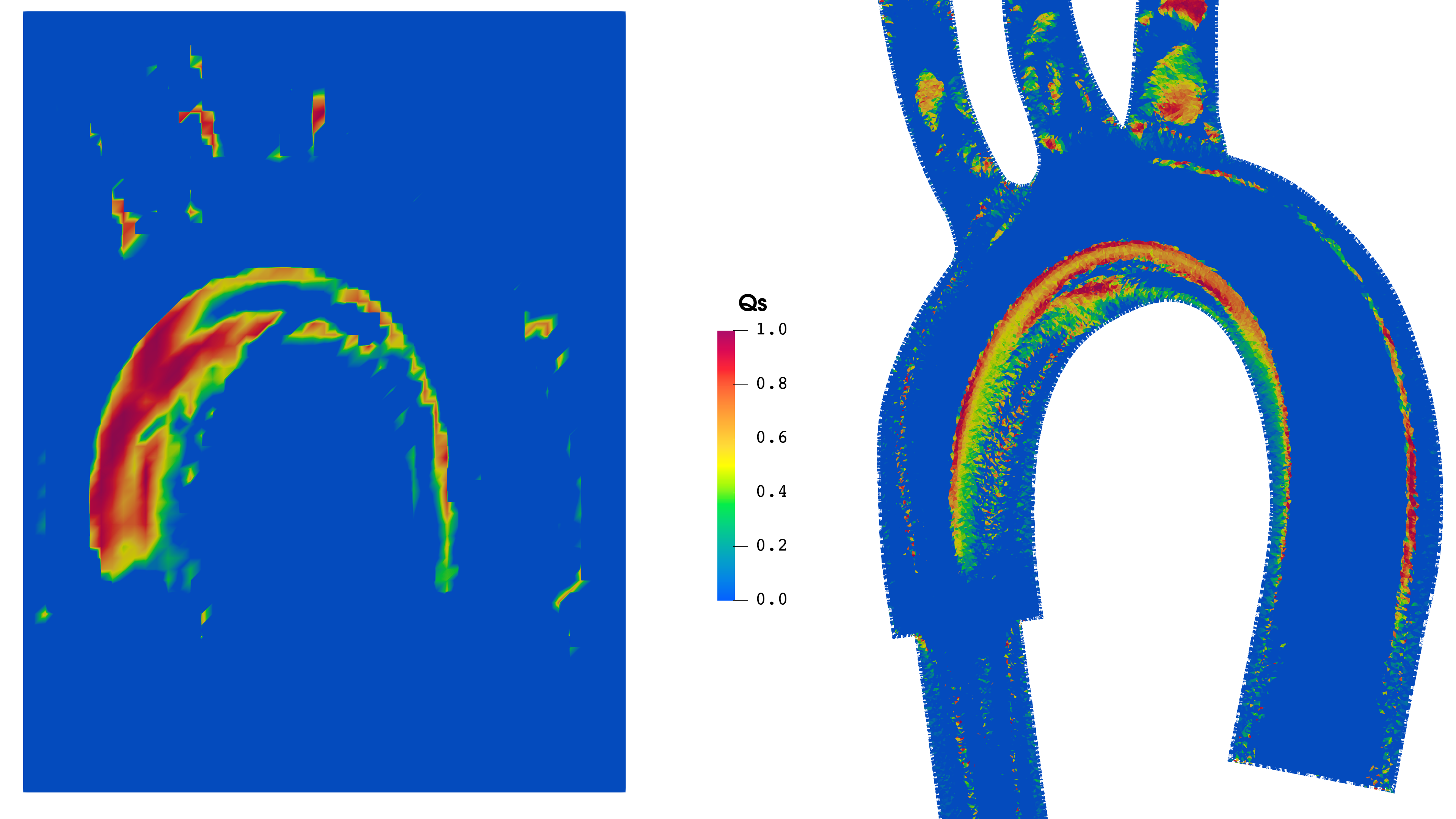}  
  \caption{$\mathrm{Re} = 3000$}
  \label{fig:AKH:3000}
\end{subfigure}
\begin{subfigure}{0.5\textwidth}
  \centering
  \includegraphics[width=\linewidth]{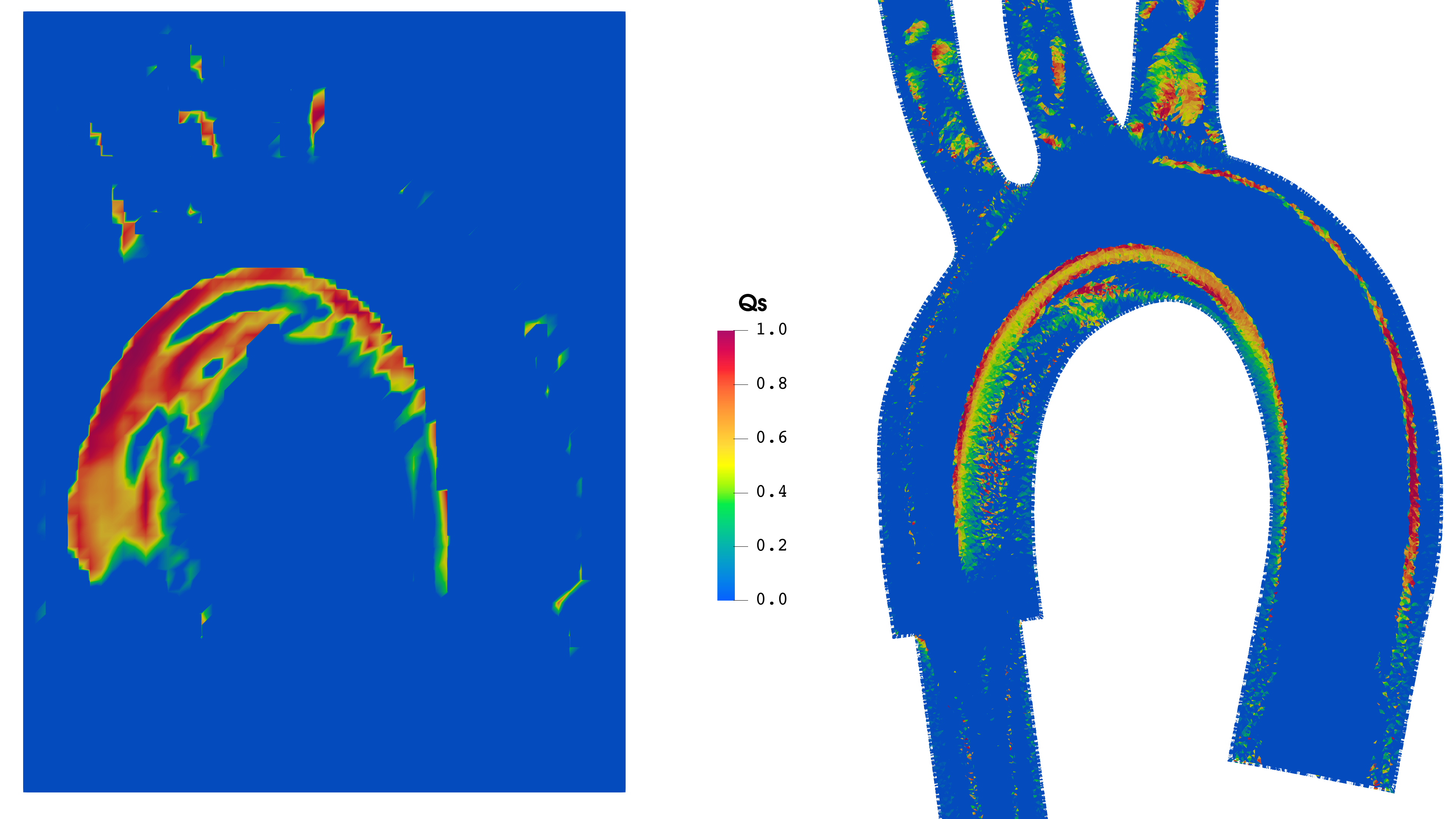}  
  \caption{$\mathrm{Re} = 4000$}
  \label{fig:AKH:4000}
\end{subfigure}
\begin{subfigure}{0.5\textwidth}
  \centering
  \includegraphics[width=\linewidth]{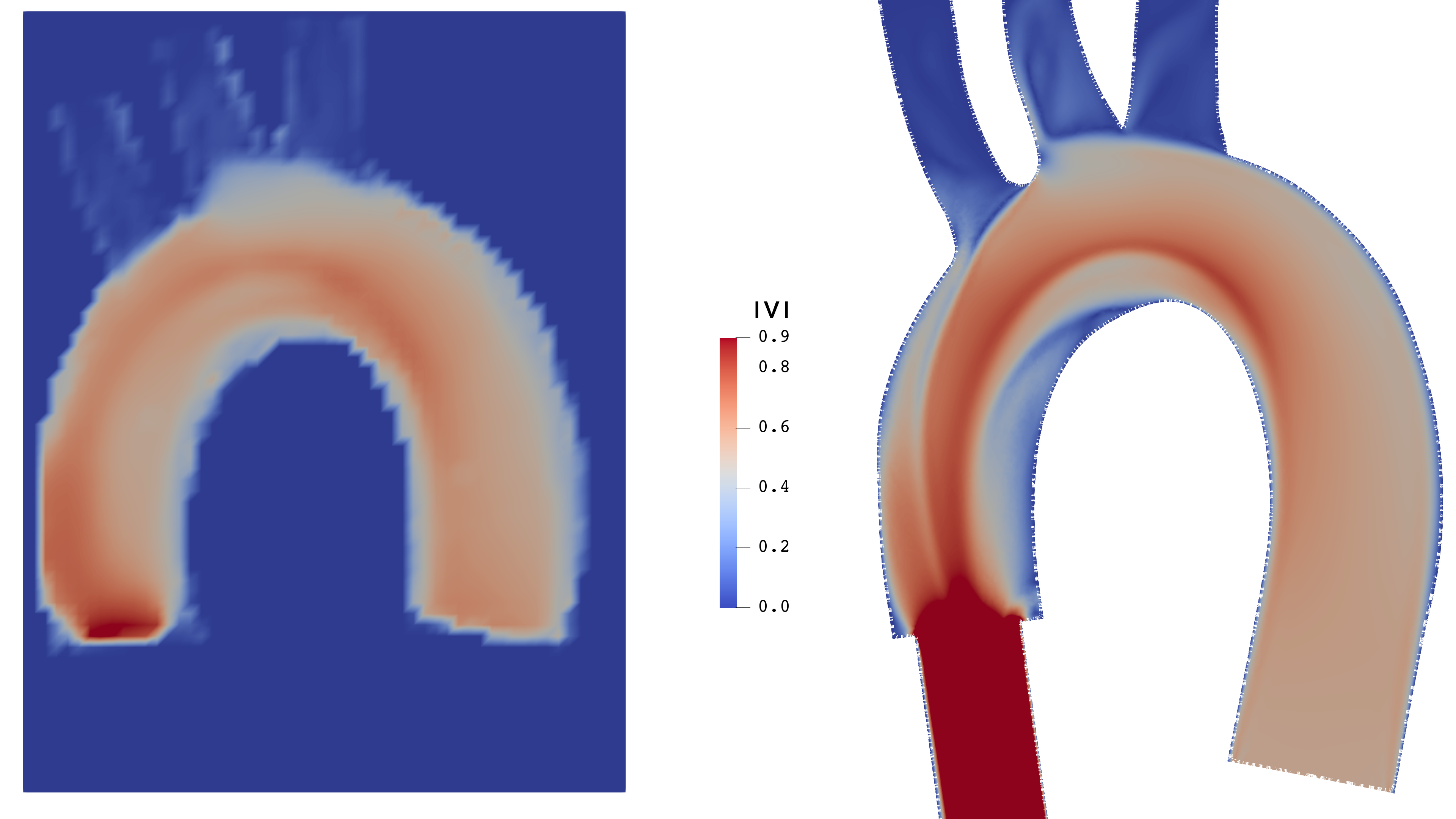}  
  \caption{Comparison of velocity magnitude for $\mathrm{Re} = 4000$.}
  \label{fig:AKH:4000:velcomp}
\end{subfigure}
\caption{Quantitative comparison between PIV data and numerical simulation of scaled $Q$-criteria for stationary cases.}
\label{fig:AKH:results}
\end{figure}

We created a virtual setup with the provided CAD files used for 3D printing the experimental setup in Section\,\ref{appendix:piv:exp}.
The geometry is depicted in Figure\,\ref{fig:AKH:geometry}.
The parameters, density $\rho =\SI{1060}{\kilo\gram\per\cubic\metre}$, and viscosity $\mu =\SI{4.0e-3}{\pascal\second}$, for the CFD simulation were chosen in agreement with the PIV experiment.
The boundary conditions were chosen according with the experimental setup, see color coding in Figure\,\ref{fig:AKH:geometry}.
For the stationary cases a parabolic inflow profile was chosen as inflow boundary condition. 
For the transient cases two dimensional inflow profiles were recovered by polar interpolation of the one dimensional velocity data sampled from the two orthogonal PIV imaging planes at the inflow region as described in \cite{Fuchsberger2019}.
A mixed mesh consisting of tetrahedral and prismatic elements with an average edge length of $\SI{0.25}{\milli\metre}$ was generated with \emph{Meshtool} from the provided CAD corresponding to $\approx 1200k$ nodes and $\approx 7000k$ elements. 
Both stationary and transient simulations simulations were ran from $t = 0$ to $t = \SI{2}{\second}$ with a time step size of $\Delta t = \SI{0.5}{\milli\second}$.
In the stationary cases the inflow rate was ramped up to its peak value, see Table~\ref{tab:AKH:cond}, and then left constant.
In the transient cases the inflow profile was scaled to match the measured flow rate values and 4 periods were simulated.
Simulation times for all experiments ranged at maximum up to $\SI{24}{\hour}$. 
All simulations were executed on VSC4 using 1200 MPI processes with an average computation time per time step of $\approx \SI{18}{\second}$.
As the computational geometry includes a very sudden narrowing, stationary inflow conditions with Reynolds numbers 2000 and higher displayed non stationary behavior after the sudden expansion of the artery model.
However for the PIV comparison only flow phenomena occurring in the physiological part of the model were studied.
Here, all stationary inflow conditions converged to a stationary flow field.
Due to limited resolution and the reduced dimensionality of measured data a quantitative validation of the transient cases remained inconclusive.
While computed flow patterns appear plausible and showed qualitative agreement with measurements, a full quantitative analysis requires further research.
In the supplementary material we provide videos display the temporal evolution of isosurfaces for the scaled $Q$-criterion for all pulsatile conditions.
Figures \ref{fig:AKH:0750}--\ref{fig:AKH:4000} show an eyeball comparison of the scaled $Q$-criterion \cite{Kao2015} for the stationary flow case defined as $Q_\mathrm{s} := \frac{1}{2}(\frac{\norm{\tensor\Omega}{\mathrm{F}}^2}{\norm{\tensor S}{\mathrm{F}}^2}+1)$ showing a good agreement for the lower Reynolds numbers and deteriorating with increasing inflow, see also \cite{Fuchsberger2019}.
Yet, as can be seen in Figure \ref{fig:AKH:4000:velcomp}, the simulated velocity profiles follow the trend of the PIV data.

\section{Methods for Studying Mesh Convergence}\label{sec:appendix:meshconv}

In this section we summarize the most important methods used for assessing mesh convergence.
This methods are well known and widely used in the CFD community, see for example \cite{Longest2007,Scuro2018,Jin2017,Hodis2012,Chnafa2014,Dede2019}.

Following \citep{Stern2001}, we use the convergence ratio $R$, \eqref{eq:R}, to guarantee mesh independence.
For defining $R$ one needs three mesh refinement levels.
Here we use the index $0$ for the coarse mesh, $1$ for the medium mesh, and $2$ for the fine mesh.
\begin{equation}
R(t):=\frac{\varepsilon_{12}(t)}{\varepsilon_{01}(t)}\label{eq:R},
\end{equation}
where
\begin{equation}
\varepsilon_{ij}(t):=S_{m_i}(t)-S_{m_j}(t). \label{eq:epsilon}.
\end{equation}
In \eqref{eq:epsilon}, $S_{m_i}$ stands for the quantity used to study mesh convergence. 
This includes any derived variable from a CFD simulation on mesh refinement level $m_i$, like flux, pressure drop at a certain location, but can also be used with the primal fields directly.
In the latter, \eqref{eq:R} is calculated for every point in space yielding a convergence ratio field $R(t,\vec{x})$.
Point-wise calculation of $R$ can be problematic, as $\varepsilon_{12}$ and $\varepsilon_{01}$ can both go to zero. 
To handle this problem \citep{Stern2001} proposes two strategies:
\begin{enumerate}
    \item Consider $R(t,\vec{x})$ only in regions, where the solution changes are both non-zero.
    \item Utilize the global convergence ratio $\langle R (t)\rangle$
    \begin{align}
        \langle R(t) \rangle &=\frac{\norm{\varepsilon_{12}(t)}{2}}{\norm{\varepsilon_{01}(t)}{2}} \label{eq:Rglob}
    \end{align}
\end{enumerate}

The interpretation of $R$ and $\langle R \rangle$ is noted in \eqref{eq:Rinterpretation2}, however one has to keep in mind, that oscillatory convergence cannot be differentiated from monotonic convergence with $\langle R \rangle$.
\begin{equation}
    \begin{array}{cl}\label{eq:Rinterpretation2}
    0<R<1 & \hspace{1cm}\text{monotonic convergence} \\
    -1<R<0 & \hspace{1cm}\text{oscillatory convergence}\\
    \text{else} & \hspace{1cm}\text{divergence}
    \end{array}
\end{equation}

For a constant refinement ratio $r$ one can calculate the order of grid convergence following \citep{deVahlDavis1983}:
\begin{equation}
    \alpha(t):=\nicefrac{\mathrm{ln} \left|\frac{\varepsilon_{01}(t)}{\varepsilon_{12}(t)} \right|}{\mathrm{ln}(r)}\label{eq:alpha} 
\end{equation}
For unstructured grids, the refinement ratio $r$ may be replaced by the effective refinement ratio $r_\mathrm{eff}$, see \citep{Roache1994}, which is calculated from the total number of elements $N_{m_i}$ of the respective grids $m_i$ in the following fashion:
\begin{equation}
    r_\mathrm{eff}:= \left( \frac{N_{m_{i+1}}}{N_{m_{i}}}\right)^{\frac{1}{D}},\label{eq:reff}
\end{equation}
where $D$ is the space dimension.

Furthermore, one would like to give an error estimate concerning the mesh resolution error.
To this end, we use the well known grid convergence index (GCI) following \citep{Roache1994}, which is calculated from two different mesh resolutions using the relative solution change $\tilde{\varepsilon}$
\begin{align}
    \mathrm{GCI}_{i,i+1}(t) &= F_s\frac{\tilde{\epsilon}_{i,i+1}(t)}{r^{\alpha(t)} -1}, \label{eq:GCI} \\
    \tilde{\epsilon}_{ij}(t)&= \left|\frac{\epsilon_{ij}(t)}{S_{m_{j}}(t)}\right| \label{eq:epsrel}.
\end{align}
Roache introduces the fudge factor $F_s$ and recommends a value of $F_s=1.25$, for mesh convergence studies using three or more mesh resolutions, see \citep{Roache1997}.

It is important to keep in mind, that the GCI is certainly not a bound on the error which cannot be exceeded, but a tolerance on the accuracy in which one may have a practical level of confidence, as Roache puts it in \citep{Roache1994}. 

\subsection{Pope's Criterion as a Measure of Turbulence Resolution}\label{sec:pope_crit}
In LES type formulations the resolved velocity field is fundamentally linked to the numerical method used, hence there is no notion of convergence to the solution of a partial differential equation \citep{Pope2004}.
This leads to the problem that mesh convergence often cannot be established by the methods discussed above for turbulent flow.
To remedy this problem \citep{Pope2004} proposes the use of a measure of turbulence resolution $M$, see \eqref{eq:TKEfrac}, utilizing the fraction of turbulent kinetic energy resolved by the grid in question.
In order to obtain a point-wise measure, rather than the kinetic energy itself, the kinetic energy density $K(\bm{x},t)$ is considered:
\begin{equation}
    K(\bm{x},t)=\frac{\rho}{2}\bm{u}(\bm{x},t)^2. \label{eq:KE}
\end{equation}
The resulting point-wise measure of turbulence resolution $M$ reads:
\begin{equation}
    M(\bm{x},t)= \frac{K'(\bm{x},t)}{K_{tot}(\bm{x},t)}. \label{eq:TKEfrac}
\end{equation}
Here $K'$ is the turbulent kinetic energy of the residual motions, hence of the motions not resolved by the grid, and $K_{tot}$ stands for the total kinetic energy.
$K_{tot}$ may be written as the sum of the resolved turbulent kinetic energy $K^h$ and the not resolved turbulent kinetic energy $K'$:
\begin{equation}
    K_{tot}(\bm{x},t)=K^h(\bm{x},t)+K'(\bm{x},t)
\end{equation}
The resolved turbulent kinetic energy $K^h$ is calculated from \eqref{eq:KE} using the fluctuating part of the resolved fluid velocity $\bm{u}_f$, which is given by:
\begin{equation}
    \bm{u}_f(\bm{x},t)=\overline{\bm{u}^h}(\bm{x},t)-\bm{u}^h(\bm{x},t),
\end{equation}
where $\overline{\bm{u}^h}$ is an average with respect to time.
When considering a constant inflow $\overline{\bm{u}^h}$ is given by the standard mean over all time steps ($t=1 \dots T$, hence $\overline{\bm{u}^h}$ is not time-dependent:
\begin{equation}
    \overline{\bm{u}^h}(\bm{x})=\frac{1}{T}\sum_{t=1}^{T}\bm{u}^h(\bm{x},t)
\end{equation}
In the case of a pulsatile inflow however a phase average is considered:
\begin{equation}
    \overline{\bm{u}^h}(\bm{x},t)=\frac{1}{n}\sum_{k=0}^{n-1}\bm{u}^h(\bm{x},t+k \tau),
\end{equation}
where $n$ is the number of cycles and $\tau$ is the period.
By the use of \eqref{eq:TKEfrac} a criterion for sufficient mesh resolution is given:
\begin{equation}
       M(\bm{x},t) \leq \epsilon_M
\end{equation}
In \cite{Pope2004} a value of $\epsilon_M=0.2$ is proposed, which corresponds to requiring a minimum of 80\% of the total turbulence energy to be resolved. 

\section{Computational Details Regarding Obstacle Representation}
\label{sec:appendix:pseudocode}
Algorithm\,\ref{algo_volfracs} schematically shows the calculation of the volume fraction distribution $v_f(t,\tau)$ for one point in time. 
To capture moving obstacles, this procedure is repeated for every time-step.
Note that, for a moving obstacle we need knowledge of the obstacle velocity $u_s$, that can be computed from consecutive positions of the obstacle surface in a straight forward manner. 
Due to the assumption of a rigid obstacle, the interior velocity values can be obtained using radial basis function interpolation \cite{Lazzaro2002}.
{\scriptsize
\begin{algorithm}
\SetKwData{Mesh}{mesh}\SetKwData{Surfmesh}{surfmesh}\SetKwData{Fracs}{fractions}
\SetKwData{Tree}{tree}
\SetKwData{Inside}{inside}
\SetKwData{Outside}{outside}
\SetKwData{Split}{split}
\SetKwData{Enclosed}{enclosed\_volume}
\SetKwData{Volume}{volume}
\SetKwData{Subtets}{subtets}
\SetKwData{InsideN}{inside\_nodes}
\SetKwData{OutsideN}{outside\_nodes}
\SetKwFunction{TetEnclosedVolume}{TetEnclosedVolume}
\SetKwFunction{FindCompress}{FindCompress}
\SetKwFunction{FillKDtree}{FillKDtree}
\SetKwFunction{Classify}{ClassifyElements}
\SetKwFunction{TetVolume}{TetVolume}
\SetKwFunction{ElemVolume}{ElemVolume}
\SetKwFunction{SplitElem}{SplitElem}
\SetKwFunction{InsideOutsideNodes}{InOutNodes}
\SetKwInOut{Input}{input}\SetKwInOut{Output}{output}
\Input{Mesh \Mesh, surface mesh \Surfmesh}
\Output{Vector \Fracs containing a value $v_f$ for each mesh element}
Build a $k$-$d$ tree of the surface mesh elements\;
\Tree$\leftarrow$\FillKDtree{\Surfmesh}\;
Classifiy elements of \Mesh in inside, outside, or split\;
Inside elements get $v_f=1$, outside elements get $v_f=0$\;
\Classify{\Tree,\Mesh,\Inside,\Outside,\Split}\;
Loop over elements that are split by at least one surface element\;
\For{$\tau \in$ \Split}
{
    \If{$\tau=$ Tetrahedron}
    {
      Split nodes of $\tau$ into inside and outside nodes\;      {\InsideN,\OutsideN}$\leftarrow$\InsideOutsideNodes{$\tau$,\Inside,\Outside}\;
      Calculate enclosed volume with Möller-Trumbore ray-triangle intersection \cite{Moeller1997}\;
      \Enclosed$\leftarrow$\TetEnclosedVolume{\InsideN,\OutsideN,\Tree}\;
      \Volume$\leftarrow$\TetVolume{$\tau$}\;
    }
    \Else 
    {
      Split element into subtetrahedra and sum up the individual subvolumes\;
      \Subtets$\leftarrow$\SplitElem{$\tau$}\;
      \For{$\pi\in$ \Subtets}
      {
        {\InsideN,\OutsideN}$\leftarrow$\InsideOutsideNodes{$\pi$,\Inside,\Outside}\;
        \Enclosed$+=$\TetEnclosedVolume{\InsideN,\OutsideN,\Tree}\;
      }
      \Volume$\leftarrow$\ElemVolume{$\tau$}\;
    }
    \Fracs$[\tau]$ = \Enclosed / \Volume\;
}
\caption{volume\_fractions}\label{algo_volfracs}
\end{algorithm}
}

\color{red}
\section{\textcolor{red}{Testing Darcy Behavior in the Low Permeability Regime}}
\label{sec:appendix:darcytest}
In the following we examine the low permeability behavior of the NSB model. 
The theoretical analysis of the NSB equations \cite{Angot1999feb}
shows, that the Darcy drag term increasingly dominates the solution behavior as low values of permeability are approached.
Darcy's law is known to be valid for laminar flows and negligible inertia effects, for further information on the subject we refer to \citep{Scheidegger1974}. 
Therefore, it is important to choose an appropriate test case concerning the permeability value as well as Reynolds number in order to test low permeability characteristics of the NSB equations.
Fortunately, one of the most popular applications of Darcy's law is the experimental determination of permeability of a porous material, which makes for plenty of experimental validation data to choose from.
\subsection{Validation Experiment}
\begin{figure}[htp]
    \centering
    \includegraphics[scale=1]{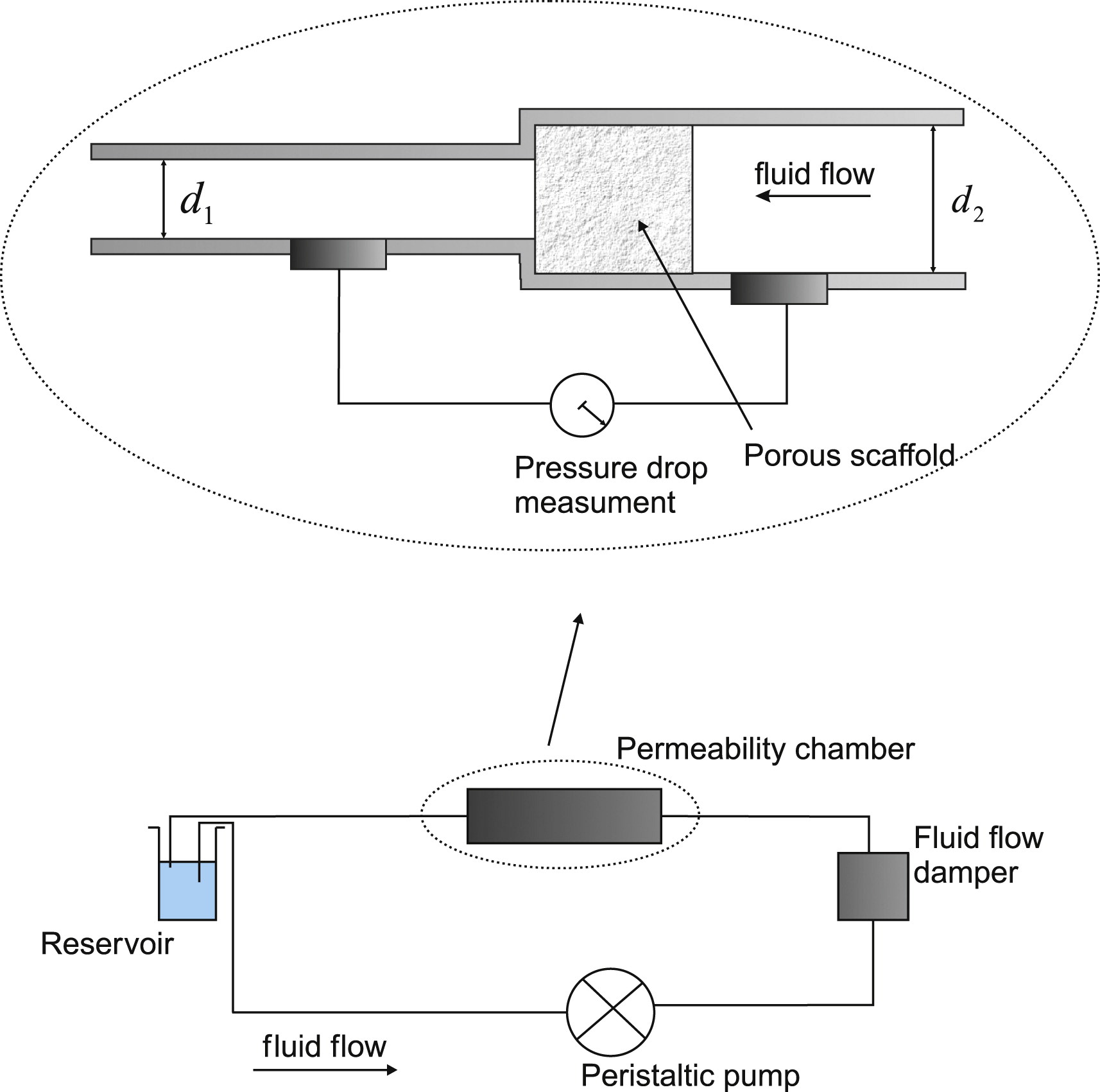}
    \caption{Experimental rig used by \cite{Ochoa2009}\footnotemark to measure permeability.}
    \label{fig:DarcyExperimentSetup}
\end{figure}

\begin{figure}[htp]
    \centering
    \includegraphics[scale=0.85]{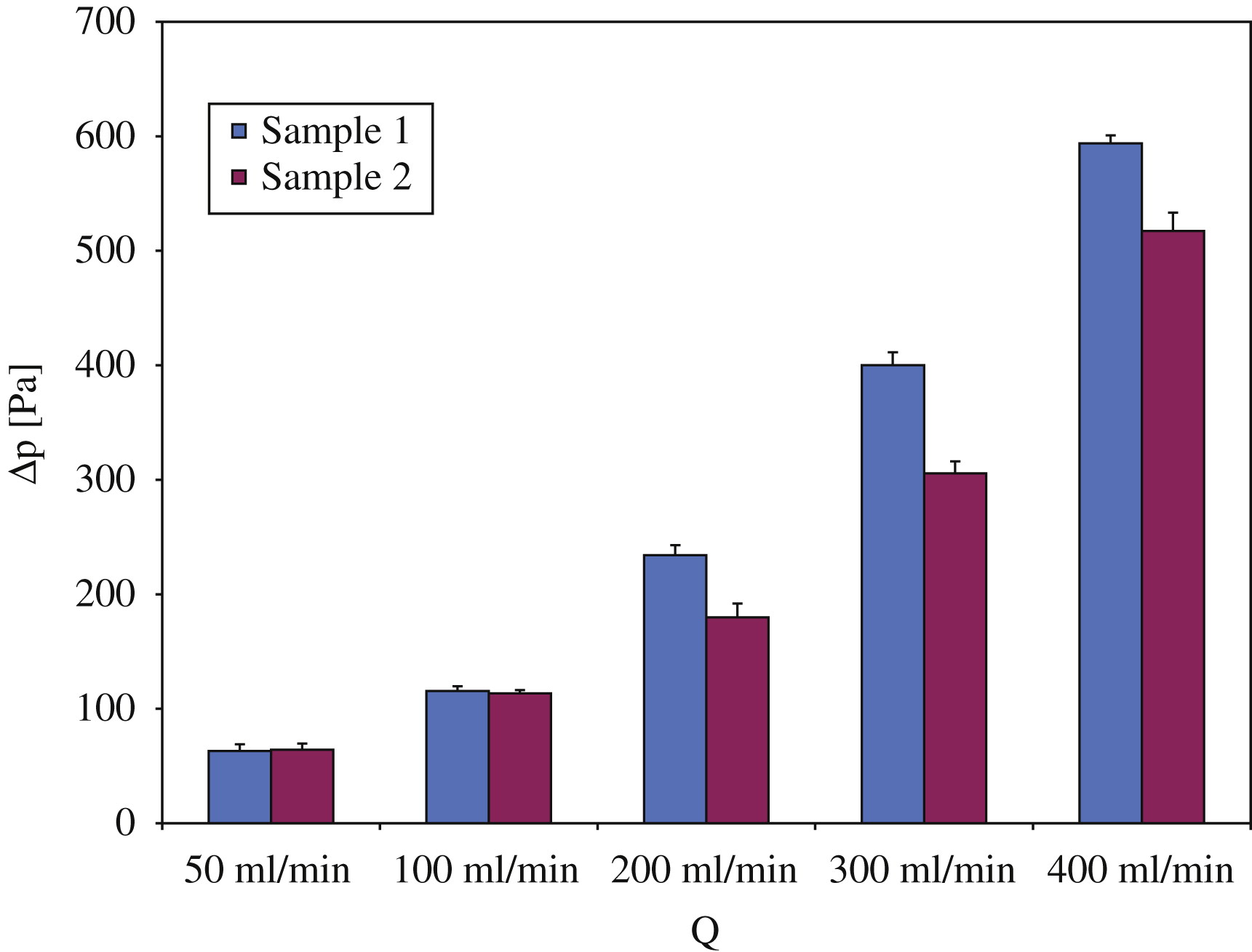}    
    \caption{Experimental pressure drop measurements for different inflow rates from \citep{Ochoa2009}\footnotemark.}
    \label{fig:DarcyExperimentResults}
\end{figure}

The experiment chosen for validation of the linear behavior in the Darcy regime was carried out in \cite{Ochoa2009}.
They set up an experimental rig (see \cref{fig:DarcyExperimentSetup}) to measure the permeability of 45S5 Bioglass\textregistered-based scaffolds for bone tissue engineering.
Two different samples of porous scaffold $s_1$ / $s_2$ of width $d=$ \SI{8}{\milli\metre} and thickness $d_{s_1}=$ \SI{8.52}{\milli\metre} / $d_{s_2}=$ \SI{8.11}{\milli\metre} respectively were placed in the permeability chamber.
A continuous flow of deionized water ($\rho=$ \SI{1e3}{\kilo\gram\per\metre\cubed}, \hspace{0.1cm} $\mu=$ \SI{1e-3}{\pascal\per\second}) was pumped trough the chamber employing different inflow rates ($Q =$ 50 / 100 / 200 / 300 / 400 \SI{}{\milli\litre\per\minute} ). 
Subsequently the pressure drop was measured and utilizing the integral form of Darcy's law  \cref{eq:DarcyIntegral} permeability estimates were computed from the experimental $\Delta p$-$Q$ curves.
\begin{equation} 
Q = -\frac{K A}{\mu L} \Delta p \label{eq:DarcyIntegral}
\end{equation}
The resulting permeability values were $K_{s_1} = $ \SI{1.85e-9}{\metre\squared} for sample one and $K_{s_2} = $ \SI{2.07e-9}{\metre\squared} for sample two.
The original pressure drop measurements are depicted in \cref{fig:DarcyExperimentResults}.
\footnotetext{Reprinted from \cite{Ochoa2009} with permission from Elsevier.}
\subsection{Computational Experiment and Validation Results}
\begin{figure}[htp]
    \centering
\begin{tikzpicture}[spy using outlines={black,circle,magnification=6,size=25*2,connect spies, style=thick}]
\node[inner sep=0.0] (pic) at (0,0){ \includegraphics[scale=0.04]{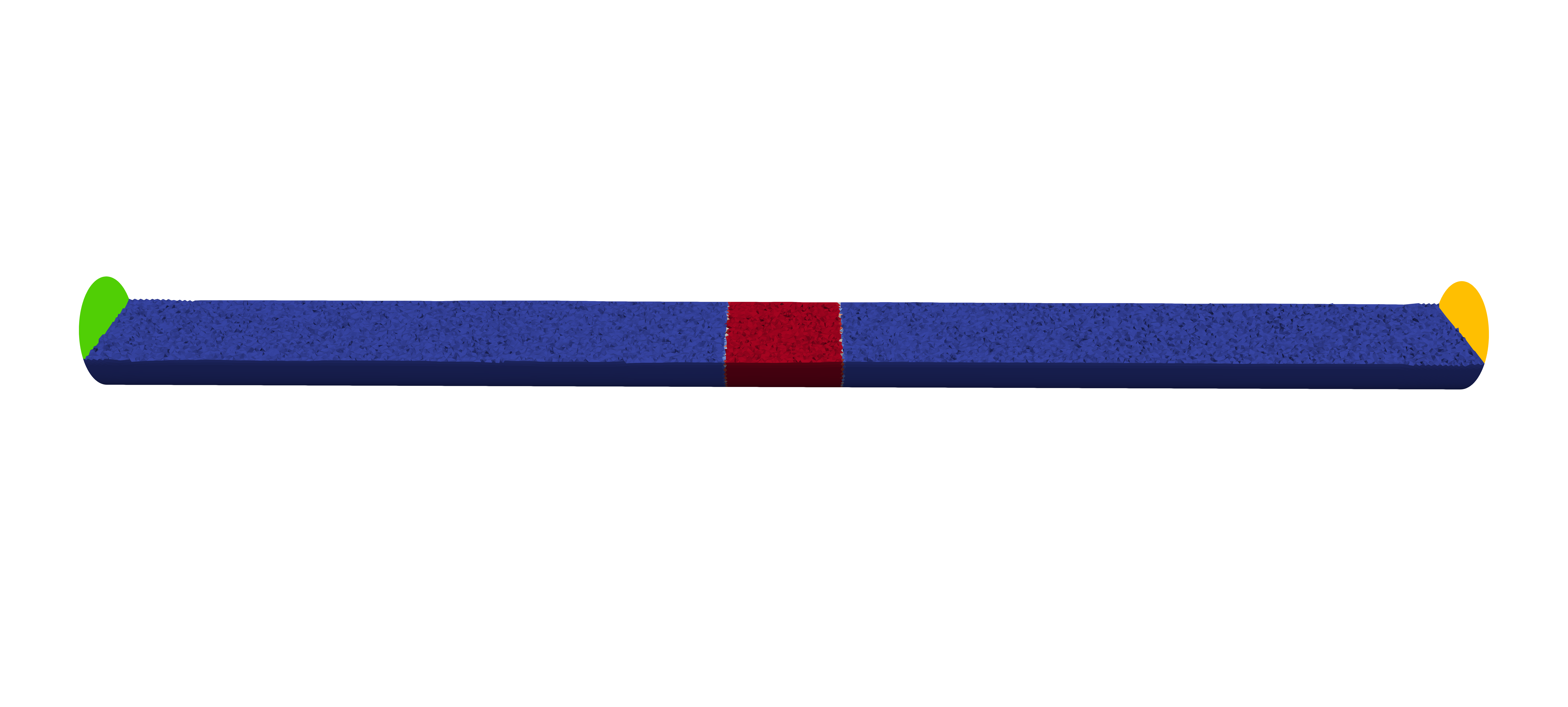}};
\spy on (0.5,0.2) in node at (2,2);
\node[draw=none,text=inflow] at (-5,1) {Inflow};
\node[draw=none,text=outflow] at (5,1) {Outflow};
\end{tikzpicture}
    \caption{In-silico permeability chamber containing a porous scaffold (red) with inflow (green) and outflow (orange) boundaries.}
    \label{fig:DarcySimulationSetup}
\end{figure}

To recreate the validation experiment an in-silico model of the permeability chamber was constructed yielding a finite element mesh containing approximately \SI{2.7e6}{} elements and \num{5e5}{} nodes, see \cref{fig:DarcySimulationSetup}.
The narrowing of the pipe used in the original experiment to hold the scaffold in place is not required for the simulation and therefore omitted in the model.
At the inflow a Dirichlet boundary condition was used to prescribe a constant inflow rate, while at the outflow a do-nothing outflow condition was used to emulate the free flow into the reservoir.
The NSB simulation was set up using all parameters matching the validation experiment and furthermore taking the experimentally obtained permeability values as permeability input parameters for both samples respectively. 
All calculations reached a steady state after a maximum of 15 time steps, wherein the computation time for one time step was 1.3 seconds on average.
\footnotetext{Reprinted from \cite{Ochoa2009} with permission from Elsevier.}
After a steady state was reached, measurements were taken from the simulated pressure field to determine the pressure drop over the porous scaffold.
The pressure measurement was carried out by taking an average over solution values from a randomly selected ensemble of nodes in the area of interest, which was a plane just in front of/behind the scaffold in the present case.

\begin{figure}[htp]
    \centering
    \includegraphics[width=\linewidth,trim=1.5cm 0.5cm 1.5cm 1.1cm, clip]{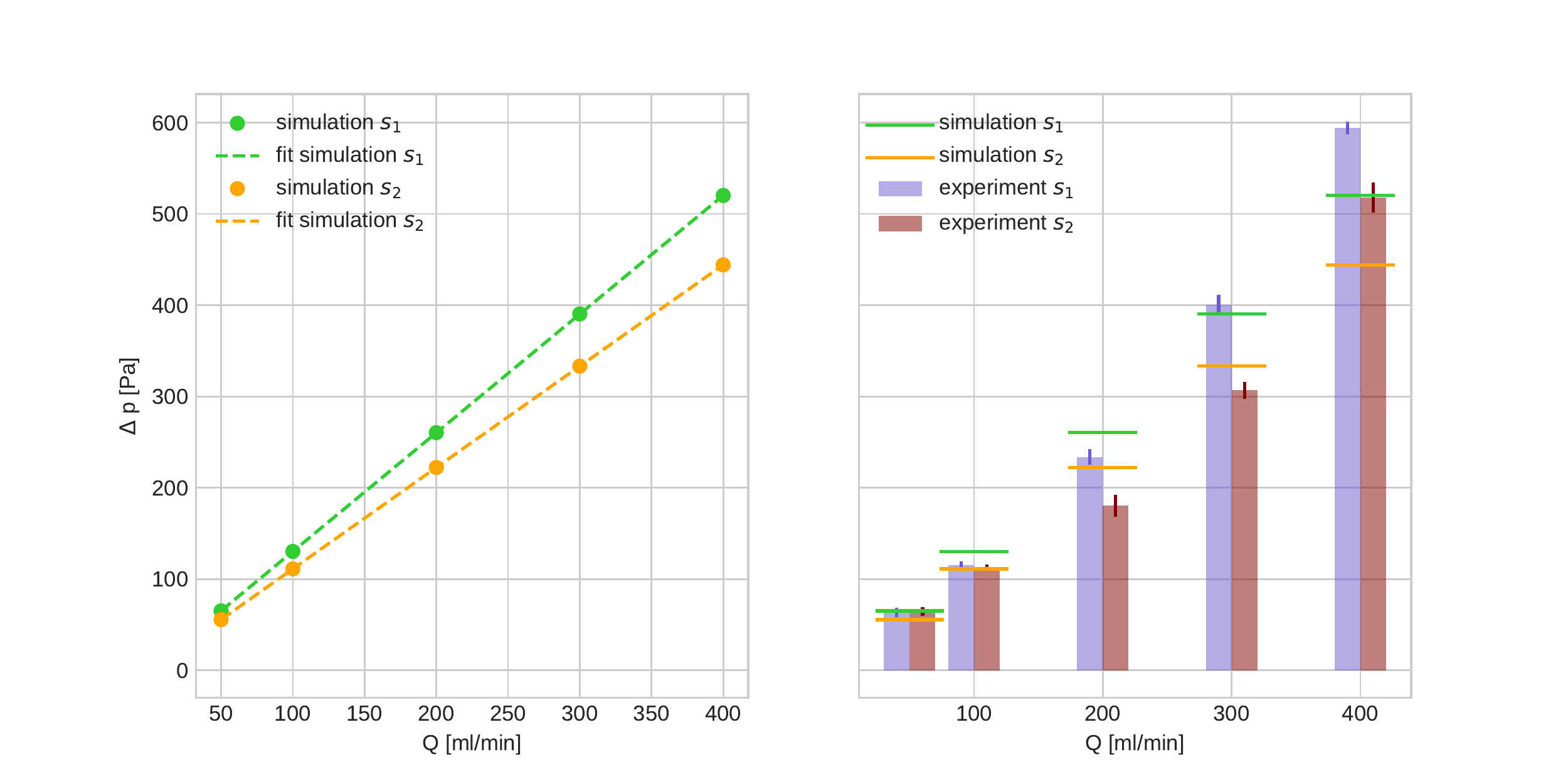}
    \caption{left: Simulated pressure drop versus inflow rate plotted together with a least squares linear fit for both samples. right: Comparison of simulated and experimentally measured \citep{Ochoa2009} pressure drop.}
    \label{fig:DarcySimulationResults}
\end{figure}

A least squares linear fit was performed using the simulated data points for both scaffold samples.
The resulting $\Delta p$--$Q$ curves are plotted in \cref{fig:DarcySimulationResults} and clearly show the linear relation between pressure drop and flow rate that was sought to find. 
Furthermore a comparison between simulation results and experimental data is given, that shows good agreement.
When compared by inspection one might suspect a slightly steeper slope of the experimental $\Delta p$--$Q$ curve compared to the slope of the simulated curve, which implies differing permeability values.
Estimating the permeability using Darcy's law \cref{eq:DarcyIntegral} from the simulated $\Delta p$--$Q$ curve yields permeability values, that are slightly higher than the experimentally found ones, which were used as input parameters.
This discrepancy can be interpreted as a measure for the dominance of the Darcy drag term over the inertia and diffusion contributions of the NSB model.
In practice, that means that the NSB model exhibits a slightly lower resistance to flow compared to the pure Darcy model using the same permeability value.

{
\color{red}
\begin{table}[htp]
    \centering
    \begin{tabular}{l|ccc}
        \toprule
         & Estimate from & Experiment / & rel. Error\\
         & Simulation Data &  Simulation Input & \\
        \midrule
        $K_{s_1} \; [\si{\meter\squared}]$ & \num{2.1e-9} & \num{1.85e-9} & \SI{12.65}{\percent}\\ 
        $K_{s_2} \; [\si{\meter\squared}]$ & \num{2.42e-9} & \num{2.07e-9}& \SI{15.59}{\percent}\\ 
        \bottomrule
    \end{tabular} 
    \caption{Comparison of permeability values estimated from simulation and the permeability values originating from the validation experiment \citep{Ochoa2009} for both samples respectively.}
    \label{tab:my_label}
\end{table}
}
\color{black}
\end{document}